\documentclass[iicol,sn-mathphys]{sn-jnl}

\usepackage{comment}
\usepackage{csquotes}
\usepackage{dsfont}
\usepackage{cleveref}
\usepackage{mathrsfs}
\usepackage{subcaption}
\usepackage{amsmath}
\usepackage{upgreek}
\usepackage[switch,columnwise]{lineno}
\usepackage{cuted}
\begin{document}

\title[Estimation of the number of counts on a particle counter detector with full time resolution]{Estimation of the number of counts on a particle counter detector with full time resolution}

\author*[1,2]{\fnm{Flavia} \sur{Gesualdi}}
\email{flavia.gesualdi@iteda.cnea.gov.ar}

\author[1]{\fnm{Alberto} \fnm{Daniel} \sur{Supanitsky}}

\affil*[1]{\orgname{Instituto de Tecnolog\'{i}as en Detecci\'{o}n y Astropart\'{i}culas (CNEA, CONICET, UNSAM)}, \orgaddress{\street{Av. General Paz 1499}, \city{San Mart\'{i}n}, \postcode{B1650KNA}, \state{Buenos Aires}, \country{Argentina}}}

\affil[2]{\orgdiv{Karlsruhe Institute of Technology (KIT)}, \orgname{Institute for Astroparticle Physics}, \orgaddress{\street{Hermann-von-Helmholtz-Platz 1}, \city{Eggenstein-Leopoldshafen}, \postcode{76344}, \state{Baden-W\"{u}rttemberg}, \country{Germany}}}

\abstract{We present a general method for estimating the number of particles impinging on a segmented counter or, in general, on a counter with sub-units. We account for unresolved particles, i.e., the effect of two or more particles hitting the same sub-unit almost simultaneously. To achieve full time resolution we account for the dead time that occurs after the first time-bin of a particle signal. This general counting method can be applied to counting muons in existing detectors like the Underground Muon Detector of the Pierre Auger Observatory. We therefore use the latter as a study case to test the performance of our method and to compare it to other methods from literature. Our method proves to perform with little bias, and also provides an estimate of the number of particles as a function of time (as seen by the detector) to a single time-bin resolution. In this context, the new method can be useful for reconstructing parameters sensitive to cosmic ray mass, which are key to unveiling the origin of cosmic rays.}

\maketitle

\section{Introduction}\label{sec:intro}

In this work we introduce a new strategy to count particles hitting a counting detector that has counting sub-units. This general method could be used in existing or new particle or astroparticle detectors. The proposed statistical model provides useful count estimates if: (1) the signal processing of each sub-unit is based on a discrimination threshold (not on the signal amplitude or charge) and is thus subject to the effect of unresolved particles, (2) there is a subset (larger than one) of counting sub-units that expect the same particle rate, and (3) not all the sub-units of the subset have signal simultaneously. The new strategy is particularly useful for estimating the particle counts as a function of time, even when single particles cannot always be time-resolved. 

We find a concise application of the counting strategy in the Underground Muon Detector of the Pierre Auger Observatory \cite{AugerPrime2016,AugerUMD2021}, which we take as study case. Therefore, in the following paragraphs we describe the importance of this new strategy in the context of cosmic ray physics, within which it was developed. We delay the discussion on other possible applications of the counting strategy to Sec.~\ref{sec:conclusions}.

In spite of extensive research, the origin, nature, and acceleration mechanisms of ultra-high-energy cosmic rays are still not fully understood \cite{Batista2019}. To unravel these mysteries, cosmic rays are studied mainly through three observables: the energy spectrum, the distribution of arrival directions, and the mass composition as a function of the energy \cite{Mollerach2021}. 

Low-energy cosmic rays are abundant and can thus be measured directly. However, the flux drops steeply with the energy, so cosmic rays with energies above $\sim 10^{15}\,\text{eV}$ are measured only indirectly using ground-based observatories. The large areas the observatories cover provide enough exposure to detect the extensive air-showers (EASs) \cite{Mollerach2021}. The latter are conformed by the particles that result from the chains of interactions and decays initiated by a cosmic ray \cite{Engel2011}. 

The mass composition of cosmic rays is essential to understanding their origin, the transition energy between the galactic and extragalactic components, the flux suppression at the highest energies, and also for improving current high-energy hadronic interaction models. We know that high-energy cosmic rays consist mainly of nuclei ranging from proton (light) to iron (heavy) \cite{Blumer2009}. In their journey from their sources to Earth, these charged nuclei are subject to deflection by magnetic fields. Lighter, less charged nuclei are less deflected than heavy nuclei \cite{Aramburo2021}. Then identifying the mass of cosmic rays would allow one to use the high-energy, light component to infer what the sources are \cite{AugerPrime2016}. Furthermore, a change in composition as a function of the energy around the transition from galactic to extragalactic cosmic rays is expected. The transition energy would be related to the acceleration limits of galactic sources. The magnetic fields of galactic sources are capable of accelerating heavy nuclei to higher energies than light nuclei. The transition would occur presumably between the highest-energy galactic iron component to the lowest-energy extragalactic proton component \cite{PetersCycle}. It would also be expected that the highest-energy cosmic rays are mainly extragalactic intermediate mass or even heavy nuclei \cite{CombinedFitAuger}. Finally, the primary energies to which cosmic rays accelerate are inaccessible to human-made accelerators like the Large Hadron Collider. High-energy hadronic interaction models can only be tested at the highest energies through measurements of the ultra-energetic EASs \cite{Prado2018}. These models are typically tested by analyzing the consistency of the composition implications of different EAS observables \cite{Prado2018}. Composition studies help to improve these models, which in turn improve the precision of the inferred mass composition \cite{AugerInclinedMuons2015,AugerMuonFluctuations2021}.

There are two EAS observables that are specifically sensitive to the mass composition for a given energy: the depth of the shower maximum and the number of muons \cite{Supanitsky2008}. The first is measured using fluorescence telescopes. These telescopes measure the ultra-violet light emitted from the deexcitation of nitrogen in air, that is excited by charged air-shower particles \cite{Arqueros2008}. The second is measured using underground or surface detectors. In surface detectors, the muons impinge on the detector together with the much more abundant electromagnetic component of air-showers. Isolating the muonic component in this case is typically only possible using inclined events, in which most of the electromagnetic component is absorbed by the atmosphere \cite{AugerInclinedMuons2015}. On the other hand, buried or shielded detectors are able to detect mainly muons regardless of the inclination of the event, because the other components are mostly absorbed in the earth or shielding material above the detector. 

Segmented particle detectors are widely used in particle and astroparticle physics experiments \cite{Book}. They can be designed to operate as calorimeters or as counters. In the first case, the integral of the output signal is converted to a total deposited energy, that is in turn converted to a number of particles by knowing the energy deposited by a single particle. This is the case, for example, of the Surface Scintillator Detector of the Pierre Auger Observatory \cite{AugerPrime2016,Taboada2020,Cataldi2021}, or the surface detectors of Telescope Array \cite{TelescopeArray2012}.
For the second case, to use the segmented detectors as muon counters, the output signal of a segment is compared against a threshold using a discriminator. The resulting binary signal is matched to a known muon signal. For the Underground Muon Detector (UMD) of the Pierre Auger Observatory \cite{AugerPrime2016,AugerUMD2021}, the single muon signal is a pattern of 0s and 1s. The Auger UMD and the shielded detectors of the AGASA experiment \cite{Hayashida1995} are examples of segmented detectors used both as counters and as calorimeters. 

For each EAS event, an estimate of the muon density can be obtained at each module of the detector by dividing the reconstructed counts by the effective area of the module. The muon densities as a function of the distance to the shower axis, measured on the shower plane, constitute a sample of the so-called muon lateral distribution function. Customarily the sampled distribution is fitted, and the fitted function is evaluated at a fixed distance to the shower axis. This provides an event-wise measure of the size of the muon component which is then used for further analyses (see for example Refs.~\cite{KASCADEmuonsize2006,Ravignani2014,AugerMueller2020}).

Estimating the number or density of muons impinging a segmented detector without bias is not trivial. If two or more particles hit one scintillator strip almost simultaneously, they will be read out as only one particle. This effect, referred to as pile-up, constitutes a source of undercounting \cite{Ravignani2014,Ravignani2016}. There are several existing methods that attempt to provide pile-up-unbiased estimates of the number of muons. A simple yet powerful strategy is the one used by the AGASA collaboration \cite{Hayashida1995}. It is based on counting the number of occupied channels out of all available ones. The advantage of this strategy is that it presents little to no bias in the non-saturation region. The disadvantage is that it does not use the time information of the signal. Another strategy is the one used for the photo-multiplier data of the Auger UMD \cite{AugerMueller2020,Mueller2018,Ravignani2016}. In this case, the output signals of each channel are tested for matches to a single-muon pattern. Then the signal is divided in time windows of a length equal to that of the single-muon pattern, and for each window the number of starting pattern matches across all channels is counted. The final estimate is then the sum over all windows. 
This strategy trades an increased use of the temporal information of the signal for increased biases. We analyze a modification of this strategy based on centering one of the windows around the peak of the signal. We will show that this modification reduces biases in some cases. 
Finally, we present a new strategy that exploits the full time resolution of the detector, that provides an estimate of the number of muons at each time-bin, and that has also little bias in the non-saturation region. The key of this strategy to achieve full time resolution is considering that a channel becomes inhibited after the first time-bin of a particle pattern. 

\begin{figure}[!ht]
 \centering
  \includegraphics[width=0.45\textwidth]{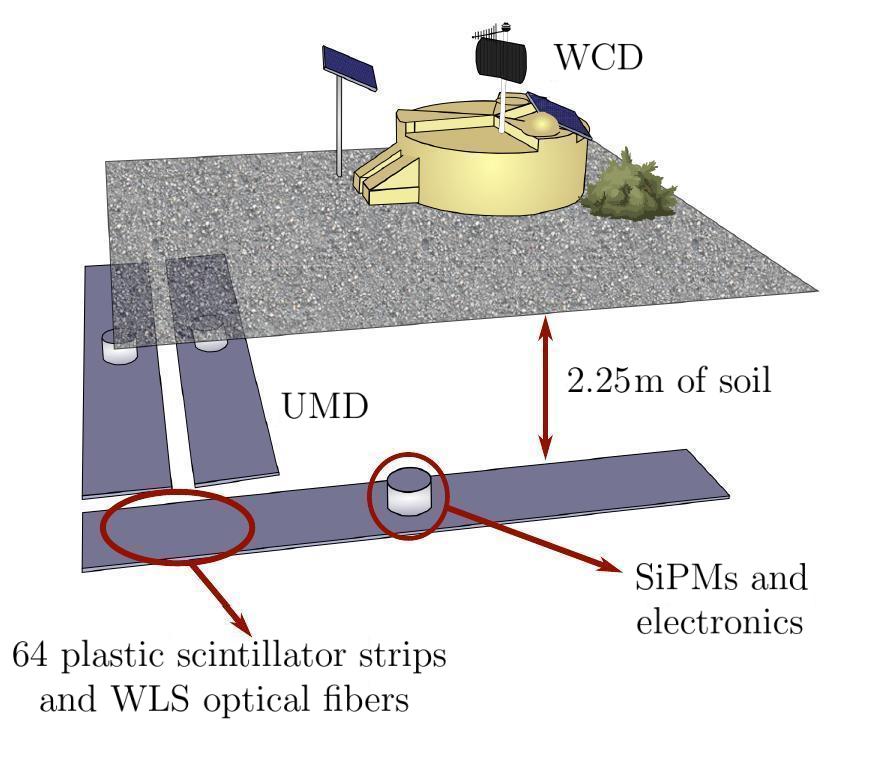}
  \caption{Scheme of a water-Cherenkov detector (WCD) of the Pierre Auger Observatory with its three paired UMD modules. The modules lie $2.25\,\text{m}$ underground and have an area of $10\,\text{m}^{2}$ each. They consist of $64$ plastic scintillator strips embedded with wavelength-shifting (WLS) optical fibers. The SiPMs and the electronics are placed in the middle of the modules below an access tube. Figure adapted from Ref.~\cite{BottiThesis}.}
  \label{fig:umd}
\end{figure}

In this work we simulate the Auger UMD to use it as a test scenario and as a point of comparison among the different counting strategies. The Auger UMD will consist of an hexagonal array of 219 modules at 73 locations when its construction is finished. 61 locations will be separated by $750\,\text{m}$, and 12 separated by $433\,\text{m}$ \cite{AugerUMD2021}. Each location has a water-Cherenkov detector (WCD) paired to three $10\,\text{m}^{2}$ modules, each of which is segmented into $64$ plastic scintillator strips with embedded wavelength-shifting (WLS) optical fibers. Figure \ref{fig:umd} shows a scheme of a WCD with its three paired UMD modules. A previous implementation of the electronics used multipixel photo-multipliers, but now almost all modules are equipped with silicon photo-multipliers (SiPMs). The modules are buried $2.25\,\text{m}$ underground implying a vertical muon energy threshold of $\sim\!1\,\text{GeV}$. The UMD of Auger is designed to measure showers with energies between $10^{16.5}\,\text{eV}$ and $10^{19}\,\text{eV}$ \cite{AugerUMD2021,AugerPrime2016}. 

In Sec.~\ref{sec:strat} we explain the different statistical models or algorithms that aim to provide unbiased estimates of the impinging muons. In Sec.~\ref{sec:sim} we describe how we simulate the detector and the air-shower library that we generated. In Sec.~\ref{sec:results} we show the comparison of the performance of the different algorithms. Finally, in Sec.~\ref{sec:conclusions}, we summarize the main conclusions and provide an outlook.

\section{Counting strategies}\label{sec:strat}
The problem of counting particles on a segmented detector can be compared, on a first approximation, to the problem of counting \enquote{balls in boxes}\footnote{This problem is also referred to as \enquote{classical occupancy problem} or \enquote{classical shot problem} \cite{ONeill2021}.}. In this problem, a finite number of balls (the impinging particles) are randomly uniformly allocated in a finite number of boxes (the segments of the detector). After one realization, each box can contain zero (with no pattern match), or one or more balls (with pattern match or occupied). The occupancy is then defined as the number of occupied boxes. 

Considering a muon counter, the balls represent the number of impinging muons on the counter $N_{\mu}$. We assume that $N_{\mu}$ originates from one realization of a Poisson distribution \cite{Gaisser1985} with mean $\mu$, the average number of particles expected on the counter. $\mu$ depends on the primary identity, energy, and zenith angle, and on the distance to the shower axis on the shower plane. For a flat detector, it can be expressed as $\mu = \rho_{\mu} A \cos\theta$, where $\rho_{\mu}$ is the average muon density of an EAS on the shower plane, $A$ is the active area of the detector, and $\theta$ is the zenith angle of the shower. It is important to understand the difference between $N_{\mu}$ and $\mu$: $N_{\mu}$ is the number of particles actually impinging the detector, a property of the event, while $\mu$ is the average number, a property of the EASs. Both $N_{\mu}$ and $\mu$ can be used to reconstruct the muon lateral distribution function \cite{Supanitsky2008,Ravignani2014,Ravignani2016}.

The aim of the counting strategies is to answer two questions. The first one is: Knowing the occupancy, what is the estimated number of impinging particles $N_{\mu}$? And the second one is: Assuming that $N_{\mu}$ is a realization of a Poisson distribution of mean $\mu$, what is the estimated mean number of particles $\mu$? 

The simplest estimator of both $N_{\mu}$ and $\mu$ is the sum of the number of pattern matches over all the scintillator strips of one module. Yet it is evident that the obtained estimates would be biased due to the pile-up effect, since one pattern match can account for more than one muon. Therefore, the goal of these counting strategies is to use the information of the event trace to provide a pile-up-unbiased estimate of $N_{\mu}$ and $\mu$. 

We now introduce the four counting strategies that we consider in this work.
\vspace{1em}

\noindent\textbf{Infinite window strategy:} This is the simplest strategy since it does not make use of the temporal structure of the event trace (hence its name). Instead, it only uses the number of occupied channels $k$ (the occupancy), which is computed as \cite{Supanitsky2008,Ravignani2014,Supanitsky2021}
\begin{equation}
k = \sum_{i=1}^{n_s}\Theta(m_i),
\label{eq:nchon}
\end{equation}
where $n_s$ is the number of active segments of the detector, $\Theta$ is the Heaviside step function, and $m_i$ is the number of starting pattern matches of the $i$-th channel. $\Theta(m_i) = 0$ if $m_i=0$ and $\Theta(m_i) = 1$ otherwise.

The number of impinging muons on a scintillator strip follows a Poisson distribution of mean $\mu/n_s$. 
It follows that the probability of an empty channel is $q=\exp(-\mu/n_s)$, and the probability of an occupied channel is $p=1-q$ \cite{Ravignani2014,Ravignani2016}. The probability of having $k$ occupied channels (successes) out of $n_s$ channels (trials) given $\mu$ (which determines the success probability) follows a binomial distribution \cite{Ravignani2014,Ravignani2016}
\begin{equation}
B(k \vert\, \mu) = \binom{n_s}{k} p^{k} q^{n_s - k} = \binom{n_s}{k} e^{-\mu} (e^{\mu/n_s} -1)^{k}.
\label{eq:binom}
\end{equation}
This is also the expression for the likelihood of $\mu$ given $k$ occupied channels. For $k < n_s$ the maximum likelihood estimator of $\mu$ is \cite{Ravignani2014,Ravignani2016}
\begin{equation}
\widehat{\mu} = -n_s \ln \left( 1 - \frac{k}{n_s}\right).
\label{eq:muiw}
\end{equation}

Moreover, the probability of having $k$ occupied channels given $N_{\mu}$ impinging muons and $n_s$ scintillation bars is given by the occupancy distribution \cite{Supanitsky2021}
\begin{eqnarray}
\label{eq:occ}
Occ\,(k \vert\, N_{\mu}, n_s) &=& \binom{n_s}{k} \frac{S(N_{\mu}, k)}{n_s^{N_{\mu}}} \\
&& \ \forall \, k \in \mathds{Z} \mid 1 \leq k \leq n_s, \nonumber 
\end{eqnarray}
where $S(N_{\mu}, k)$ are the Stirling numbers of the second kind (see also Ref.~\cite{ONeill2021}). As shown in Refs.~\cite{Supanitsky2021,Supanitsky2008}, a good approximation of the maximum likelihood estimator of $N_{\mu}$ is
\begin{equation}
\widehat{N}_{\mu} = \frac{\ln \left( 1 - {\mathop{\displaystyle \frac{k}{n_s}}}\right)}{\ln \left( 1 - {\mathop{\displaystyle \frac{1}{n_s}}} \right)}. 
\label{eq:nmuiw}
\end{equation}

The similarity between Eq.(\ref{eq:muiw}) and Eq.(\ref{eq:nmuiw}) is evident, and in fact $\widehat{N}_{\mu} \rightarrow \widehat{\mu}$ in the limit when $n_s \rightarrow \infty$.

It is relevant to add that when $k = n_s$ we say that the module is saturated, and both $\widehat{\mu}$ and $\widehat{N}_{\mu}$ tend to infinity. 

This strategy was the one employed by the AGASA collaboration (see for example Ref.~\cite{Hayashida1995}).

\vspace{1em}

\noindent\textbf{N-bin window strategy:} In this strategy the event trace is divided into time windows of N bins, where N is the number of bins of a single-muon pattern (12 time-bins of $3.125\,$ns in the case of SiPMs of the Auger UMD). If the number of bins of the inhibition window is not an exact divider of the number of bins of the trace, the last window is actually shorter, yet still included in the computation. Then, for the $j$-th window, the number of occupied channels $k_{j}$, the estimated number of muons $\widehat{N}_{\mu,j}$, and the estimated average number of muons $\widehat{\mu}_j$ are computed using \cref{eq:nchon,eq:muiw,eq:nmuiw} repectively. The overall $\widehat{N}_{\mu}$ and $\widehat{\mu}$ are estimated from the sum of those of each window
\begin{equation}
\widehat{N}_{\mu} = \sum_{j=1}^{n_w} \widehat{N}_{\mu,j},
\label{eq:nmuoff}
\end{equation} 
\begin{equation}
\widehat{\mu} = \sum_{j=1}^{n_w} \widehat{\mu}_j,
\label{eq:muoff}
\end{equation}
where $j$ runs over the $n_w$ number of windows of the trace.

This strategy was introduced in Ref.~\cite{Ravignani2016} and has been used for analyses of PMT data of the Auger UMD in Refs.~\cite{Mueller2018,AugerMueller2020}\footnote{In Sec.~\ref{sec:results} we show that this strategy introduces significant biases. These biases are compensated in the analyses of Refs.~\cite{Mueller2018,AugerMueller2020}, at least to first order, by correcting them against simulations in a later step.}.

\vspace{1em}

\noindent\textbf{N-bin centered window strategy:} This strategy originated as an attempt to solve the biases created in the N-bin window strategy (A.M. Botti, private communication, 2021), and is used in this work to enlighten the origin of said biases. It is very similar to the N-bin strategy, but it determines the position of the windows such that the center of one of those coincides with the peak of the signal. 

For the purpose of the explanation, let us take the case of the 12 time-bin windows corresponding to SiPMs of the Auger UMD. The idea is that there exists at least one 12 time-bin window $j^*$ where the number of occupied channels is maximal. If there is more than one, we take the one that starts earliest. To find it, we slide a 12 time-bin window over all the trace, computing $k_j$ for every $j$-th possible window. The \enquote{centered} window is the earliest that fulfills $k_{j^*} = \max_{j}(k_{j})$. This window determines the way to complete the partition of the trace into 12 time-bin windows. For the case of the Auger UMD, since the event signal in the trace starts typically at around $3300\,\text{ns}$ or $3800\,\text{ns}$ \cite{BottiThesis}, we discard the first bins of the trace that do not complete a full window. The method can be generalized for other time-window widths, optionally keeping the first bins of the trace.

Having divided the trace into windows, $\widehat{N}_{\mu}$ and $\widehat{\mu}$ are computed as in \cref{eq:nmuoff,eq:muoff}.

\vspace{1em}

\textbf{1-bin window strategy:} This is the strategy developed in this work. Rather than taking the trace as a whole, or in many-bin windows, this strategy uses the complete time structure of the signal. The idea is to compute for each $j$-th time-bin of the trace, not only the number of occupied channels $k_{j}$, but also the number of inhibited channels $n_{\text{inhib},j}$. The latter is the number of channels that have a pattern match which started at an earlier time-bin. We consider these as inhibited channels because if a muon fell within the inhibition window (i.e, the single-muon pattern match) it would not be possible to measure or resolve it; effectively, inhibited channels are equivalent to dead channels. In the analogy to balls in boxes, having inhibited channels is equivalent to having less boxes. The number of non-inhibited channels is then $n_s - n_{\text{inhib},j}$.

Furthermore, having less available (non-inhibited) channels reduces the detector area, which leads to a smaller number of detectable muons. It becomes evident that the detector area varies bin by bin. To obtain the number of muons that would be observed for a constant detector area equal to the active area, we need to multiply by a factor equal to the number of active segments $n_s$ divided by the number of non-inhibited segments $n_s - n_{\text{inhib},j}$. We therefore compute 
\begin{align}
\widehat{\mu}_j =& -(n_s - n_{\text{inhib},j}) \ln \left( 1 - \frac{k_{j}}{n_s - n_{\text{inhib},j}}\right)\nonumber\\
 &\times \frac{n_s}{n_s - n_{\text{inhib},j}} \nonumber\\
 =& -n_s \ln \left( 1 - \frac{k_{j}}{n_s - n_{\text{inhib},j}}\right) ,
\label{eq:muour}
\end{align}
and
\begin{equation}
\widehat{N}_{\mu,j} = \frac{\ln \left( 1 - {\mathop{\displaystyle \frac{k_{j}}{n_s - n_{\text{inhib},j}}}}\right)}{\ln \left( 1 - {\mathop{\displaystyle \frac{1}{n_s - n_{\text{inhib},j}} }}\right)} \times \frac{n_s}{n_s - n_{\text{inhib},j}},
\label{eq:nmuour}
\end{equation}
for each bin of the trace. The overall $\widehat{N}_{\mu}$ and $\widehat{\mu}$ are computed as their sum over all the bins of the trace (this is, as in \cref{eq:nmuoff,eq:muoff}, with the number of windows $n_w$ being equal to the number of bins in the trace).

\vspace{1em}
Table \ref{tab:summ} summarizes the characteristics of the four strategies. In the table, we consider the trace-length (2048 time-bins) as well as the single-muon pattern length (12 time-bins) of the Auger SiPM UMD signals to calculate the number of windows $n_w$ of each strategy.

\begin{table*}[!ht]
\begin{center}
\begin{minipage}{\textwidth}
\caption{Summary of the characteristics of the four strategies considered in this work. $n_w$ is the number of windows of each strategy considering the characteristics of the Auger UMD trace. The column \enquote{Centered} details whether the position of the windows is chosen such that one of those is centered in the peak of the signal.}
\label{tab:summ}
\begin{tabular*}{\textwidth}{@{\extracolsep{\fill}}lcccc@{\extracolsep{\fill}}}
\toprule%
Strategy         & $\widehat{\mu}$ & $\widehat{N}_{\mu}$ & $n_w$ & Centered \\
\hline
\parbox[t]{1.5cm}{Infinite}  & $-n_s \ln \left( 1 - \frac{k}{n_s}\right)$ & $\frac{\ln \left( 1 - \frac{k}{n_s}\right)}{\ln \left( 1 - \frac{1}{n_s}\right)}$              & 1       & -                                     \\
N-bin          &  $-\displaystyle\sum_{j=1}^{n_w} n_s \ln \left( 1 - \frac{k_j}{n_s}\right)$     &  $\displaystyle\sum_{j=1}^{n_w} \frac{\ln \left( 1 - \frac{k_j}{n_s}\right)}{\ln \left( 1 - \frac{1}{n_s}\right)}$         & 171       & No                                    \\
\parbox[t]{1.5cm}{N-bin\\centered} &  $-\displaystyle\sum_{1}^{j=n_w} n_s \ln \left( 1 - \frac{k_j}{n_s}\right)$     &  $\displaystyle\sum_{j=1}^{n_w} \frac{\ln \left( 1 - \frac{k_j}{n_s}\right)}{\ln \left( 1 - \frac{1}{n_s}\right)}$         & \parbox[t]{0.65cm}{170-\\171}       & Yes                                   \\
1-bin        &  $-\displaystyle\sum_{j=1}^{n_w} n_s \ln \left( 1 - \frac{k_{j}}{n_s - n_{\text{inhib},j}}\right)$     &           $\displaystyle\sum_{j=1}^{n_w} \frac{n_s}{n_s - n_{\text{inhib},j}}\frac{\ln \left( 1 - \frac{k_{j}}{n_s - n_{\text{inhib},j}}\right)}{\ln \left( 1 - \frac{1}{n_s - n_{\text{inhib},j}}\right)}$ & 2048         & -                                    
\end{tabular*}
\end{minipage}
\end{center}
\end{table*}


\section{Simulations of the response of the underground muon detector to air-showers}\label{sec:sim}

\subsection{Simulation of the detector}\label{sec:detsim}
In our aim to recreate a realistic scenario, we develop an end-to-end simulation chain for the counter mode of the Auger UMD. In this section we describe the idea of how the detector works and how we simulate it, and we leave the complete technical details for the Appendix \ref{sec:app}. The model of the detector follows Refs.~\cite{Botti_2021,BottiThesis}. Assuming said model, we derive an analytical solution of the detector response to one or many muons. For this purpose we neglect noise, which is anyway intrinsic to the detector and irrelevant for assessing the performance of the counting strategies. It is also relevant to add that we do not consider muons transversing two scintillator strips (commonly referred to as corner-clipping muons) as this effect constitutes an independent source of bias that can be corrected afterwards \cite{Mueller2018}.

The principle of detection of the UMD of Auger is as follows. The energy of the muons that transverse the soil above the detector has to be larger than $\sim 1\,\text{GeV}/\cos(\theta_\mu)$, where $\theta_\mu$ is the angle between the direction of motion of the muon and the vertical direction. When such a muon interacts with the scintillator material, it produces photons that excite the wave-length shifting optic fiber of a segment, producing photons inside of it. These photons propagate through the optic fiber, and some of them reach the SiPM connected to that segment. The photons produce photo-electrons at the SiPM (with a certain efficiency) and this current is processed with the electronics, which outputs a binary signal \cite{Platino2011,BottiThesis}. If the water-Cherenkov station paired to the muon counter triggered, the digital signals of the 64 binary channels of the module are stored as part of the Auger UMD event trace. For SiPMs, the trace is $6.4\,\upmu\text{s}$ long with bins of $3.125\,\text{ns}$. Typically one muon creates a binary signal of $\sim 8$ consequtive 1s (more precisely $7.8 \pm 1.5$) \cite{BottiThesis}. However, to optimize the signal-to-noise ratio, the matching strategy of the Auger UMD uses a 12 time-bin inhibition window, and it consists of identifying patterns of the kind \enquote{1111xxxxxxxx}, where x can be 0 or 1 \cite{BottiThesis}. 

The detector simulation takes the number and impinging times of the muons as input, and outputs the event trace of the module, i.e., the binary signals of the 64 channels of the Auger UMD. 

A feature of the UMD traces is that the signal does not always start at the same time-bin. This is related to the fact that the paired surface detector triggers the data acquisition in the UMD. Therefore, as a first step we determine the start-time of the signal in the Auger UMD trace by sampling the distribution of the time-delays with respect to the trigger in the paired surface detector. All muon impinging times are then taken as relative to the signal start-time. 

Afterwards, we randomly assign a scintillator strip to each impinging muon, as well as an impinging position within the strip (i.e., the distance to the SiPM). 

Each muon generates several photons in the optic fiber, some of which reach the SiPM and generate photo-electrons with a certain efficiency. The average number of photo-electrons that are generated in the SiPM $\left\langle N_{\text{PE}} \right \rangle$ can be expressed as a function of the distance between the impinging position of the muon and the SiPM. We use said function, and sample the actual number of generated photo-electrons $N_{\text{PE}}$ from a Poisson distribution of mean $\left\langle N_{\text{PE}} \right \rangle$. 

For each photo-electron corresponding to a muon, there is one common time delay that comes from the propagation through the optic fiber from the impinging point to the SiPM. But there are also additional time delays due to the scintillator and the optic fibers excitation and de-excitation. We sample each of these additional delays from exponential distributions. The resulting time of a photo-electron $t_{\text{PE}}$ is the addition of the signal start-time, the impinging time of the muon (measured with respect to the signal start time), the propagation in the optic fiber, and the scintillator and fiber delays. 

At this point, the muon number and impinging times are \enquote{translated} to a photo-electron number and arrival times at the SiPMs. We then use a model of the pulse generated by a single photo-electron at the input of the electronics. It is relevant to add that non-linearities due to pile-up in cells of the Geiger-avalanche photodiode arrays of the SiPMs are only relevant for high occupancy and are not considered here \cite{BottiThesis}.

The electronics of a Auger UMD module has four elements at each of its 64 channels: a pre-amplifier, a fast shaper, a discriminator, and a Field-Programmable Gate Array (FPGA). We model the pre-amplifier as a low-pass filter, and the fast shaper as a practical differentiator. We then simply add the fast-shaper signal generated by each photo-electron within a same channel. Having this, we model the discriminator by simply imposing a threshold on the signal after the fast shaper, outputting a fixed voltage if the threshold is passed, and zero voltage otherwise. Finally, like the FPGA, we sample the signal in $3.125\,\text{ns}$ time-intervals, with a total duration of $6.4\,\upmu\text{s}$ (2048 samples). The binary signal of each channel is then matched to patterns of the kind \enquote{1111xxxxxxxx}, as explained in Sec.~\ref{sec:intro}. Repeating the process for all the channels, we obtain the final event trace for a module, as well as the pattern matches. 

\begin{figure}[!ht]
 \centering
  \includegraphics[width=0.45\textwidth]{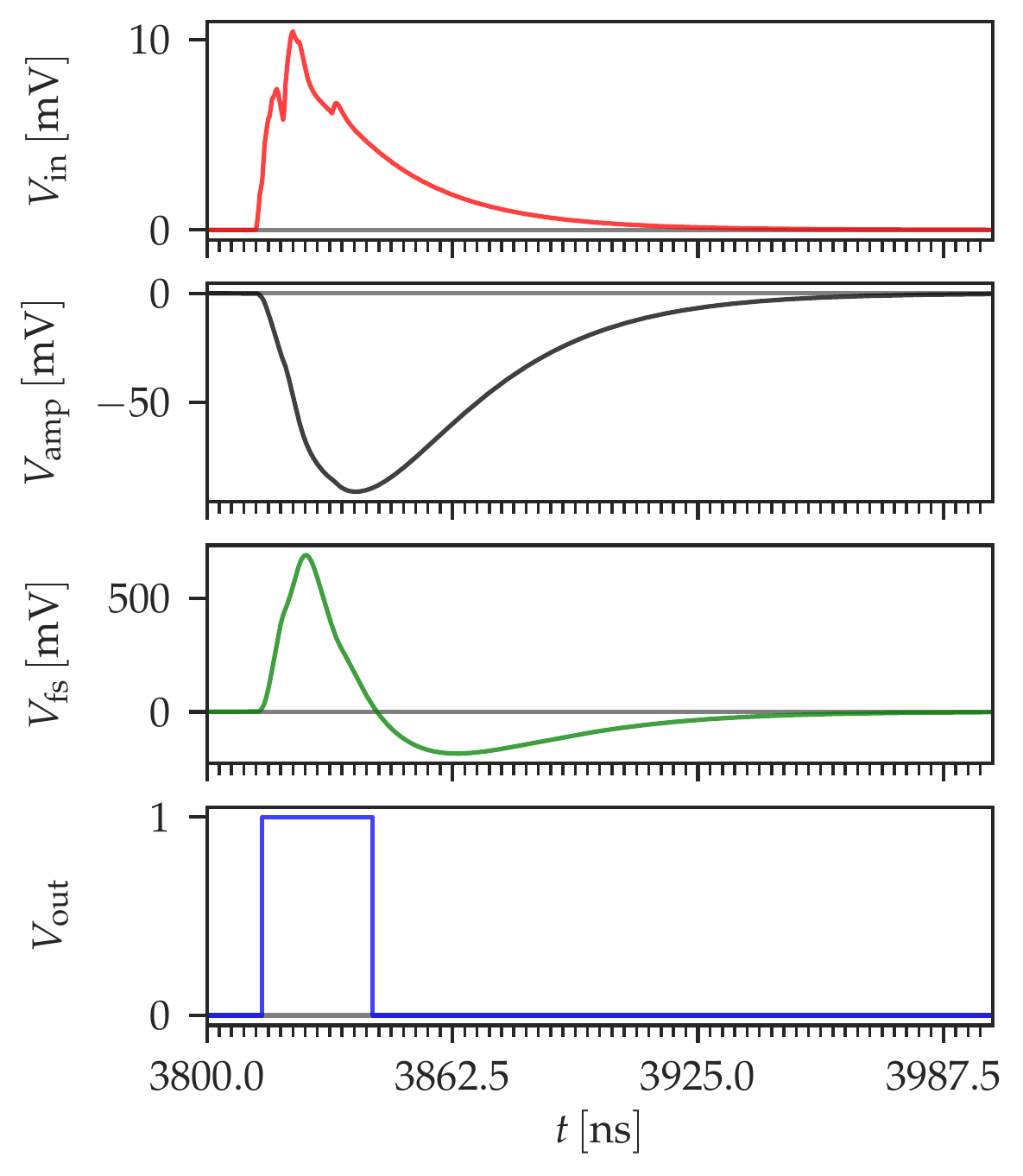}
  \caption{Simulation of the pulse generated by one muon. The input signal (top) generated at the SiPM goes through a pre-amplifier (second), a fast shaper (third), and a discriminator and FPGA (bottom). The output signal is digitized into $3.125\,\text{ns}$-wide time intervals by the FPGA.}
  \label{fig:onemuonpulse}
\end{figure}

Figure \ref{fig:onemuonpulse} shows the input (at SiPM), pre-amplified, after fast-shaper, and output (after discriminator and FPGA) signals as a function of time, for one simulated muon. In the input signal it is easy to distinguish the single photo-electron pulses. The pre-amplifier amplifies and inverts the input signal, which is further amplified and again inverted by the fast-shaper. Finally, the discriminator and FPGA output a digital binary signal. 

\begin{figure}[!ht]
 \centering
  \includegraphics[width=0.45\textwidth]{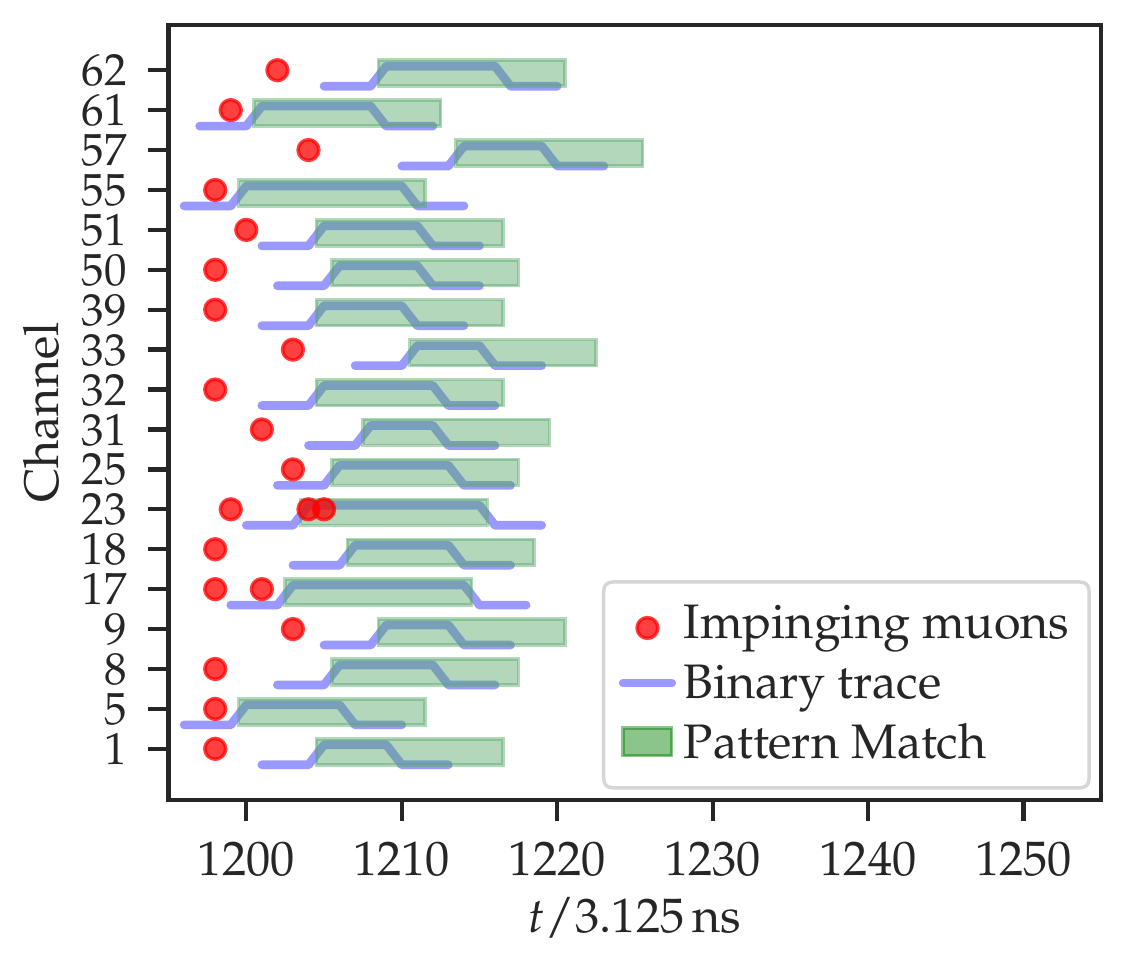}
  \caption{Simulation of the signal generated by muons of an air-shower simulated event on a UMD module. The muons (circles) impinge at a certain time (x-axis) on the different segments of the detector, which are coupled to a channel (y-axis), generating a binary signal (lines), which are then matched to a single-muon pattern (rectangles).}
  \label{fig:allchannels}
\end{figure}

Finally, Fig. \ref{fig:allchannels} shows, in one module-level event, the impinging muons as a function of time, the binary traces, and the matched patterns for all active channels. In this example we can see pile-up in channels 17 and 23, where the signal of two and three impinging muons respectively is matched to only one pattern.

A validation test of the complete simulation chain can be found in Appendix \ref{sec:appsat}, where we compare the fraction of saturated events in simulations against the expected one as computed from the statistical models in Sec.~\ref{sec:strat}.

\subsection{Detector effects affecting the counting strategies}\label{sec:deteff}

The $\widehat{\mu}$ and $\widehat{N}_{\mu}$ estimated as explained in Sec.~\ref{sec:strat} are subject to two sources of bias that are not intrinsic to the counting strategies, and not related to the pile-up either.

In first place, it is possible that one muon does not create an output signal strong or long enough such that it matches the pattern \enquote{1111xxxxxxxx}. This is an inefficiency of the detector and of the pattern matching strategy that leads to undercounting, and affects all strategies to the same extent.

In the second place, as can be seen in Fig.~\ref{fig:onemuonpulse}, the signal after the fast-shaper presents an undershoot. It can happen that a signal of a later muon ends up mounted on the undershoot caused by previous muons in a same channel. If the amplitude of the undershoot is large (very negative) or the amplitude of the later muon signal too small, a pattern match can be missed. The infinite window strategy is the only one insensitive to this effect, because it is enough to match the early muons in the channel to tag it as occupied during the whole trace. All other strategies are subject to undercounting due to undershoot to the same extent. This undercounting effect is due to the design of the electronics, and its impact depends also on the time resolution of the detector and on the length of the single-muon pattern.

\begin{figure}[!ht]
 \centering
  \includegraphics[width=0.45\textwidth]{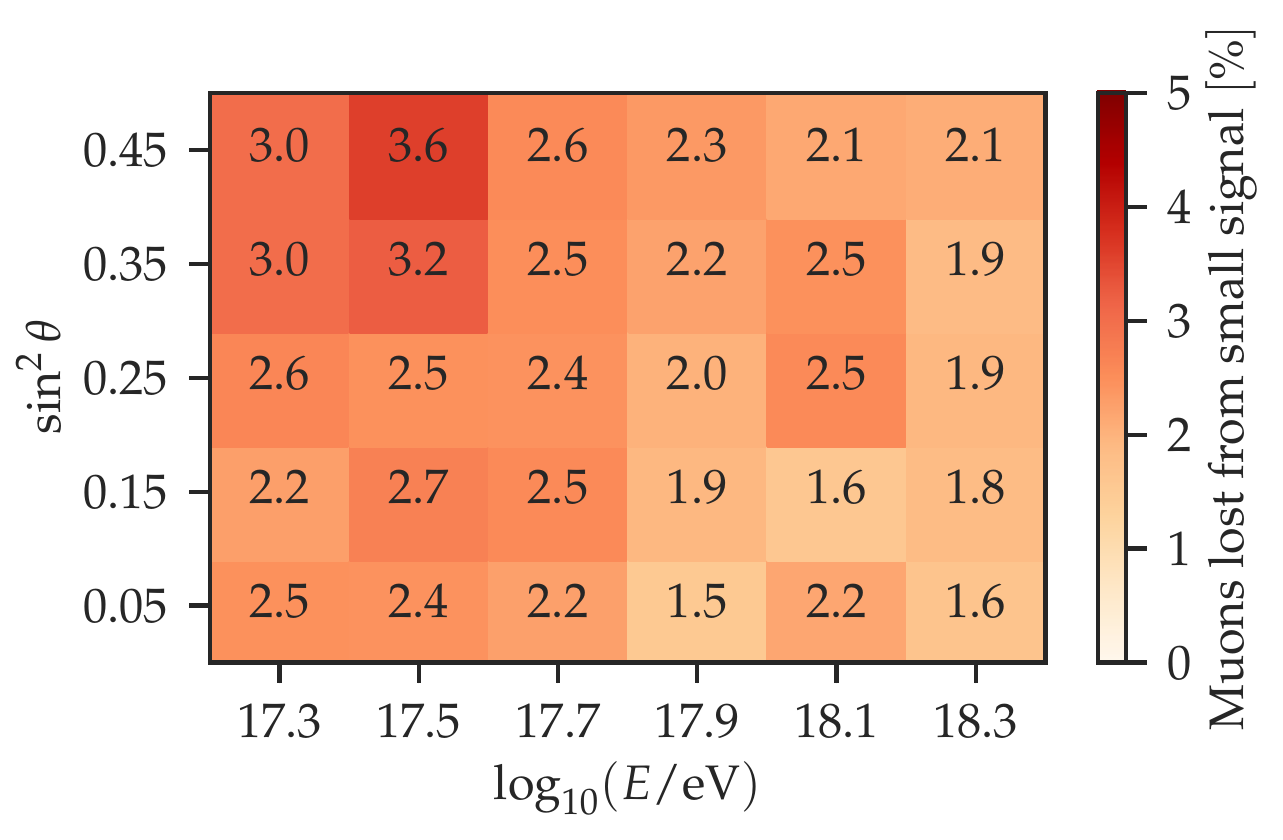}
  \includegraphics[width=0.45\textwidth]{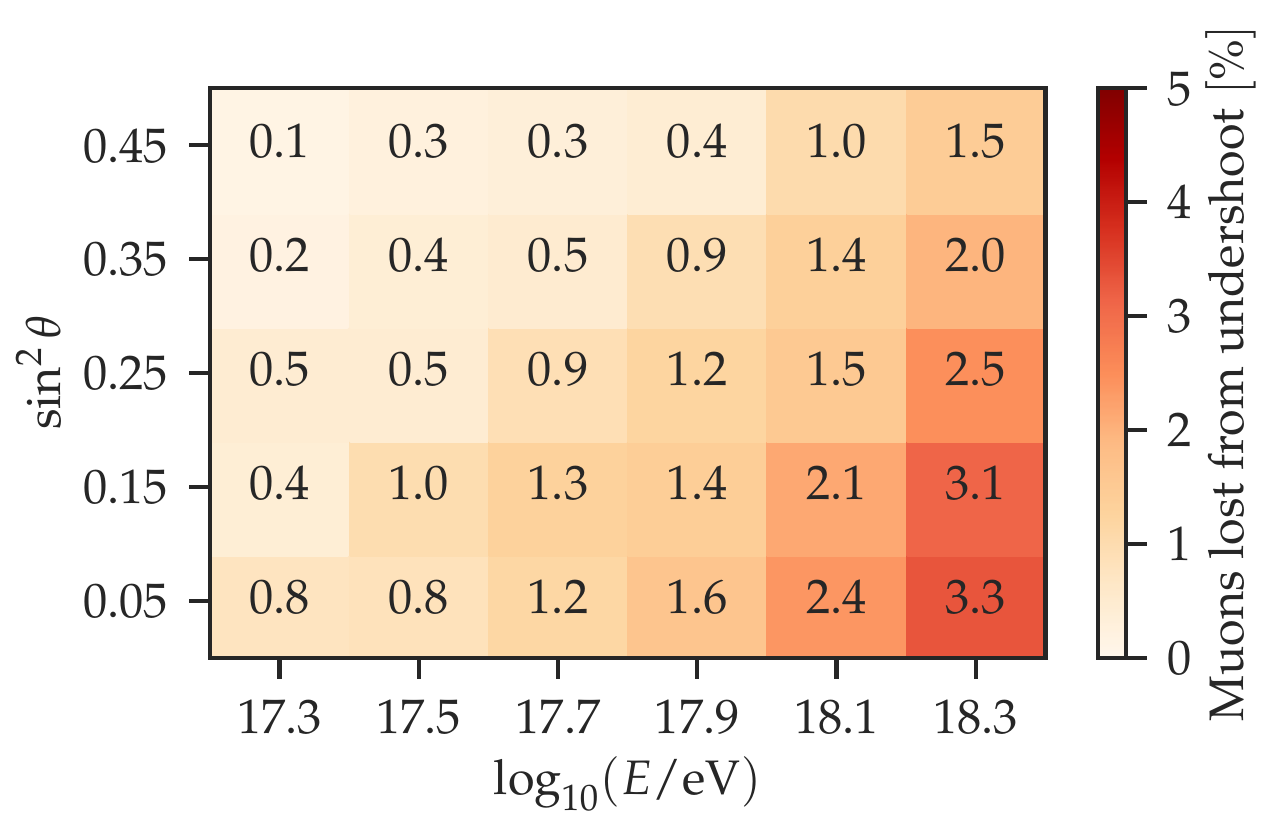}
  \includegraphics[width=0.45\textwidth]{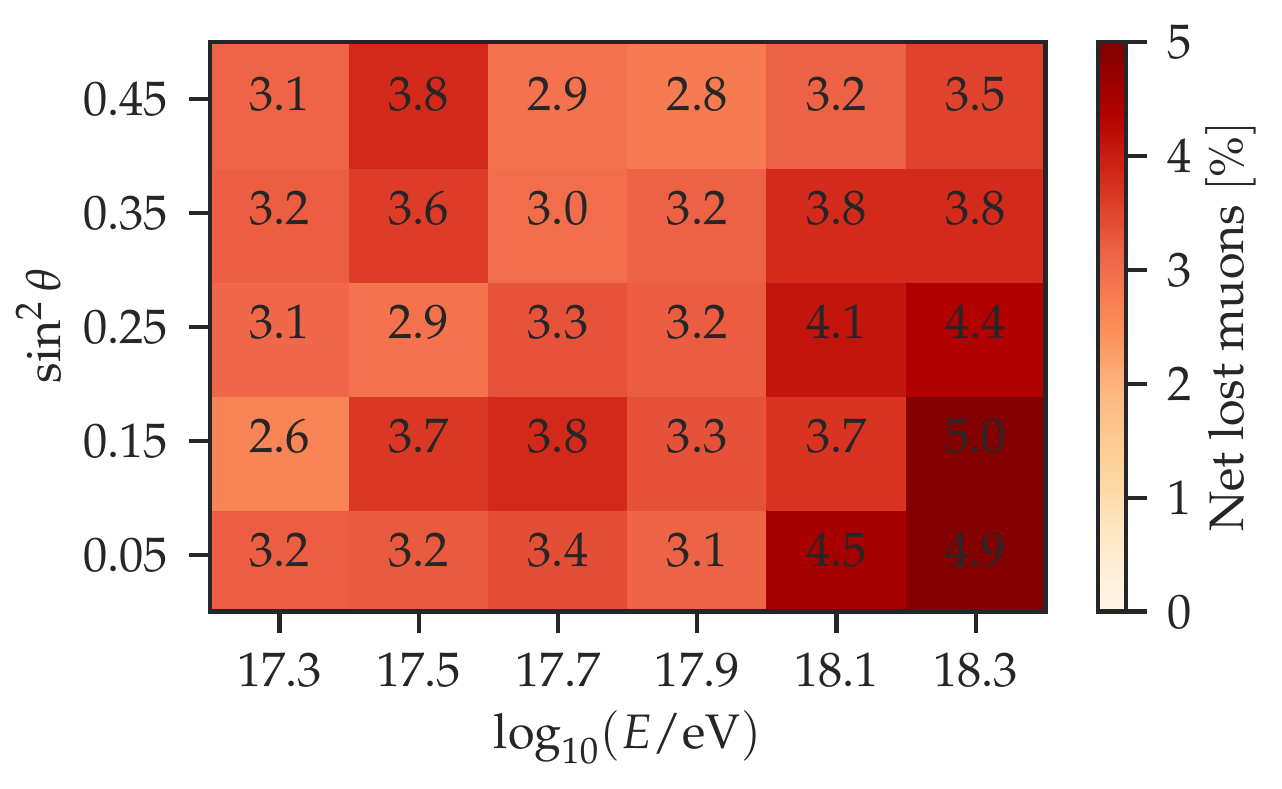}
  \caption{Average percentage of missed muon counts due to signals being to small or too short (upper panel), due to undershoot (middle panel), and the sum (lower panel), as a function of the logarithmic energy and of the sine square of the zenith angle. We use proton initiated air-shower simulations with the distribution of distances of an hexagonal array.}
  \label{fig:undershoot}
\end{figure}

To understand the impact of both effects in our test case (signals of SiPMs of the Auger UMD), we quantify the pattern matches lost due to \enquote{small signals} and due to undershoot. For this, in each channel we identify whether the muon signal would be matched to a pattern if it were the only muon signal in the channel. If not, then the muon is in principle lost for having a small signal. Approximately $3.6\,\%$ muons would be lost in this way, with no dependence on the energy or zenith angle of the originating cosmic ray. However, if the pattern match that the muon would have generated\footnote{We take the time where the fast-shaper signal is maximum as the start time of the pattern that the muon would have generated.} overlaps that of other muon of the channel, we consider the muon recovered. The latter happens more often with higher muon rates, this is, for more vertical and more energetic air-showers. The net loss of muons due to small signals can be seen in the upper panel of Fig.~\ref{fig:undershoot}. In the figure shown, we used proton initiated air-showers with the typical signals in an hexagonal array (for more details see Sec.~\ref{sec:sets}). As expected, the net loss is smaller for higher-energy and more vertical air-showers. The net effect is below $4\,\%$.

If a single muon signal matches a pattern, as if it were the only muon signal in the channel, we analyze whether that pattern match overlaps with the pattern matches of the total signal of the channel. If this is not the case, the muon is lost due to undershoot. The results are shown in the middle panel of Fig.~\ref{fig:undershoot}. The muons lost because of the undershoot can amount from $0.1\,\%$ for low-energetic, inclined air-showers, to $3.3\,\%$ for high-energetic, vertical air-showers.

The total detector effects are simply the sum of the two contributions (i.e. the sum of what is displayed in the upper and middle panels of Fig.~\ref{fig:undershoot}), and it is shown in the lower panel of Fig.~\ref{fig:undershoot}. We can see that the dominating effect is the undershoot, thus there are more lost muons for high-energy, more vertical air-showers. The total lost muons due to detector effects range from $2.6\,\%$ to $5.0\,\%$, and it can reach up to $5.3\,\%$ for iron-initiated air-showers.

\subsection{Air-shower simulations}
To continue with our aim of simulating a realistic scenario, we created a library of $\sim\! 7600$ EASs of proton and iron primaries, using EPOS-LHC \cite{Epos} and UrQMD \cite{UrQMD1,UrQMD2} as high- and low-energy hadronic interaction models, respectively. The showers were generated using CORSIKA v7.7402 \cite{Corsika}. The logarithm of the primary energy of the showers is uniformly distributed in $17.2 \leq \log_{10}(E/\text{eV}) \leq 18.4$ and the arrival directions of the showers correspond to an isotropic distribution with zenith angles in $0^{\circ} \leq \theta \leq 48^{\circ}$. 

We divide the simulations by primary, in $\log_{10}(E/\text{eV})$-bins centered at $17.3$, $17.5$,..., $18.3$ with widths of $0.2$, and in $\sin^2\theta$-bins centered at $0.05$, $0.15$,..., $0.45$ with widths of $0.10$. For each primary and $(\log_{10}(E/\text{eV}), \sin^2\theta)$-bin, we compute an average profile of the number of muons as a function of the (logarithmic) distance to the shower plane and of the time $\textrm{d}\mu/\textrm{d}t \!\times\! \Delta t$. In order to achieve this, we first compute the profile for each shower by retrieving, for each muon that reaches ground with sufficient energy ($\geq 1\,\text{GeV}/\cos(\theta_\mu)$ for the Auger UMD), the distance to the shower axis on the shower plane and the time it reaches the shower plane. Then, for all showers that fall into the $(\log_{10}(E/\text{eV}), \sin^2\theta)$-bin, we make a weighted average of the $\textrm{d}\mu/\textrm{d}t \!\times\! \Delta t$ profiles, where the weight of the $i$-th shower is its energy $E_i$ times the cosmic ray flux at that energy $J(E_i)$. By weighting the showers in this way we obtain the average profile corresponding to a distribution of the shower energies that follows the cosmic ray flux. We model the flux following Ref.~\cite{AugerSpectrum2020}.

\begin{figure}[!ht]
\centering
\includegraphics[trim={0 0 0 1.9cm},clip,width=0.45\textwidth]{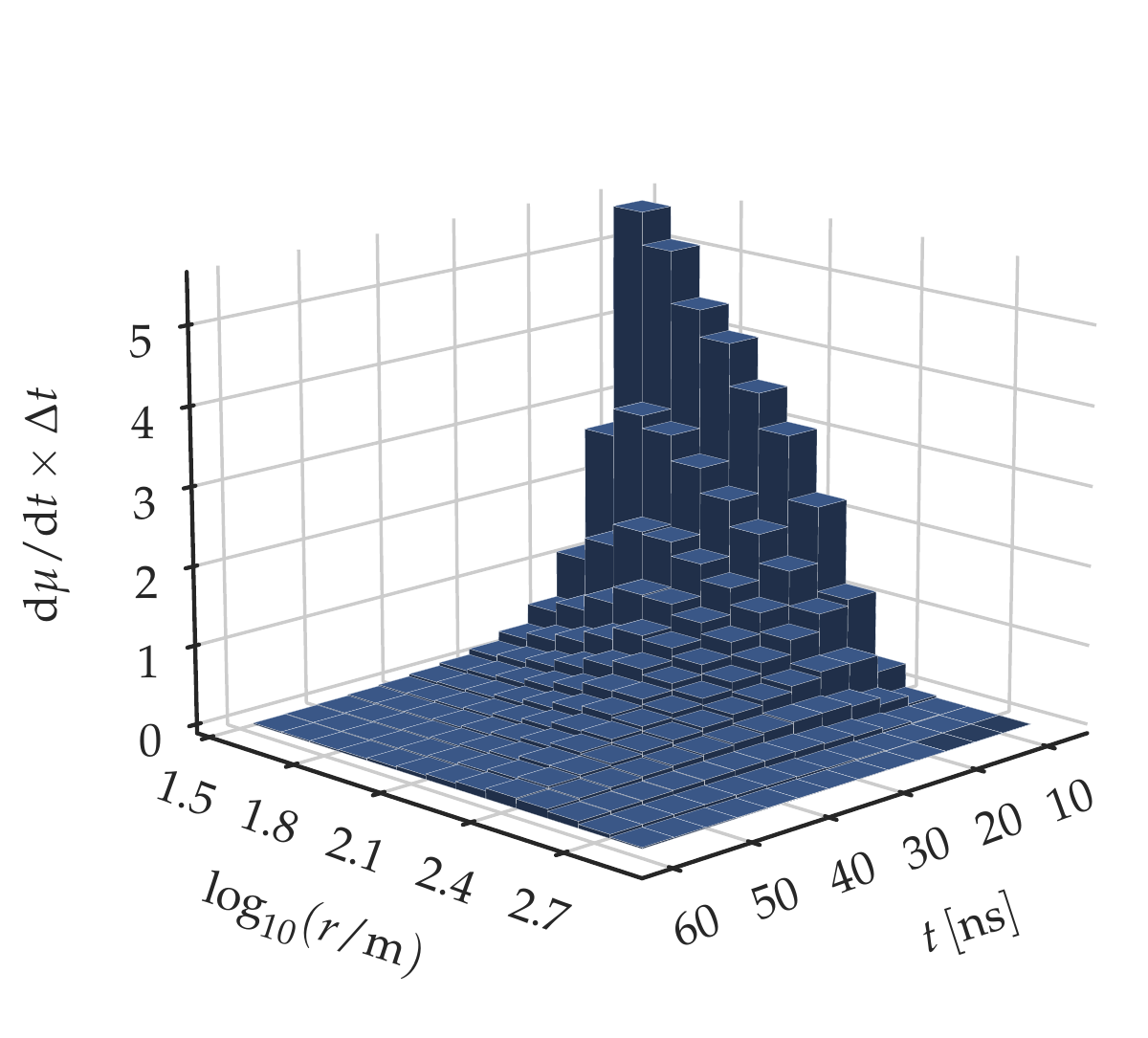}
\caption{Average number of muons per time-bin as a function of the logarithm of the distance to the shower axis (measured on the shower plane) and of the shower plane arrival time. The average corresponds to proton air-showers with $18.0 \leq \log_{10}(E/\text{eV}) \leq 18.2$ and $27^\circ \lesssim \theta \lesssim 33^\circ$ ($0.20 \leq \sin^2\theta \leq 0.30$).}\label{fig:murt}
\end{figure}

Figure \ref{fig:murt} shows an example of the profile $\textrm{d}\mu/\textrm{d}t \!\times\! \Delta t$ for proton EASs with $18.0 \leq \log_{10}(E/\text{eV}) \leq 18.2$ and $27^\circ \lesssim \theta \lesssim 33^\circ$ ($0.20 \leq \sin^2\theta \leq 0.30$).

\subsection{End-to-end simulations}\label{sec:sets}

We generate the first detector-simulation set as follows: For each primary and $(\log_{10}(E/\text{eV}), \sin^2\theta)$-bin, we evaluate $10,000$ times the profile $\textrm{d}\mu/\textrm{d}t \!\times\! \Delta t$ at a random logarithmic distance $\log_{10}(r^{*}/\text{m})$ which we sample from a uniform continuous distribution $\mathcal{U}\{1.0, 3.5\}$. This determines the input average number of muons as a function of time $\textrm{d}\mu/\textrm{d}t \!\times\! \Delta t \, \vert\,_{r=r^{*}}$, the integral of which is $\mu = \mu\,(\log_{10}(r/\text{m}))\,\vert\,_{r=r^{*}}$. We then obtain the number of impinging muons $N_{\mu}$ from sampling a Poisson distribution of mean $\mu$. Finally, we sample the times of the $N_{\mu}$ muons from $\textrm{d}\mu/\textrm{d}t \!\times\! \Delta t \,\vert\,_{r=r^{*}}$. This information is the input of the detector simulation, which assigns random scintillator strips/channels to the muons, simulates the response of the electronics, and outputs the binary signals of the 64 channels, as explained in Sec.~\ref{sec:detsim}. Then we match the pattern \enquote{1111xxxxxxxx} to the binary signals. With this information, we reconstruct the estimated $\widehat{\mu}$ and $\widehat{N}_{\mu}$ with each of the four strategies presented in Sec.~\ref{sec:strat} (see Table \ref{tab:summ}).  

\begin{figure}[!th]
\centering
\includegraphics[width=0.45\textwidth]{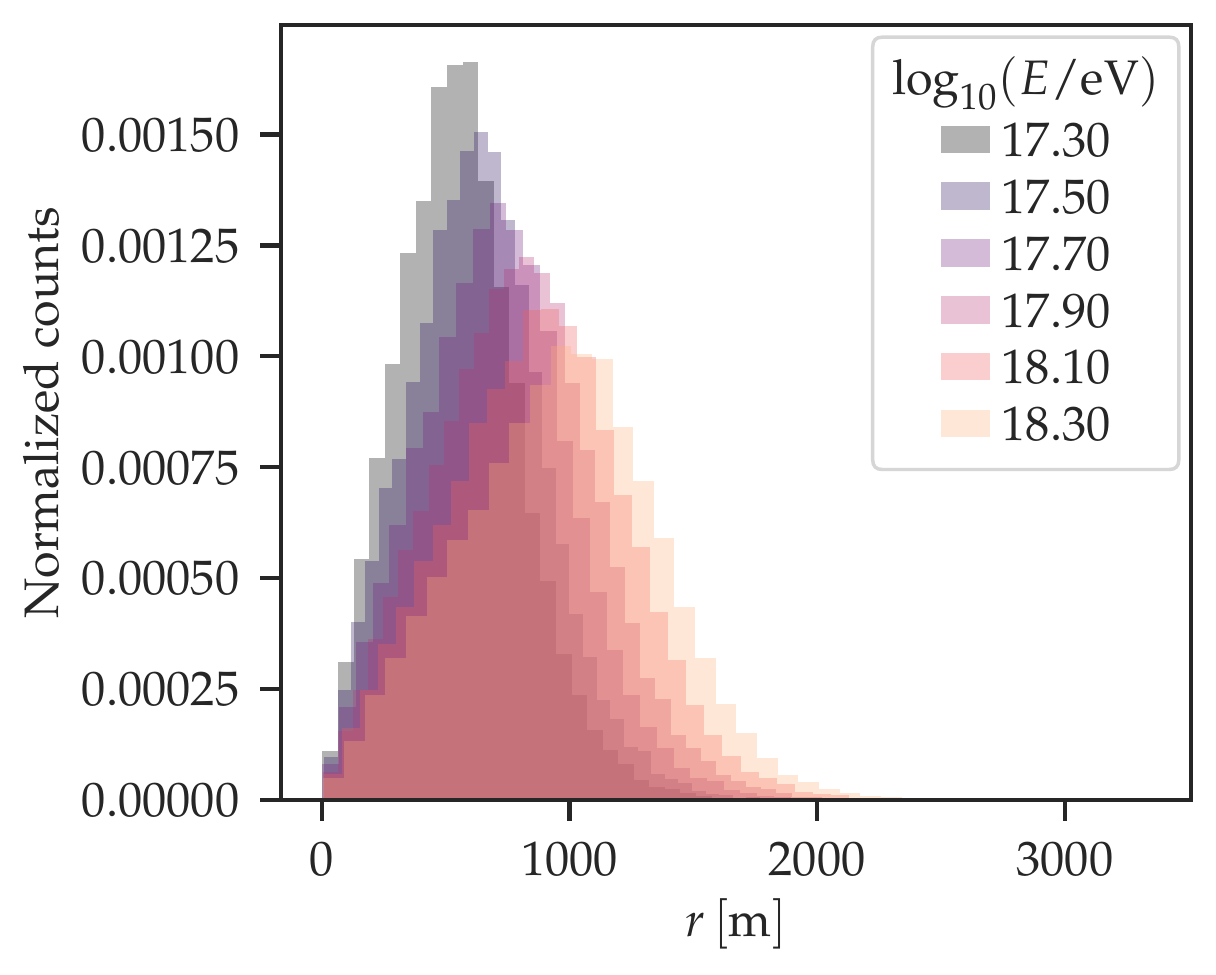}
\caption{Normalized distributions of station distances to the shower axis in a simulated hexagonal array, for a $50\,\%-50\,\%$ mixture of proton and iron, for zenith angles below $48^\circ$, and for different $\log_{10}(E/\text{eV})$-bins of width $0.2$.}
\label{fig:rdistrib}
\end{figure}

We also generate a second detector-simulations set. This one is identical to the first one, except for the distribution of distances to the shower axis. We simulate an hexagonal array of $750\,\text{m}$ spacing, as well as random event-core positions for the different primaries, energies, and zenith angles. Using the previously parameterized muon profiles integrated in time, $\mu(\log_{10}(r/\text{m}))$, we can also estimate the number of muons at the detector. We then simulate triggering by requiring that the impinging number of muons is $\geq 3$. In this way, we can obtain the distribution of the distances to the shower axis, as seen from the shower plane. We observed that the distributions of distances of a same energy are indistinguishable between different zenith angles. However, the exact distribution depends on the primary and on the energy. The latter is explained because more energetic EASs produce more muons, and therefore there are still some of them with enough energy at larger distances to the shower axis that can trigger the detector. Also, heavier primaries produce more muons than lighter primaries of the same energy and then, can trigger stations at larger distances to the shower axis. Figure \ref{fig:rdistrib} shows the distribution of station distances for the $\log_{10}(E/\text{eV})$-bins chosen in this work, for a $50\,\%-50\,\%$ mixture of proton and iron, for the hexagonal array. It can be seen that the distribution first rises linearly and then falls softly with distance, as expected. The peak of the distance distributions lie between $500\,\text{m}$ and $1000\,\text{m}$. 

It is relevant to add that we also simulated a square array of $750\,\text{m}$ spacing, obtaining indistinguishable results in the distribution of distances (cf. Fig.~\ref{fig:rdistrib}) with respect to the hexagonal array. 

\section{Evaluation of the counting strategies}\label{sec:results}

\subsection{As a function of the true and estimated number of muons}\label{subsec:res1}

For simplicity, we present the results for $N_{\mu}$ only, but the general behavior extends to $\mu$. The difference is that the resolution in the estimation of $\mu$ is also subject to the Poissonian fluctuations in $N_{\mu}$. These are dominant when $N_{\mu}$ is much smaller than the number of segments. In contrast, the detector segmentation is dominant when $N_{\mu}$ is large, meaning that the resolution in the estimators of $N_{\mu}$ and $\mu$ is similar in that case \cite{Ravignani2014}. 

\begin{figure}[!ht]
  \centering
  \includegraphics[width=0.45\textwidth]{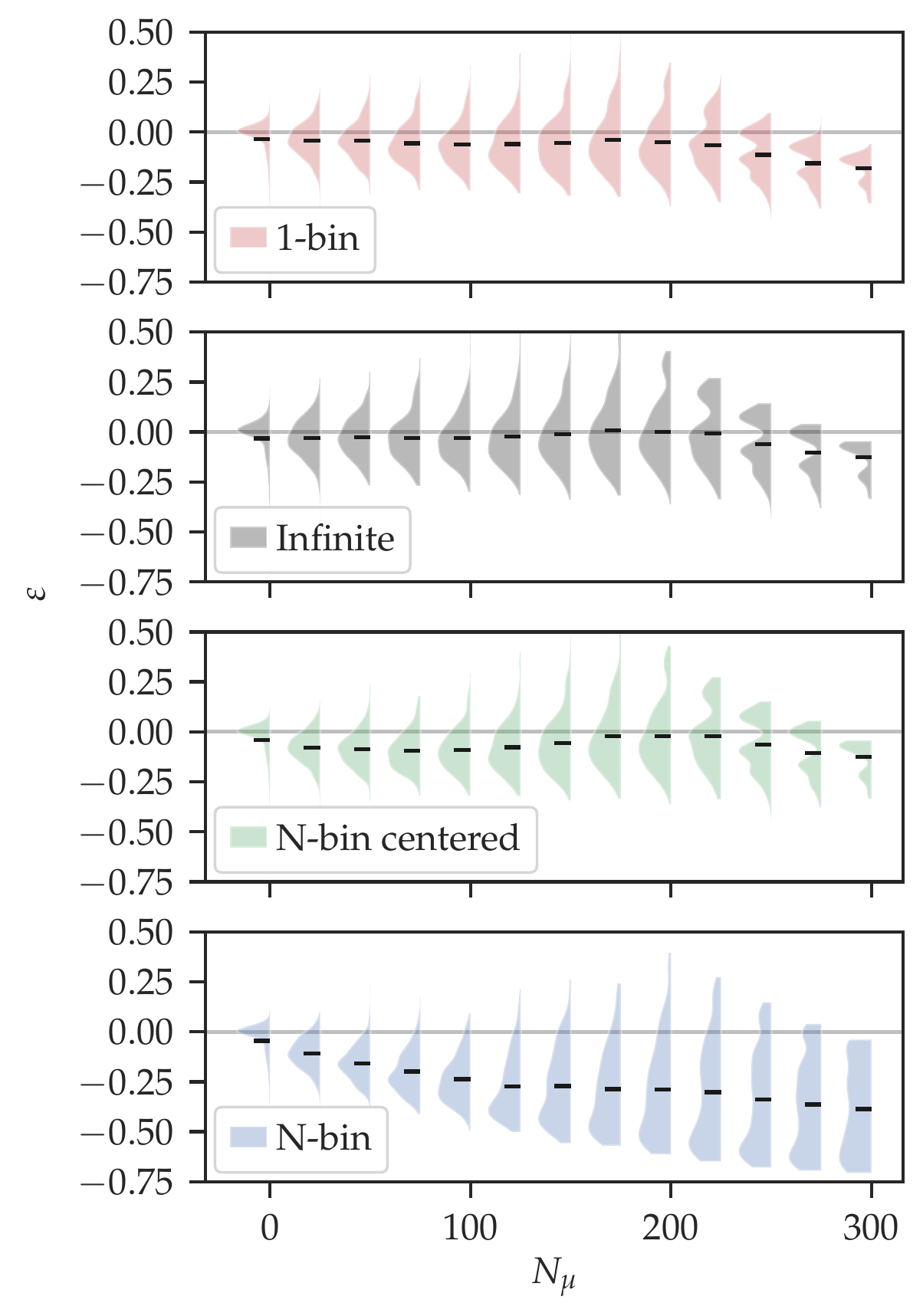}
\caption{Relative difference of the muon number estimator to the input number of muons (cf. Eq.~(\ref{eq:epsilon})) as a function of the input number of muons, for the 1-bin (top), the infinite window (second to top), the N-bin centered (second to bottom), and the N-bin (bottom) strategies. The profiles were generated from sampling uniformly in $\log_{10}(r/\text{m})$ the average muon profile of proton showers with $17.8 \leq \log_{10}(E/\text{eV}) \leq 18.0$ and $33^\circ \lesssim \theta \lesssim 39^\circ$ ($ 0.30 \leq\sin^2\theta \leq 0.40$).
}
\label{fig:violins}
\end{figure}

We start by defining the relative difference between the estimated and input (or true) number of muons as
\begin{equation}
\varepsilon = \frac{\widehat{N}_{\mu}-N_{\mu}}{N_{\mu}}.
\label{eq:epsilon}
\end{equation}  
Figure \ref{fig:violins} shows $\varepsilon$ as a function of the incident number of muons $N_{\mu}$, for proton showers with $17.8 \leq \log_{10}(E/\text{eV}) \leq 18.0$ and $33^\circ \lesssim \theta \lesssim 39^\circ$ ($0.3\leq \sin^2\theta \leq 0.4$). A small black line segment marks the mean for each distribution. Modules with saturation, for which $\widehat{N}_{\mu}$ tends to infinity, are excluded. For reference, we expect that $1\,\%$ of the modules are saturated when the number of incident muons is $N_{\mu}=176$ (see Appendix \ref{sec:appsat}). The distributions start to have multiple peaks and an increasing negative bias when saturation is significant. Furthermore, it can be seen that the N-bin strategy presents a larger variance. This is due to the random placement of the signal in the partition of the trace: a set of otherwise identical binary signals $V_{\text{out}}$ with different start times will be reconstructed as different numbers of muons simply because the signal gets partitioned differently.

\begin{figure}[!ht]
  \centering
  \includegraphics[width=0.45\textwidth]{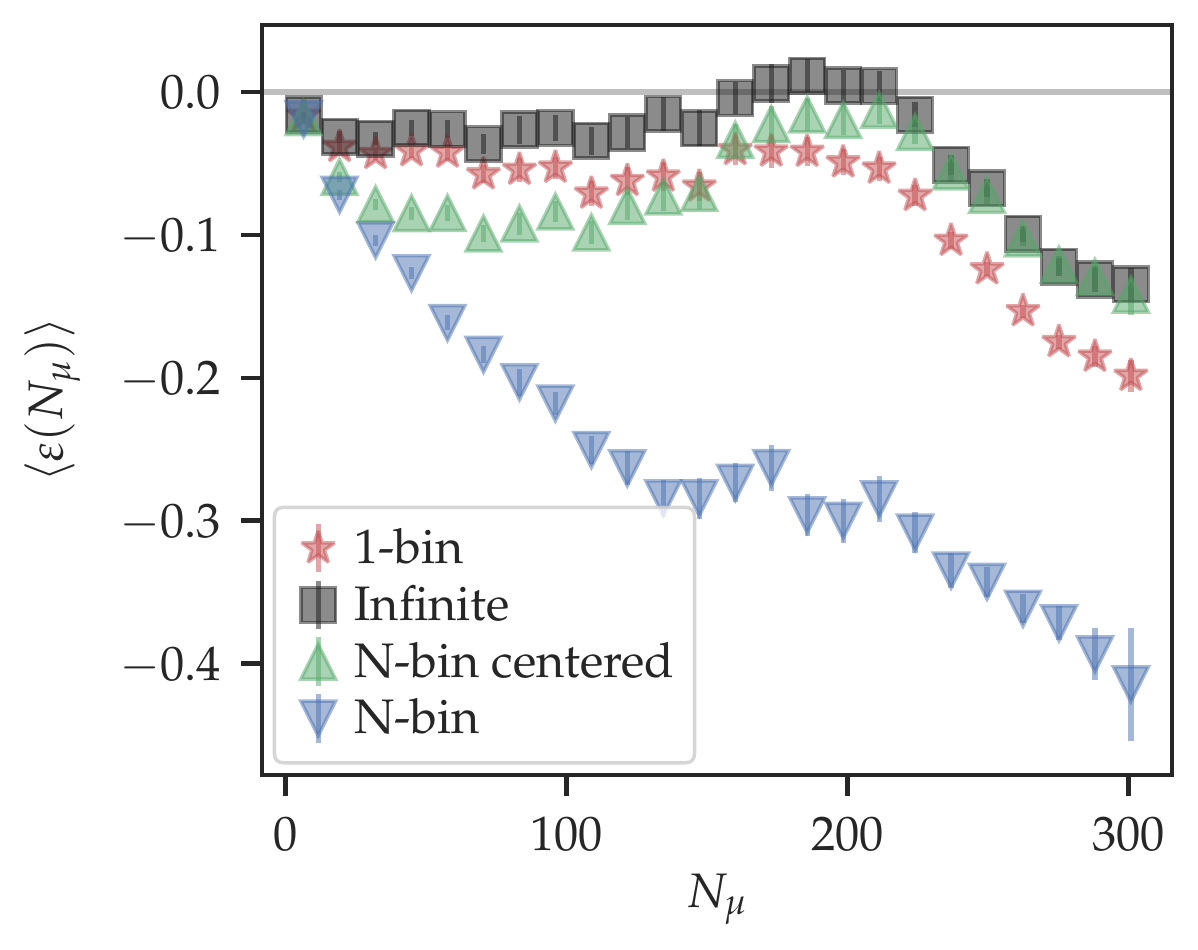}
  \includegraphics[width=0.45\textwidth]{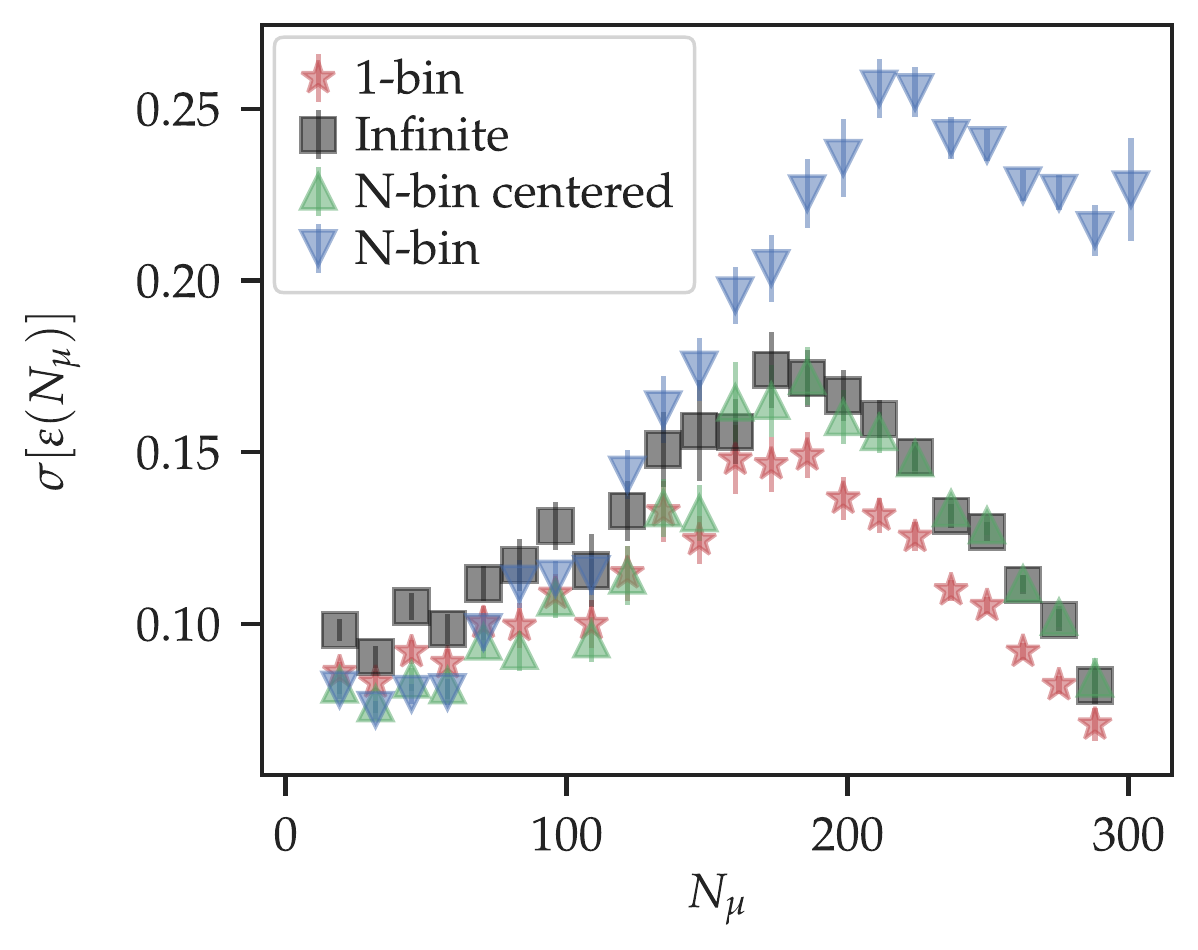}
\caption{Relative bias (top) and relative standard deviation (bottom) in the reconstructed number of muons as a function of the input number of muons for the four strategies considered. The detector-simulations set is the same as that of Fig.~\ref{fig:violins}.}
\label{fig:comparison}
\end{figure}

Figure \ref{fig:comparison} summarizes the information of Fig.~\ref{fig:violins} presenting a comparison of the relative bias $\langle \varepsilon \rangle$ (top panel) and the relative resolution $\sigma(\varepsilon)$ (bottom panel) of each strategy. The error bars on the mean $\varepsilon$ represent the standard deviation of the mean. For the case of the standard deviation of $\varepsilon$ the error bars were computed using bootstrap. At first glance it is evident that the N-bin strategy presents a significantly larger relative bias (to negative values), and also larger standard deviation than the other strategies. Moreover, the N-bin centered, the infinite window, and the 1-bin strategies have an approximately similar performance with respect to the mean relative bias: they all have a small, mostly negative bias (within $\pm 10\,\%$) dominated by detector effects for an input of $N_{\mu} \lesssim 200$, presenting an increasingly negative bias above that when the detector starts to saturate. In particular, the mean relative bias of the N-bin centered strategy tends to more negative values, like the N-bin strategy, because of not considering inhibited channels. The 1-bin strategy behaves like the infinite window strategy would with a smaller detector, but tends to slightly more negative values due to the effect of the undershoot. The infinite window strategy is not exempt from bias but it has the smallest. When comparing the standard deviations, we notice that the precision of the 1-bin strategy is approximately equal or greater than that of the other strategies.

The behavior of the bias as a function of the input number of muons of each strategy is similar when considering iron simulations or other $(\log_{10}(E/\text{eV}),$ $\sin^2\theta)$-bins. 

\begin{figure}[!ht]
  \centering
  \includegraphics[width=0.45\textwidth]{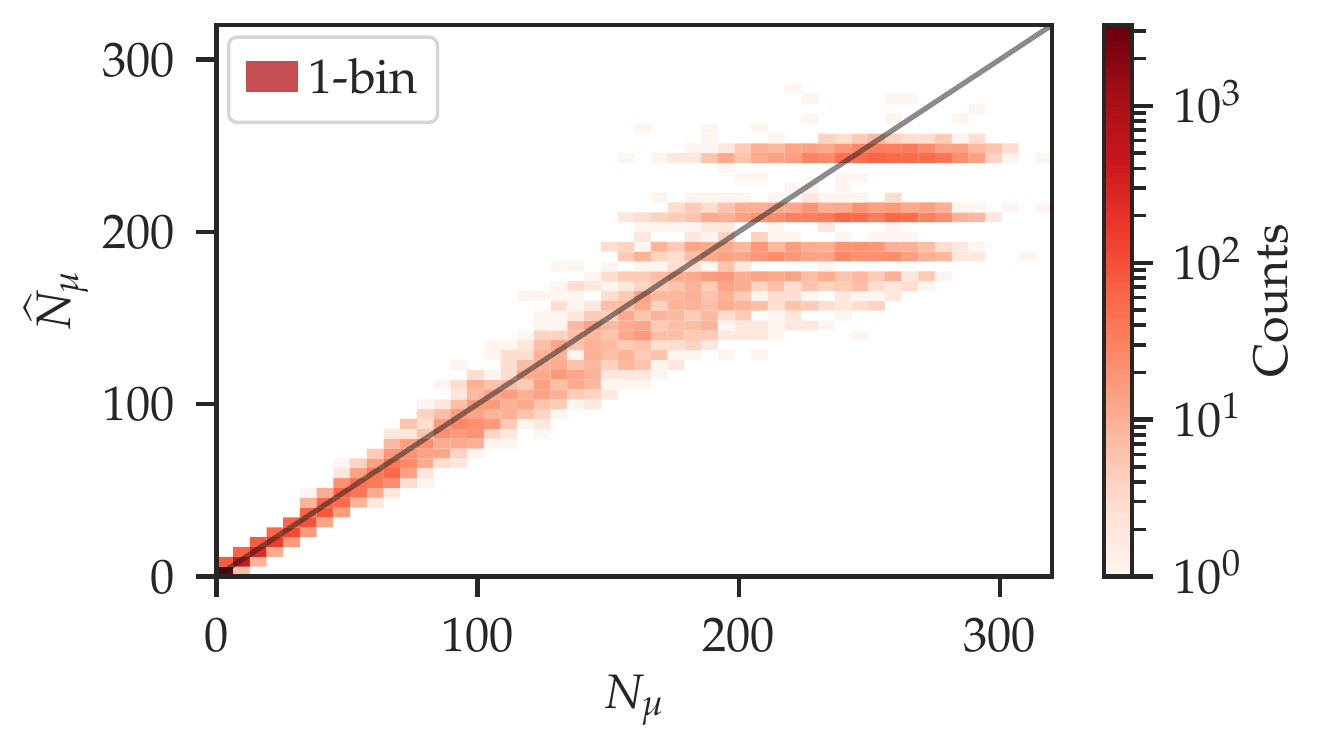}
    \includegraphics[width=0.45\textwidth]{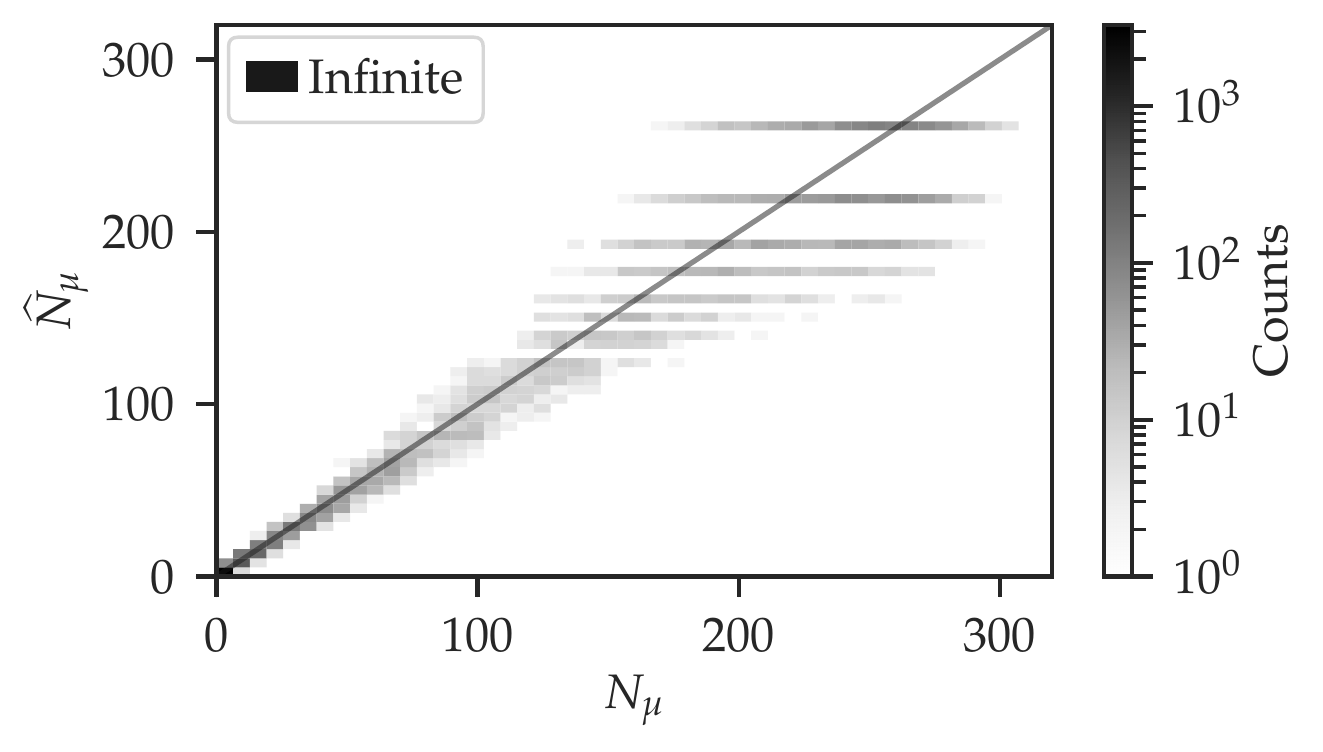}
  \includegraphics[width=0.45\textwidth]{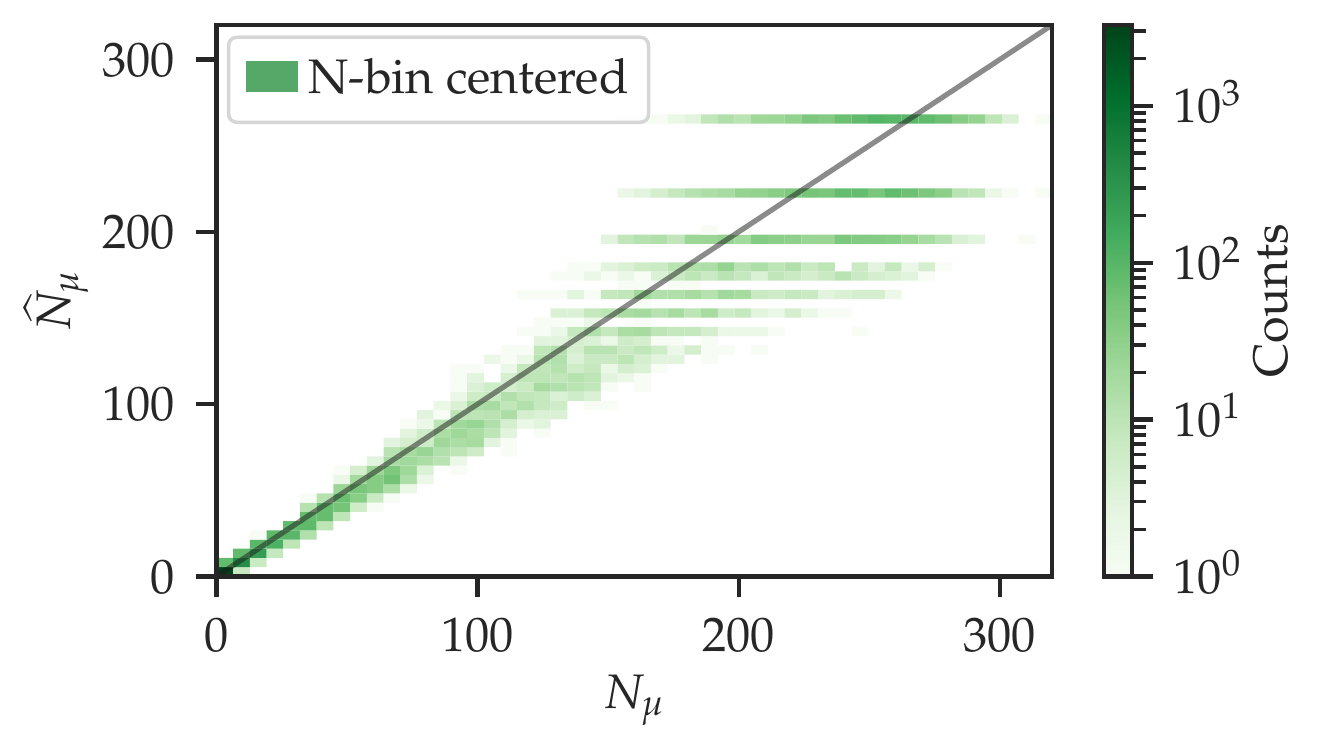}
  \includegraphics[width=0.45\textwidth]{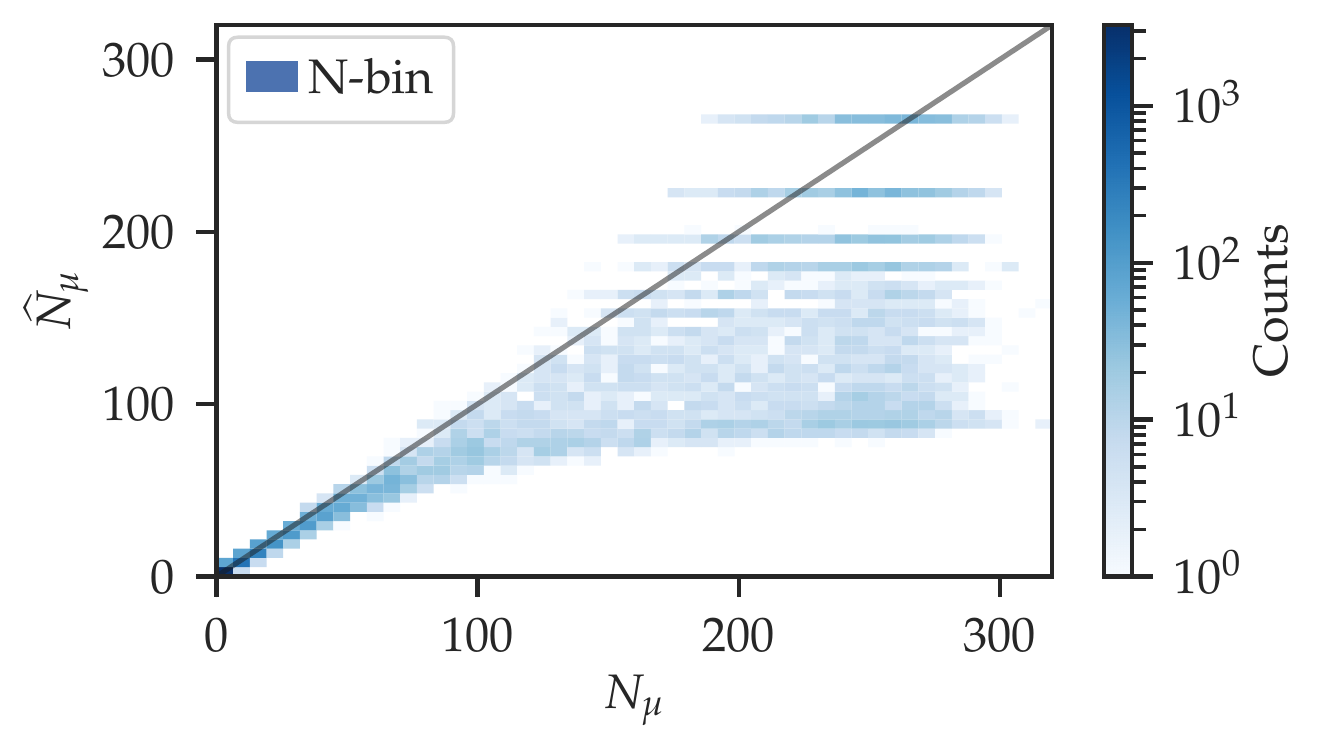}
\caption{Histograms of the estimated number of muons as a function of the input (true) number of muons for the four counting strategies considered: the 1-bin (top), the infinite window (second to top), the N-bin centered (second to bottom), and the N-bin (bottom). A grey solid line represents the identity. The detector-simulations set is the same as that of Fig.~\ref{fig:violins}.}
\label{fig:comparisonhist}
\end{figure}

It is also useful to understand how the estimated number of muons is distributed with respect to the input number of muons. Figure \ref{fig:comparisonhist} shows density histograms of the estimated against true number of muons for each of the four strategies. In general we can see that all strategies but the N-bin one distribute reasonably well around the identity (solid grey line). The N-bin strategy clearly deviates from identity. Furthermore, quantization can be noticed at large estimated values for all strategies. This is expected, and can be most easily understood taking as example the infinite window strategy: the largest non-infinite value that $\widehat{N}_{\mu}$ can take in the infinite window strategy is given just before saturation, when $k=n_s -1$. From Eq.~(\ref{eq:nmuiw}), this value is $\widehat{N}_{\mu} \sim 264$ for $n_s=64$. The next highest possible value is given when $k = n_s -2$, at which $\widehat{N}_{\mu} \sim 220$. The other possible values are given when $k=n_s-3$, $k=n_s-4$, ..., until $k=0$. For the other counting strategies, there is still quantization, but the combinations allow for more possible values of $\widehat{N}_{\mu}$. 

\begin{figure}[!ht]
  \centering
  \includegraphics[width=0.45\textwidth]{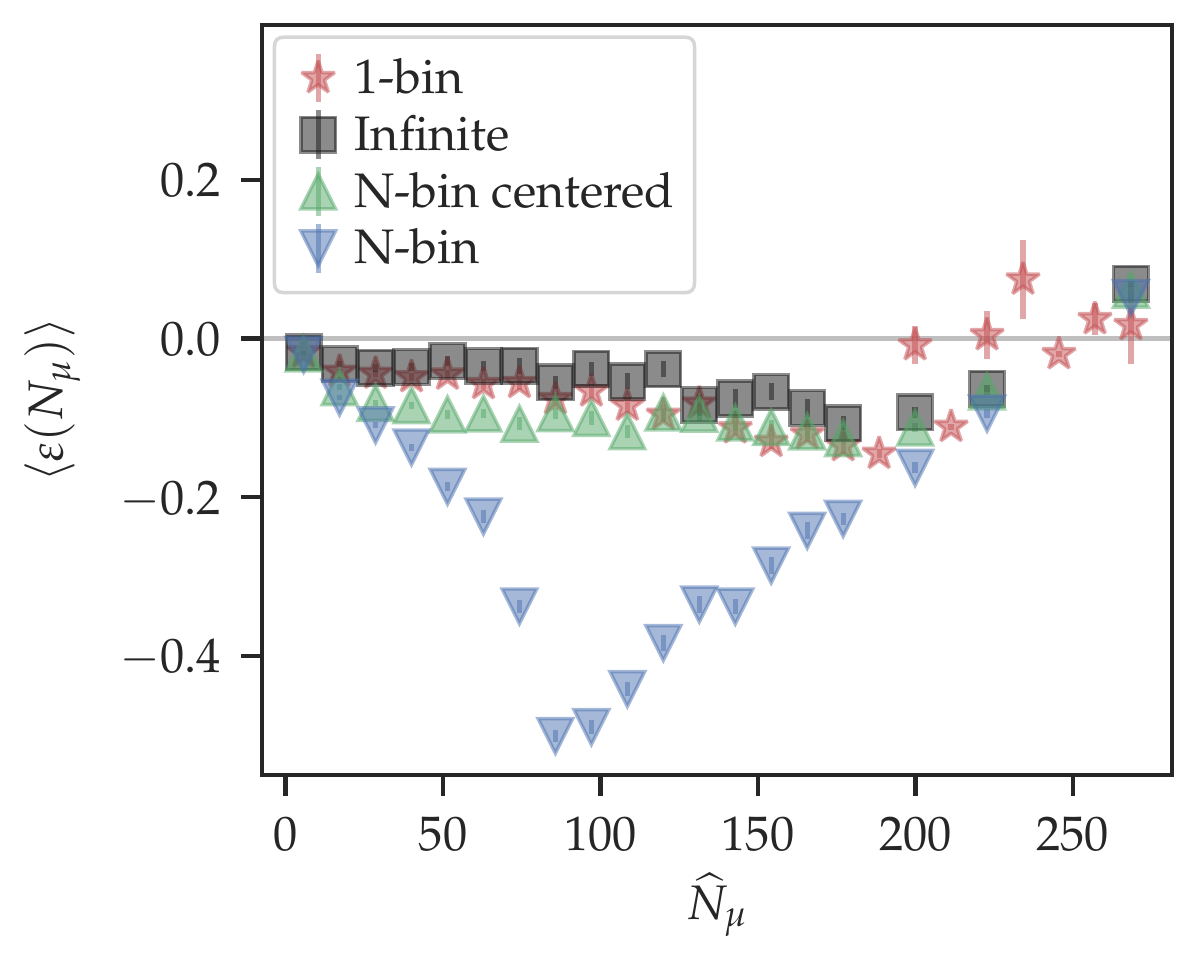}
  \includegraphics[width=0.45\textwidth]{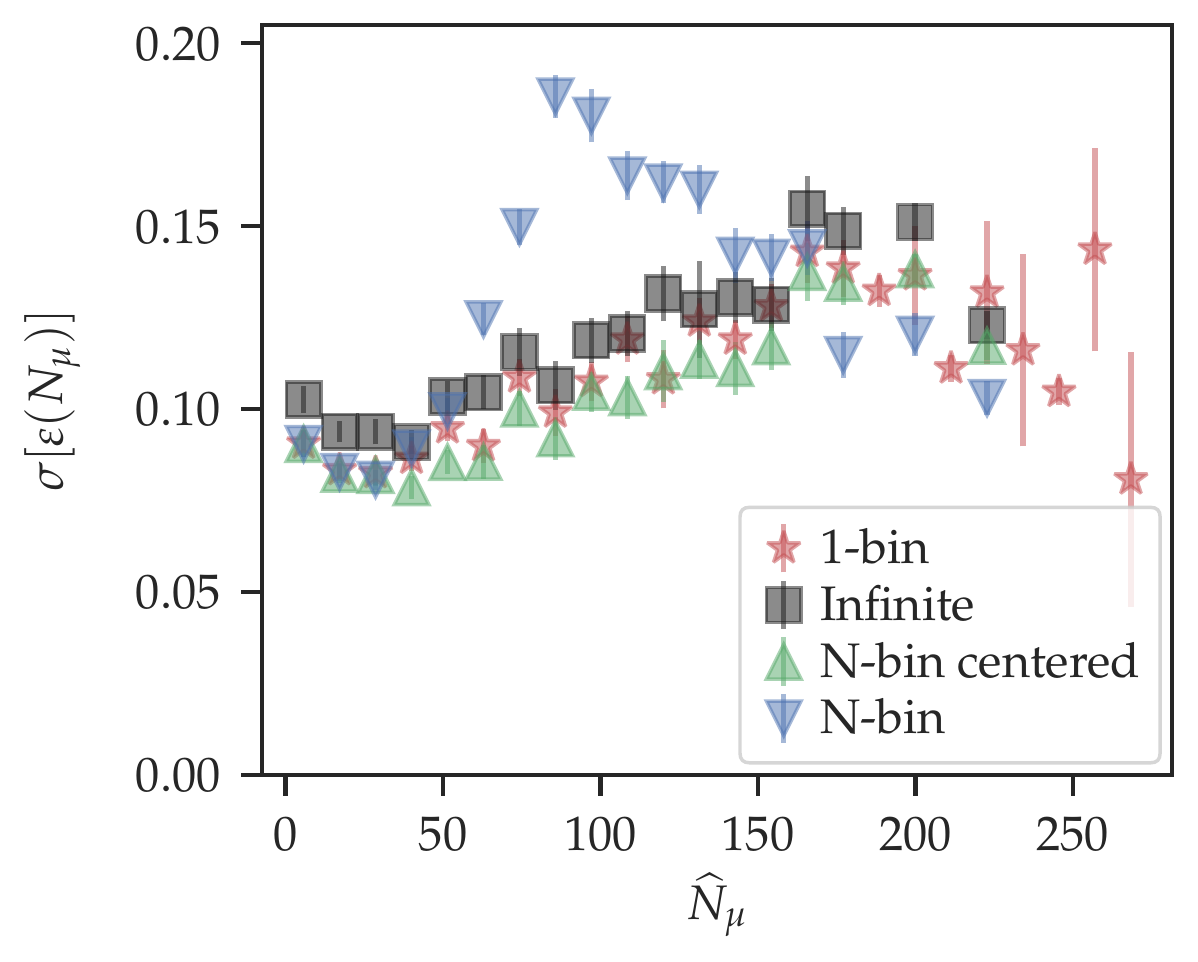}
\caption{Same as Fig.~\ref{fig:comparison} but with the estimated number of muons in the x-axis.}
\label{fig:comparison2}
\end{figure}

Since in real experimental data the true input number of muons is often impossible to know, it is of interest to analyze how the relative bias and the relative resolution depend on the estimated number of muons. This is displayed in Fig.~\ref{fig:comparison2}. The figure is equivalent to Fig.~\ref{fig:comparison} but plotting against the estimated number of muons instead of the input ones. Similar to Fig.~\ref{fig:comparison2}, all strategies behave similarly except the N-bin one. The 1-bin, infinite window, and N-bin centered strategies present a small relative bias (contained within $\pm 10\,\%$) and an increasing relative standard deviation until the point in which saturation becomes more frequent. The smallest relative bias is that of the infinite window strategy, but the 1-bin strategy follows it closely. The N-bin strategy presents a behavior that is easier to understand from Fig.~\ref{fig:comparisonhist}: At $\widehat{N}_{\mu} \sim 100$ most points satisfy $\widehat{N}_{\mu} < N_\mu$, causing the large negative average relative bias observed in Fig.~\ref{fig:comparison2}. While for larger values of $\widehat{N}_{\mu}$ the average relative bias approaches zero, from Fig.~\ref{fig:comparisonhist} or Fig.~\ref{fig:comparison}, we can see that it is not likely that the true $N_{\mu}$ maps to the estimated $\widehat{N}_{\mu}$.

It is relevant to add that, although the mean biases of the 1-bin, N-bin centered, and infinite window strategies are small, they could be further reduced using simulations. An empirical fit to the mean value of $\varepsilon$ as a function of $\widehat{N}_\mu$ could be performed, and then the fitted function could be used to correct the mean bias.

\subsection{In realistic air-shower scenarios}
The purpose of the analysis described in Sec.~\ref{subsec:res1} is to understand the performance of the different strategies as a function of the input number of muons. However, in realistic data, larger values of input number of muons $N_{\mu}$ happen with less frequency and only close to the core. 

\begin{figure}[!ht]
  \centering
  \includegraphics[width=0.45\textwidth]{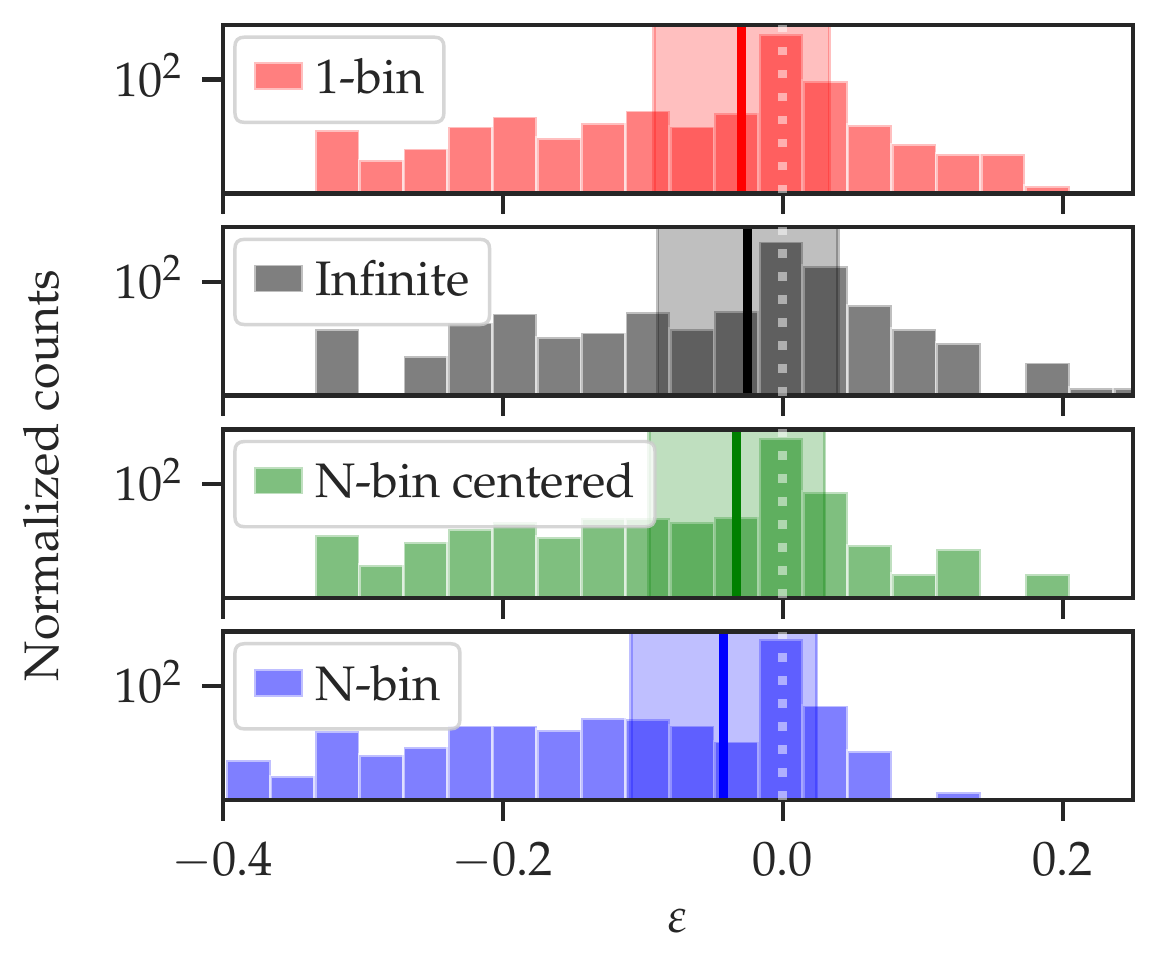}
\caption{Distributions of $\varepsilon$ for the four strategies considered. The width of the colored vertical line, centered at the mean, corresponds to the standard deviation of the mean, and the light shaded area is the standard deviation. A vertical white dashed line marks null bias. We exclude events that do not pass the distance cut $r>r_{\text{min}}$ (see text for details). 
Each count is one reconstructed trace, which was generated from sampling $r$ from the realistic  distribution (c.f. Fig.~\ref{fig:rdistrib}), and taking as input the average muon profile of proton showers in an hexagonal array with $17.6 \leq \log_{10}(E/\text{eV}) \leq 17.8$ and $ 27^\circ \lesssim \theta \lesssim 33^\circ$ ($0.20 \leq \sin^2\theta \leq 0.30$).}
\label{fig:biasdata}
\end{figure}

To understand how the different strategies affect the overall bias in the reconstructed number of muons for a given primary and $(\log_{10}(E/\text{eV}),$ $\sin^2\theta)$-bin, we use the second detector-simulations set. Figure \ref{fig:biasdata} shows the distribution of $\varepsilon$ for 1000 input proton showers with $17.6 \leq \log_{10}(E/\text{eV}) \leq 17.8$ and $ 27^\circ \lesssim \theta \lesssim 33^\circ$ ($0.20 \leq \sin^2\theta \leq 0.30$), for the four strategies considered, for the hexagonal array. The results for other energies and angles are similar. We exclude simulated (module-level) events that would not pass the lower cut $r>r_{\text{min}}$, where $r_{\text{min}}$ is the furthest distance to the shower core where there is saturation for the analyzed strategy. At these energies the exclusion of these events has a negligible effect, but at larger energies it makes $\varepsilon$ slightly more positive for all strategies. A white dashed line marks null bias. The light shaded area presents the standard deviation of $\varepsilon$ and the colored line shows its mean. The standard deviation of the mean is contained within the colored line marking the mean. We can see that in this example the mean relative bias is mostly small, dominated by detector effects, taking a value of $-2.5\,\%$ for the infinite window strategy, of $-2.9\,\%$ for the 1-bin strategy, of $-3.3\,\%$ for the N-bin centered strategy, and of $-4.2\,\%$ for the N-bin strategy. 

\begin{figure}[!h]
\begin{subfigure}{0.44\textwidth}
  \centering
  \includegraphics[width=\textwidth]{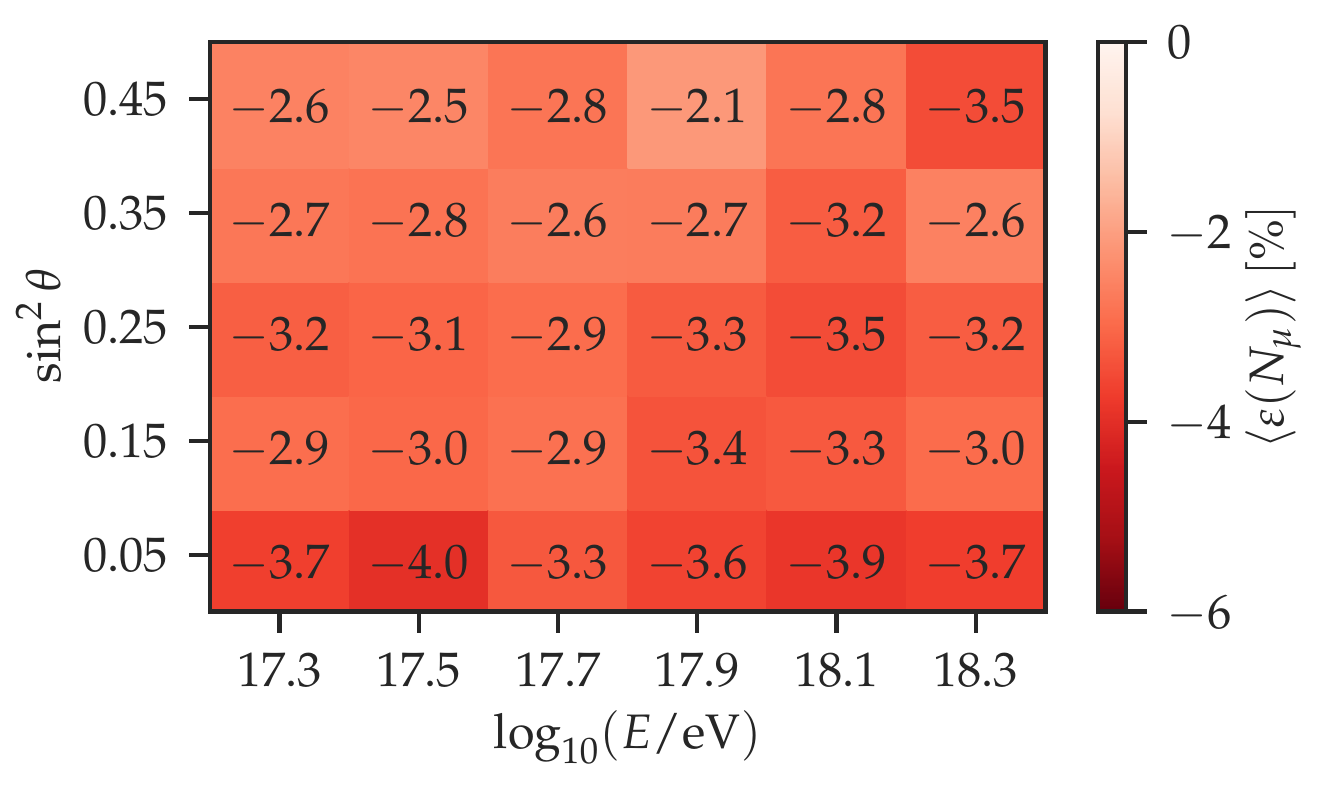}
\end{subfigure}
\begin{subfigure}{0.44\textwidth}
  \centering
  \includegraphics[width=\textwidth]{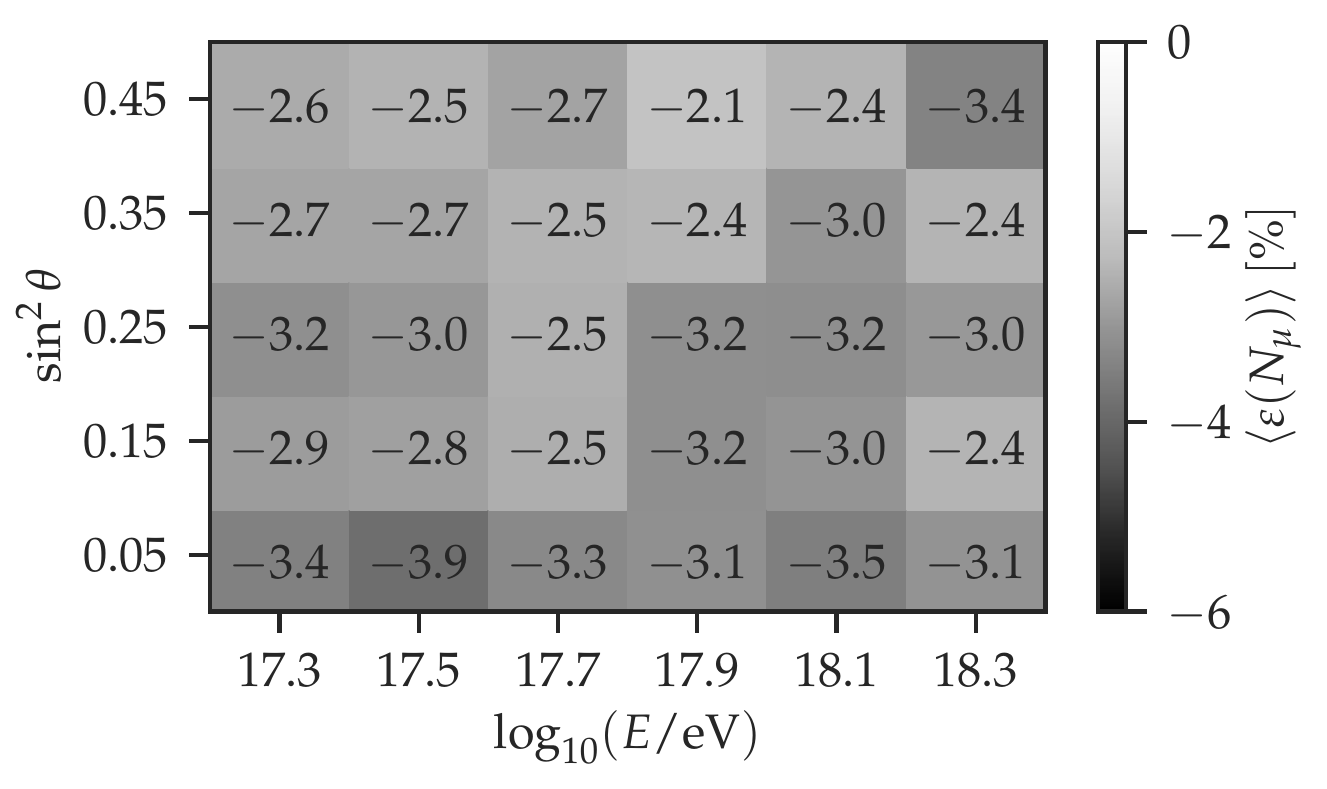}
\end{subfigure}
\begin{subfigure}{0.44\textwidth}
  \centering
  \includegraphics[width=\textwidth]{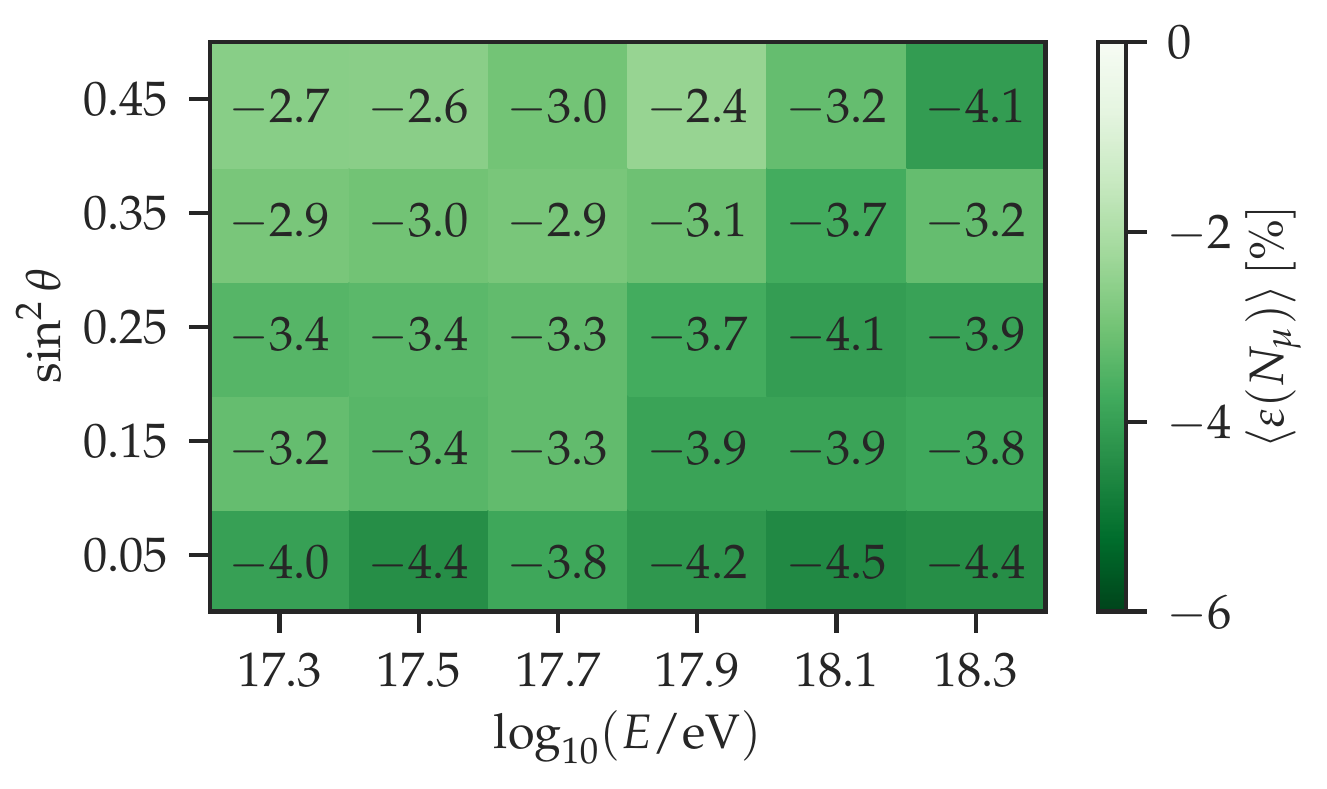}
\end{subfigure}
\begin{subfigure}{0.44\textwidth}
  \centering
  \includegraphics[width=\textwidth]{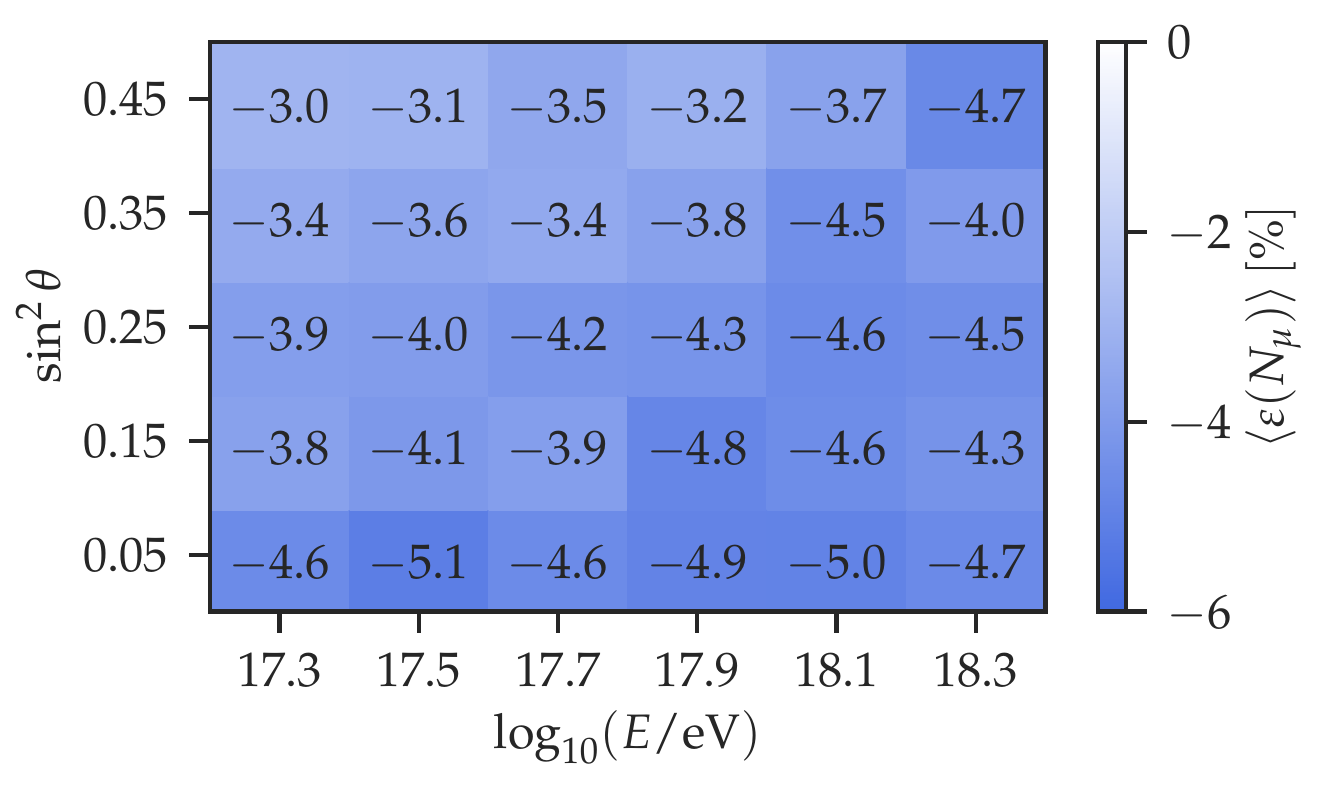}
\end{subfigure}
\caption{Mean $\varepsilon$ as a function of the logarithmic energy and the sine square of the zenith angle, for the 1-bin (top), the infinite window (second to top), the N-bin centered (second to bottom), and the N-bin (bottom) strategies. We use proton showers in an hexagonal array with distance cuts. The distance distribution is that of Fig.~\ref{fig:rdistrib}.}
\label{fig:biasall}
\end{figure}

Figure \ref{fig:biasall} summarizes the mean $\varepsilon$ as a function of $\log_{10}(E/\text{eV})$ and $\sin^2\theta$ for all strategies, considering proton showers in an hexagonal array and including the distance cuts. We can see that for the 1-bin strategy the bias is contained within $\pm 4.0\,\%$, for the infinite window within $\pm 3.9\,\%$, and for the N-bin centered strategy within $\pm 4.4\,\%$. The 1-bin strategy has a slightly larger bias, which is expected from the effects of undershoot. For the N-bin strategy, the bias is larger (more negative) reaching $-5.1\,\%$. In general, the average biases are small because, as mentioned before, most measurements happen at larger distances to the shower axis, implying a lower number of input muons for which the biases are small. Most of the bias is caused by detector effects mentioned in Sec.~\ref{sec:deteff}. We remind the reader that the infinite window strategy is insensitive to undershoot (so it is only affected by the efficiency of the detector and pattern matching strategy), while all other strategies suffer from both effects. It is the detector effects that make all biases take negative values.

We can also add that, for iron showers, the biases are in comparison slightly larger for all strategies. This is shown in Fig. \ref{fig:biasallFe} in the Appendix \ref{sec:appiron}. The distance cut in all cases does not have a significant effect: it can only make the average bias in a $(\log_{10}(E/\text{eV}),$ $\sin^2\theta)$-bin change at most by $0.2\,\%$. 

\subsection{For large or double-bump-like signals}

\begin{figure}[!ht]
  \centering
  \includegraphics[width=0.45\textwidth]{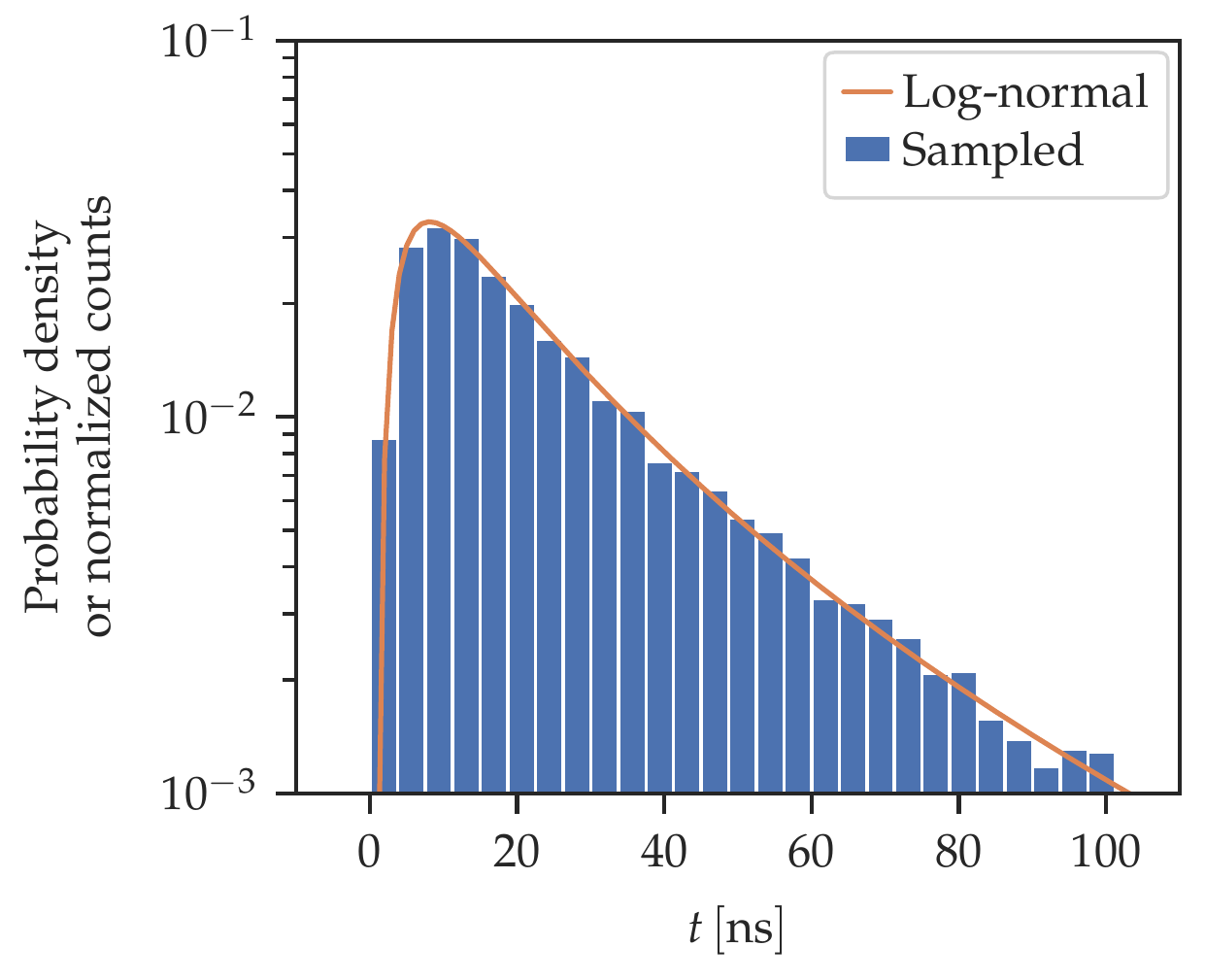}
  \includegraphics[width=0.45\textwidth]{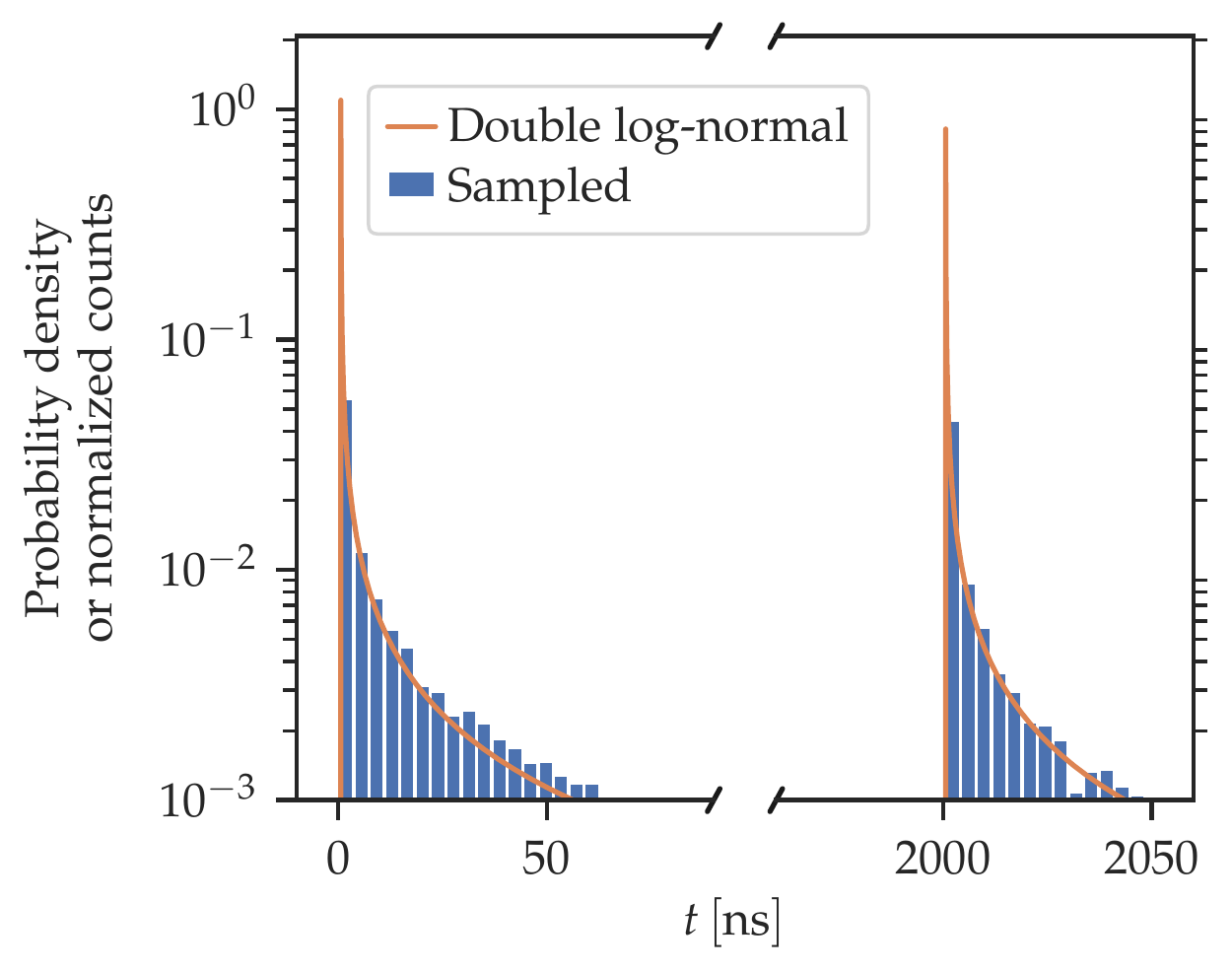}
\caption{Input signals used to test the strategies. On the top panel, a \enquote{wide} log-normal signal with a shape parameter $\sigma \approx 1$, scale parameter $m \approx 3$, and location parameter $t_0 \approx 0.6\,\text{ns}$. On the bottom panel, two log-normals with the shape and scale parameters of a proton shower of $\log_{10}(E/\text{eV}) = 18.1$ and $\theta \approx 30^\circ$, and location parameters $t_{0,1} \approx 0.6\,\text{ns}$ and $t_{0,2} \approx 2000.6\,\text{ns}$, with the second one having an amplitude $25\,\%$ smaller than the first one.}
\label{fig:inputOffFail}
\end{figure}

At first it could be thought that the biases of the N-bin strategy arise from splitting the largest part of the signal into two windows. The idea behind it is that the beginning of the matched patterns that fall into one window are mathematically treated as simultaneous, and those of consecutive windows are treated as independent. When the signal is split into two windows, the sum of the estimated muons at each window underestimates the total input number of muons. If this were indeed the problem, centering the window in the signal should be the solution. In the tested cases, the performance of the N-bin centered strategy is quite good. It has a relatively small (negative) bias, and only a larger variance than the 1-bin strategy at a large number of impinging muons. However, its relative success is actually explained because the used input signals are mostly contained within $37.5\,\text{ns}$ (one 12 time-bin window). This makes the N-bin centered strategy effectively perform as the infinite window strategy, since almost the complete signal is contained within one window. Were the window shorter or the signal wider, the N-bin centered strategy should have larger biases, as the N-bin strategy has. To test this, we design two different input signals as shown in Fig.~\ref{fig:inputOffFail}. The first one (top panel) is a log-normal
\begin{align}
\frac{\textrm{d}\mu(t)}{\textrm{d}t} =&\, \Theta((t-t_0)/\text{ns})\, \frac{A}{((t-t_0)/\text{ns})\, \sigma\, \sqrt{2 \pi}}\times \nonumber\\
& \exp\left(-\frac{(\ln((t-t_0)/\text{ns}) - m)^{2}}{2\sigma^2}\right) ,
\end{align}
with a shape parameter $\sigma \approx 1$, scale parameter $m \approx 3$, and location parameter $t_0 \approx 0.6\,\text{ns}$, with an amplitude $A$ equal to its integral $\mu \approx 337$. For comparison, the scale parameter is 4 times smaller and the amplitude 25 times larger than that of an average proton shower of $\log_{10}(E/\text{eV}) = 18.1$ and $\theta \approx30^\circ$ ($\sin^2\theta = 0.25$). It is relevant to notice that it would be equivalent to test a signal of standard width considering narrower windows (a shorter pattern), which would correspond to another detector with other electronics. The second input signal (bottom panel) is designed to have a \enquote{double bump}, and it consists of two consecutive log-normals separated by a time interval $\Delta t$ (equal to the difference between the two location parameters $t_{0,2}-t_{0,1}$) of our choice. Both log-normals have the shape and scale parameters of a proton shower of $\log_{10}(E/\text{eV}) = 18.1$ and $\theta \approx30^\circ$ ($\sin^2\theta = 0.25$), however the second one has an amplitude $25\,\%$ smaller than the first one. In total it integrates to $\mu \approx 300$.

\begin{figure}[!ht]
  \centering
  \includegraphics[width=0.45\textwidth]{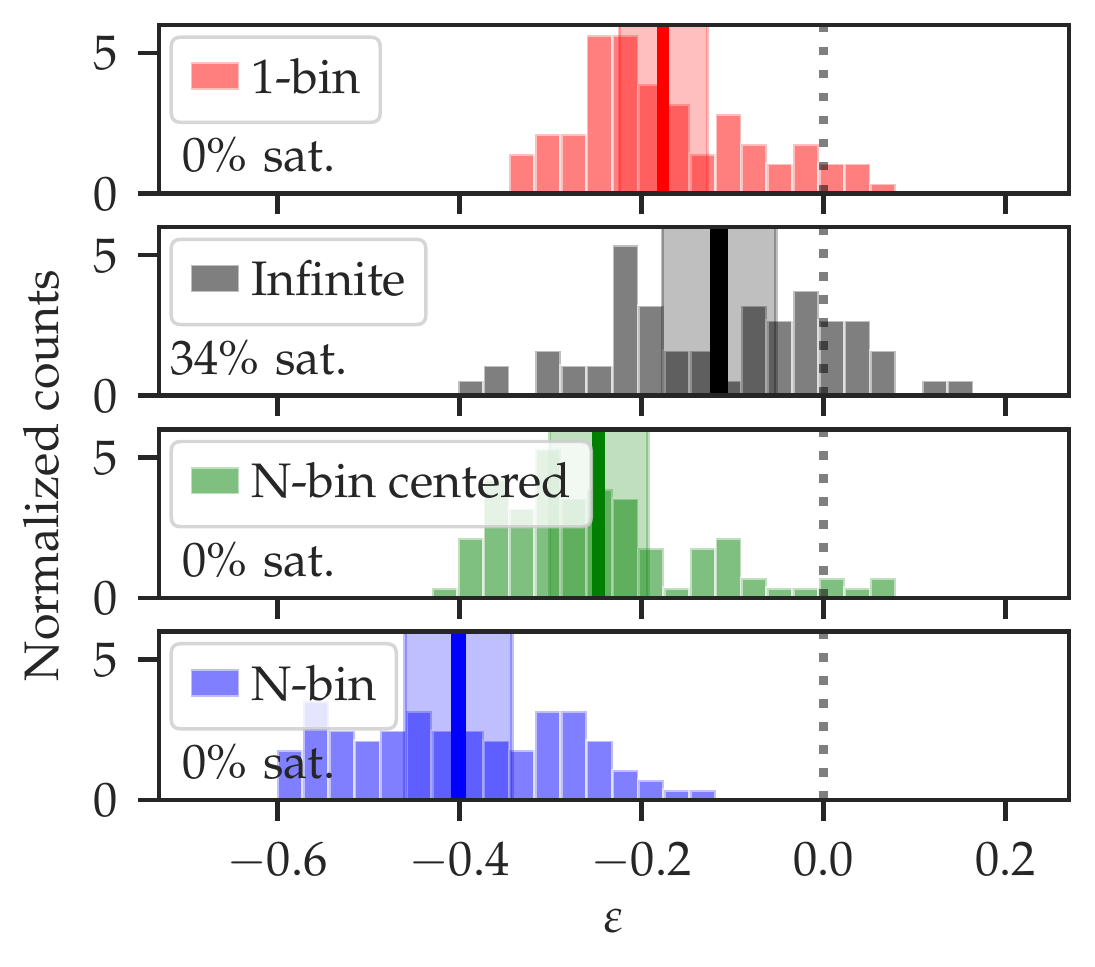}
  \includegraphics[width=0.45\textwidth]{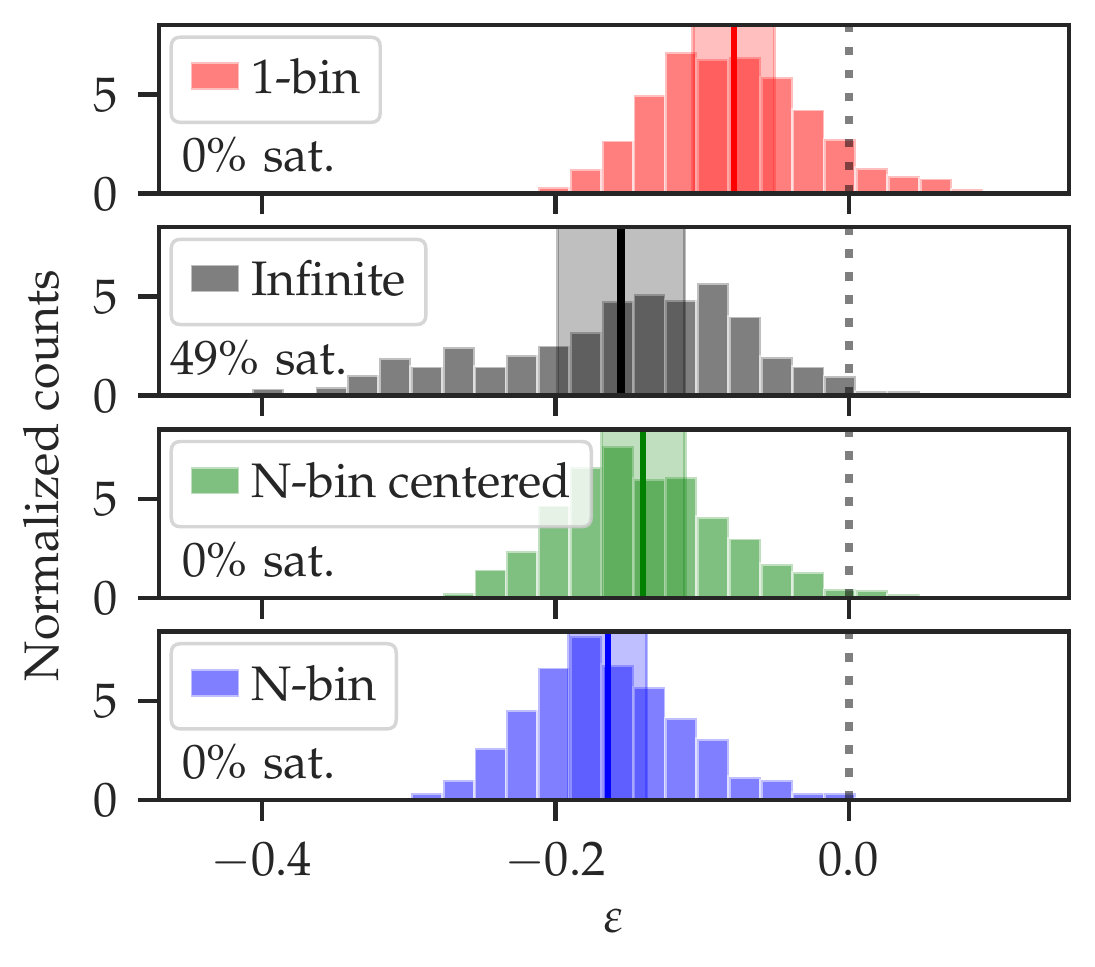}
\caption{Distributions of $\varepsilon$ for each strategy, for a wide input signal (top) and a double-bump input signal (bottom). The width of the colored vertical line, centered at the mean, corresponds to the standard deviation of the mean, and the light shaded area is the standard deviation. A vertical black dashed line marks null bias. The percentage of saturated events for each strategy and input signal is indicated in each panel.}
\label{fig:outputOffFail}
\end{figure}

We create 1048 events randomly sampling the wide signal, and 1300 randomly sampling the double-bump one. For the latter, we fix $\Delta t$ to $2000.000\,\text{ns}$, $2003.125\,\text{ns}$,..., $2037.500\,\text{ns}$, sampling 100 times each. In more detail, we sample the impinging number of muons $N_{\mu}$ from a Poisson distribution of mean $\mu$, and then the times of the $N_{\mu}$ muons from the designed signals $\textrm{d}\mu(t)/\textrm{d}t$. The distribution of $\varepsilon$ for each strategy and for the two kinds of input signals are displayed in Fig. \ref{fig:outputOffFail}. In the first case, we can see that the best performing strategy is the infinite window, despite already being in the saturation regime, followed closely by the 1-bin strategy. In the second case, the 1-bin strategy performs the best. As already mentioned, the infinite window strategy is subject to the detector inefficiencies, but is unaffected by undershoot. It is evident that it tends to saturate much more than the other strategies, which, together with the detector inefficiency, explains its bias. On the other hand, the 1-bin strategy is subject to detector inefficiencies and also to undershoot effects. The N-bin centered and the N-bin strategies have larger biases than the 1-bin strategy in both cases, as expected. This proves that the N-bin centered strategy cannot be extended to other detectors where the muon pattern is shorter, and that it underperforms in the Auger UMD when the signal is long or when there are a significant amount of late particles.

This analysis proves that the biases of the N-bin strategy are not completely solved by centering a window at the signal peak, and that therefore the biases are not explained from the signal being split into two windows. The actual reason is that the N-bin strategy does not properly model the detector, as it does not contemplate inhibited channels. This extends to the N-bin centered strategy too.

\subsection{Reconstruction of the muon signal with time resolution}

The counting strategy developed in this work takes advantage of the complete time structure of the measured signal, allowing to reconstruct the time structure of the signal as seen by the detector to a single time-bin resolution.

\begin{figure}[!ht]
  \centering
  \includegraphics[width=0.45\textwidth]{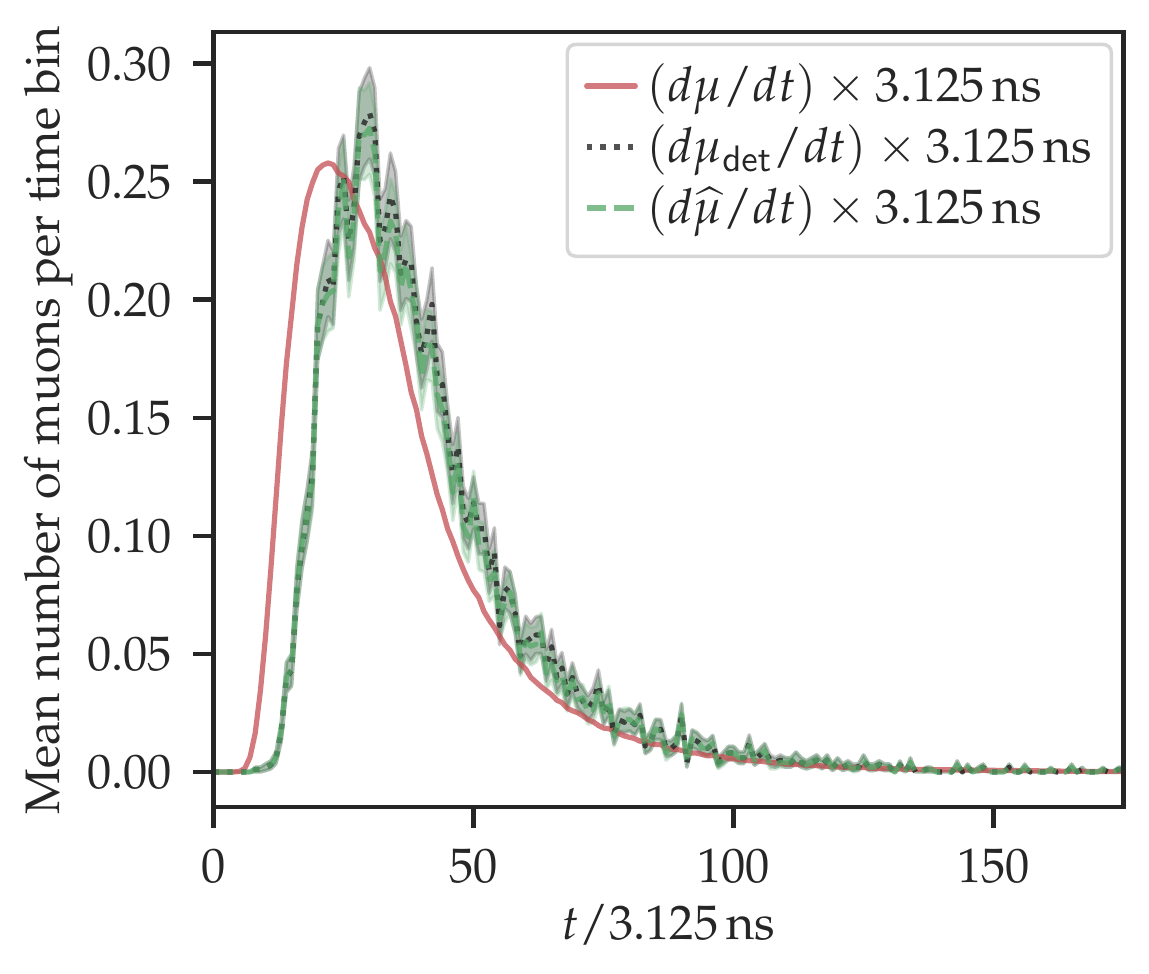}
\caption{Input number of muons (solid line), input number of muons as seen by the detector (dotted line), and reconstructed number of muons (dashed line) using the 1-bin strategy, as a function of time. The input number of muons as a function of time are the average muon profile of proton showers with $17.8 \leq \log_{10}(E/\text{eV}) \leq 18.0$ and $ \theta \lesssim 18^\circ$ ($\sin^2\theta \leq 0.10$) at $450\,\text{m}$ from the core. The average number of muons as seen by the detector is computed by introducing the time delays from the detector in the input muons. The shaded areas represent $1\,\sigma$ uncertainties.}
\label{fig:tempstruct}
\end{figure}

We present in Fig.~\ref{fig:tempstruct} an example of the input muons as a function of time, the average input number of muons as seen by the detector as a function of time, and the estimated average number of muons as a function of time, the latter using the 1-bin strategy. The input number of muons as a function of time are the average muon profile of proton showers with $17.8 \leq \log_{10}(E/\text{eV}) \leq 18.0$ and $ \theta \lesssim 18^\circ$ ($\sin^2\theta \leq 0.10$) at $450\,\text{m}$ from the core. The time delay between the input muons and the muons seen by the detector is expected from the propagation of the photons in the optic fiber, as well as from the scintillator and optic fiber delays. 

The counting strategy developed in this work opens the door to studies on the temporal structure of the muon signal. This could be used to estimate the depth-of-the-shower maximum of muons \cite{Cazon2005}, which is a great composition-sensitive parameter. Future work will be centered in deconvolving the detector effects in the reconstructed average muon signal, to obtain an estimate of the input muons as a function of time.

\section{Summary and outlook}\label{sec:conclusions} 

Segmented particle counters, or counters with sub-units in general, can be used to estimate the number of impinging particles. In doing so, the effect of pile-up needs to be accounted for. This effect occurs when two or more particles hit the same segment (or sub-unit) of the detector within a time interval so short that they cannot be individually resolved. It constitutes a source of undercounting. 

In this paper we presented and compared four counting strategies that account for the pile-up effect: the simplest \enquote{infinite window}, the N-bin strategy, the \enquote{N-bin centered} strategy (an improved version of the N-bin strategy), and the 1-bin strategy developed in this work. They are all based in the solution to the classical occupancy problem, or, as stated in Sec.~\ref{sec:strat}, the \enquote{balls in boxes problem} (where the balls are the particles and the boxes are the segments). The main difference between the strategies is the duration of the window in which the muon pattern matches of different segments are considered simultaneous: the complete trace for the infinite window strategy, the length of the single-muon pattern for the N-bin and N-bin centered strategies, and the time resolution for the 1-bin strategy. The difference between the N-bin and N-bin centered strategies is that the position of the windows is optimized for the latter, such that the center of the output signal coincides with the center of a certain window. The key difference of the 1-bin strategy with respect to the rest is the consideration of inhibited segments.

These general counting methods find a concrete application in the muon counters of the Underground Muon Detector of the Pierre Auger Observatory. We thus took the latter as our study case, to test and compare the different counting strategies.

We showed that for typical time structures of the muon signal in a detector like the Auger UMD, all strategies but the N-bin one perform well. For all strategies but the latter, the mean bias is contained within $\pm 10\,\%$ as long as saturation is not significant. At high input number of muons, the standard deviation of the estimated number of muons of the 1-bin strategy is the smallest and the one of the N-bin strategy the largest. We also observed that the average bias of each strategy for every $(\log_{10}(E/\text{eV}),$ $\sin^2\theta)$-bin for typical data are small (within $\pm 6\,\%$), but that they are larger (more negative) for the N-bin strategy. Most of the bias is actually caused by detector effects, due to the signal undershoot and the detector and pattern matching strategy efficiencies. 

When taking an input signal with a wider time structure or with a double-bump-like signal, we observed that the 1-bin and infinite window strategies perform the best. That analysis proved that the biases found in the N-bin strategy are inherent to the method, which does not take into account the inhibited channels as such. The N-bin centered strategy has the same design problem.

Lastly, we showed that we can reconstruct the time structure of the muon signal as seen by the detector to a single time-bin resolution using the 1-bin strategy. No other known strategy offers such resolution. This opens the door to new studies of the temporal structure of the muon component. Such studies can be key for composition analyses. 

Our analyses lead to the conclusion that the counting strategy developed in this work is generally as good or better than the other strategies. It only has a slightly larger bias than the infinite window one, but unlike the latter, it can be used to infer the muon time structure as seen by the detector to a single time-bin resolution.

Finally, the counting strategy developed in this work could be possibly applied for counting photons produced in liquid scintillators with photo-multiplier tubes (PMTs), a typical layout of neutrino experiments. As mentioned in Sec.~\ref{sec:intro}, the strategy is only useful when the processing of the PMT signals is performed applying a discrimination threshold (counter mode), and not by analyzing the amplitude or charge of the signal (integrator mode). In such context, the estimated number of scintillation photons produced in a neutrino event can be used to estimate the energy of the neutrino (see for example Ref.~\cite{Wu2019}). More precisely, the counting strategy could be applied to each set of PMTs lying equidistant to the interaction vertex, where the expected scintillation photon rate is the same. In spherical detectors, these set of PMTs lie in rings that correspond to the intersection of spheres centered in the vertex with the greater spherical array of PMTs. This would be the case, for example, of the Sudbury Neutrino Observatory + (SNO+) \cite{SNOplus}, and of the Jiangmen Underground Neutrino Observatory (JUNO) \cite{JUNO}. In cylindrical detectors, the group of PMTs which expect equal rate would follow more complex curves determined by the intersection of spheres centered at the vertex with the cylindrical array. 

\bmhead{Acknowledgments}

The authors would like to acknowledge the helpful discussions with Ana Martina Botti. Furthermore, the authors are grateful to the anonymous reviewer, whose comments helped to improve this paper. The authors also acknowledge support by the High Performance and Cloud Computing Group at the Zentrum f\"{u}r Datenverarbeitung of the University of T\"{u}bingen, the state of Baden-W\"{u}rttemberg through bwHPC and the German Research Foundation (DFG) through grant no INST 37/935-1 FUGG.

\noindent

\begin{appendices}

\section{Detector simulation}\label{sec:app}

We start by determining the start-time of the signal in the Auger UMD trace. For this we sample the distribution of the delays between the start-time of the Auger UMD signal and the trigger in its paired surface detector. The delay is random because the trigger time in the surface detector is determined by the timing of both the electromagnetic and muonic components, while the start-time of the UMD signal is only associated to the muonic component. Additionally, both are subject to Poission fluctuations. Furthermore, the delay depends on the type of trigger, which have different sensitivities to the different air-shower components. There are essentially two types of triggers to which the recorded data belong: threshold triggers and time-over-threshold triggers. We sample the start time of the signal at the muon counter from a double Gaussian distribution $0.27 \times \mathcal{N}(3337\,\text{ns},\, 91\,\text{ns}) + 0.73 \times \mathcal{N}(3757\,\text{ns},\, 70\,\text{ns})$ \cite{BottiThesis}. The first Gaussian corresponds to the normal delay between the time-over-threshold trigger in the paired surface detector station and the signal start-time in the Auger UMD trace, whereas the second corresponds to that of threshold triggers.

For each impinging muon, a scintillator strip and channel are assigned by sampling a discrete uniform distribution $\mathcal{U}\{1,64\}$. An impinging position in the scintillator strip $l$ is assigned sampling a continuous uniform distribution $\mathcal{U}\{0\,\text{m}, 4.01\,\text{m}\}$. The average number of photo-electrons $\left\langle N_{\text{PE}} \right \rangle$ at the SiPM generated by the muon at $l$ is computed from \cite{Botti_2021} 
\begin{equation}
\left\langle N_{\text{PE}} \right \rangle (l) = 17.4\, e^{-l/4.16\,\text{m}} + (1.0 - 17.4) \, e^{-l/0.037\,\text{m}}.
\end{equation}
Then the actual number of photo-electrons at the SiPM generated by the muon $N_{\text{PE}}$ is calculated by sampling a Poisson distribution of mean $\left\langle N_{\text{PE}} \right \rangle$.

We compute the propagation time from the impinging position in the optic fiber to the SiPM by dividing the distance $l$ by the speed of light in the optic fiber $0.60\,c$. For each photo-electron, we sample the time delays due to the scintillator and to the optic fibers from two exponential distributions of parameters $3.7\,\text{ns}$ and $3.5\,\text{ns}$ respectively \cite{Botti_2021}.

Moreover, we analytically simulate the electronics response to each photo-electron pulse. The electronics consist of a pre-amplifier, a fast shaper, a discriminator, and a Field-Programmable Gate Array (FPGA). As mentioned in Sec.~\ref{sec:detsim}, while the mathematical models for the photo-electron pulse as well as for each element of the electronics are extracted from Ref.~\cite{BottiThesis}, we present the analytical solution of the electronics response to a photo-electron signal. For this we neglect baseline noise\footnote{Accounting for all sources, background noise is estimated to be responsible for $5.5\,\%$ of the counts of a module trace \cite{BottiThesis}. The count rates from background muons and background noise are inherent to the environment and to the detector, and affect the optimal pattern-matching strategy. However, the background is not relevant for assessing the optimal counting strategy, since the latter takes as input the already matched patterns.}. 

The pulse generated by a photo-electron constitutes an input signal in the electronics $V_{\text{PE}}(t)$ modelled as \cite{Botti_2021}
\begin{align}
V_{\text{PE}}(t) =& 0.29\,\text{mV} \left( 1 - e^{-\frac{t-t_{\text{PE}}}{3.82\,\text{ns}}} \right) \, \times \nonumber \\
&  \left[ 23.22\, e^{-\frac{t-t_{\text{PE}}}{1.187\,\text{ns}}} + 1.609\, e^{-\frac{t-t_{\text{PE}}}{23.44\,\text{ns}}} + \right. \nonumber \\
& \left. \, e^{-\frac{t-t_{\text{PE}}}{0.221\,\text{ns}}}\right] \times \Theta(t-t_{\text{PE}}), \nonumber \\
 =& \sum_{i=1}^{6} A_i \, e^{-\frac{t-t_{\text{PE}}}{\tau_i}} \, \Theta(t-t_{\text{PE}}).
\end{align} 

The signal is then processed through a pre-amplifier, which is modelled as a low-pass filter \cite{BottiThesis}. To obtain the amplified signal $V_{\text{amp}}(t)$ we first Fourier transform $V_{\text{PE}}(t)$, multiply by the transfer function of the low-pass filter $H_I(\omega)$, and inverse transform
\begin{align}
V_{\text{amp}}(t) &= \sqrt{2\pi}\,\mathcal{F}^{-1} \left[ \mathcal{F}[V_{\text{PE}}(t)] \times H_I(\omega) \right], \nonumber \\
& = \sqrt{2\pi}\,\sum_{i=1}^{6} \mathcal{F}^{-1} \left[\frac{A_i}{\sqrt{2\pi}} \frac{e^{-i t_{\text{PE}} \omega}}{(1/\tau_i + i \omega)} \times \nonumber \right.\\
& \left. \hspace{1em} \frac{k_I}{\sqrt{2\pi}(1 + i \omega \tau_I)} \right],\nonumber \\
&= \sum_{i=1}^{6} \frac{k_I A_i \tau_i}{\tau_i - \tau_I} \left( e^{-\frac{t-t_{\text{PE}}}{\tau_i}} - e^{-\frac{t-t_{\text{PE}}}{\tau_I}} \right) \,\times \nonumber \\
& \hspace{1em} \Theta(t-t_{\text{PE}}) \nonumber,\\
& = \sum_{i=1}^{6} \sum_{j=1}^{2} \tilde{A}_{i,j} e^{-\frac{t-t_{\text{PE}}}{\tau_{i,j}}} \Theta(t-t_{\text{PE}}),
\end{align}
where $k_I = -17.5$ and $\tau_I = 17\,\text{ns}$ \cite{Botti_2021}. We use the fact that the Fourier transform commutes with the sum. In the last line we grouped the coefficients such that $\tilde{A}_{i,1} = k_I A_i \tau_i/(\tau_i - \tau_I)$, $\tilde{A}_{i,2} = -\tilde{A}_{i,1}$, $\tau_{i,1} = \tau_{i}$, and $\tau_{i,2} = \tau_{I}$. From the last expression it is easy to notice that the mathematical form of $V_{\text{amp}}(t)$ is equivalent to that of $V_{\text{PE}}(t)$. 

After the pre-amplifier, the signal is processed by a fast-shaper. This is modelled as a practical differentiator \cite{Botti_2021}. To obtain the signal after the fast-shaper $V_{\text{fs}} (t)$ we transform Laplace $V_{\text{amp}}(t)$, multiply by the transfer function of a practical differentiator $H_{II}(s)$, and inverse transform
%
\begin{align}
V_{\text{fs}} (t) =& \mathcal{L}^{-1} \left\lbrace \mathcal{L}[V_{\text{amp}}(t)] \times H_{II}(s) \right\rbrace,\nonumber \\ 
=& \sum_{i=1}^{6}\sum_{j=1}^{2} \mathcal{L}^{-1} \left\lbrace \frac{\tilde{A}_{i,j} e^{\frac{t_{\text{PE}}}{\tau_{i,j}}}}{s+1/\tau_{i,j}} \times \right. \nonumber \\
& \left[ \Theta(t_{\text{PE}}) \, e^{-t_{\text{PE}}(s+1/\tau_{i,j})} + \Theta(-t_{\text{PE}}) \right]  \times \nonumber \\
& \left. \frac{(-s) k_{II} \tau_{II}}{(1+s \tau_{II})^2} \right\rbrace, \nonumber \\
& = \sum_{i=1}^{6} \sum_{j=1}^{2} \frac{-\tilde{A}_{i,j} k_{II} \tau_{i,j}}{\tau_{II} (\tau_{i,j} - \tau_{II})^2} \times \nonumber \\
& \left\lbrace e^{-\frac{t -t_{\text{PE}}}{\tau_{II}}} \left[ (t -t_{\text{PE}}) (\tau_{i,j} - \tau_{II}) + \tau_{II}^2 \right] -\right. \nonumber \\
& \left.  e^{-\frac{t -t_{\text{PE}}}{\tau_{i,j}}} \tau_{II}^2 \right\rbrace \Theta(t-t_{\text{PE}}),
\end{align}
where $k_{II} = 47.1$ and $\tau_{II} = 2.4\,\text{ns}$.

After adding the contribution to $V_{\text{fs}} (t)$ of all photo-electron pulses of each of the muons falling into the same scintillator strip, we can finally simulate the response of the discriminator and FPGA, obtaining the output signal or binary trace $V_{\text{out}}(t)$ for each channel. The output of the FPGA is a 1 in the binary trace if $V_{\text{fs}}(t)$ is above the threshold ($77.5\,\text{mV}$) for more than $1.51\,\text{ns}$. The FPGA also samples the signal in $3.125\,\text{ns}$ time intervals. At last, we repeat the process for all the channels, and obtain the final event trace for a module. 

\section{Analysis of the saturation fraction}\label{sec:appsat}

In order to validate our simulations, we compare the fraction of saturated events in simulations against its expectation value. 

We predict the expected saturation fraction as a function of $\mu$ an of $N_\mu$ for the infinite window strategy just from the adopted statistical model. For all the other strategies, we would need to introduce an assumption of how $\mu$ depends on the time $t$. 

The expected saturation fraction as a function of $\mu$ follows a binomial distribution with the number of successes ($k$) equals the number of trials ($n_s$)
\begin{equation}
B(n_s \vert\, \mu) = e^{-\mu} (e^{\mu/n_s} -1)^{n_s}.
\label{eq:binom}
\end{equation}

Furthermore, the expected saturation fraction as a function of $N_\mu$ follows the occupancy distribution as described in Ref.~\cite{Supanitsky2021} (see Eq.~(\ref{eq:occ})), also evaluated at $k=n_s$
\begin{equation}
Occ\,(n_s \vert\, N_{\mu}, n_s) = \frac{S(N_{\mu}, n_s)}{n_s^{N_{\mu}}},
\end{equation}
where we remind the reader that $S(N_{\mu}, k)$ are the Stirling numbers of the second kind. 

For this analysis we used the first simulations set (see Sec.~\ref{sec:sim}) of air-showers with uniform distribution in logarithmic distance to the shower axis. Then for all air-showers of belonging to a same $(\log_{10}(E/\text{eV}), \sin^2\theta)$-bin, we computed the fraction of saturated events as a function of $\mu$ and as a function of $N_\mu$. We remind the reader that the detector is said to be saturated in an event if, for any window $j$, all channels are occupied (infinite window, N-bin, and N-bin centered strategies) or all channels are either occupied or inhibited (1-bin strategy). In this case both $\widehat{\mu}$ and $\widehat{N}_{\mu}$ tend to infinity.

\begin{figure}[!ht]
  \centering
  \includegraphics[width=0.45\textwidth]{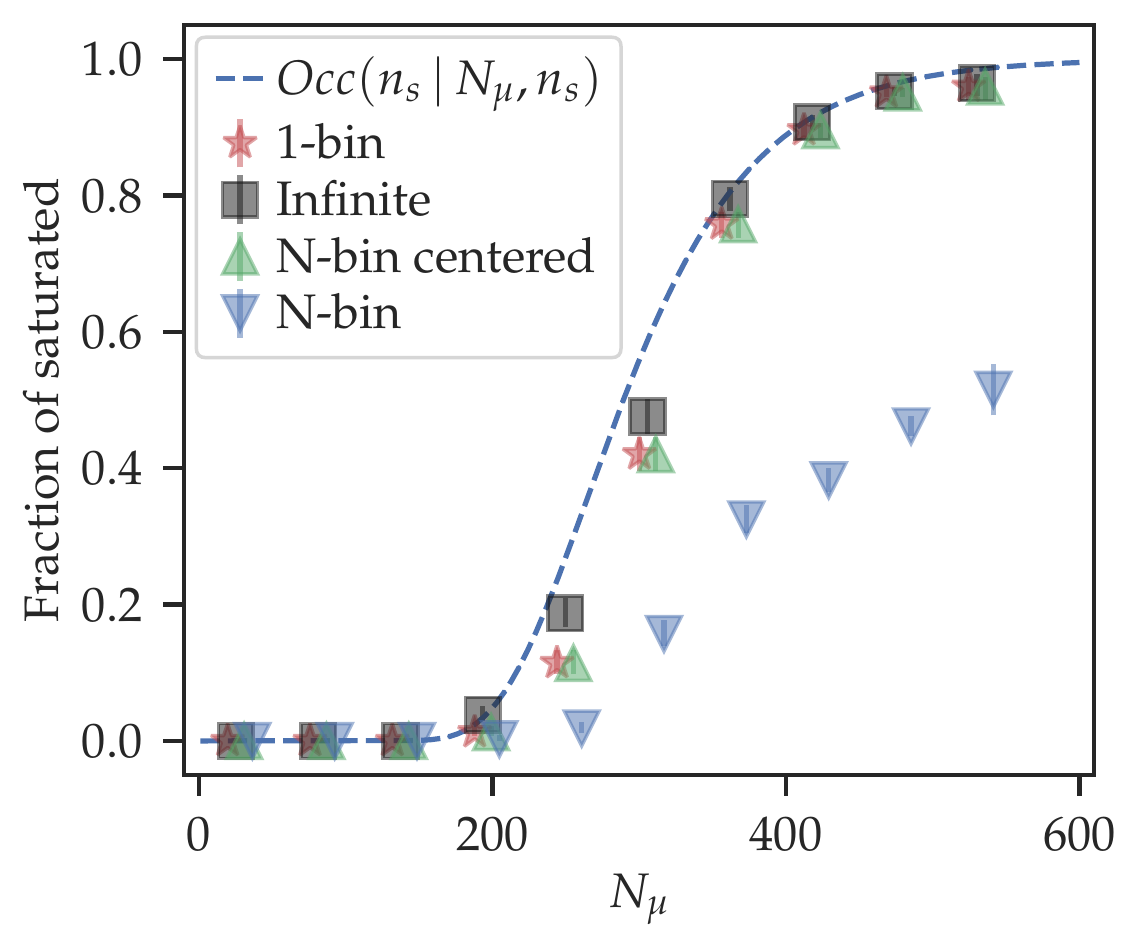}
  \includegraphics[width=0.45\textwidth]{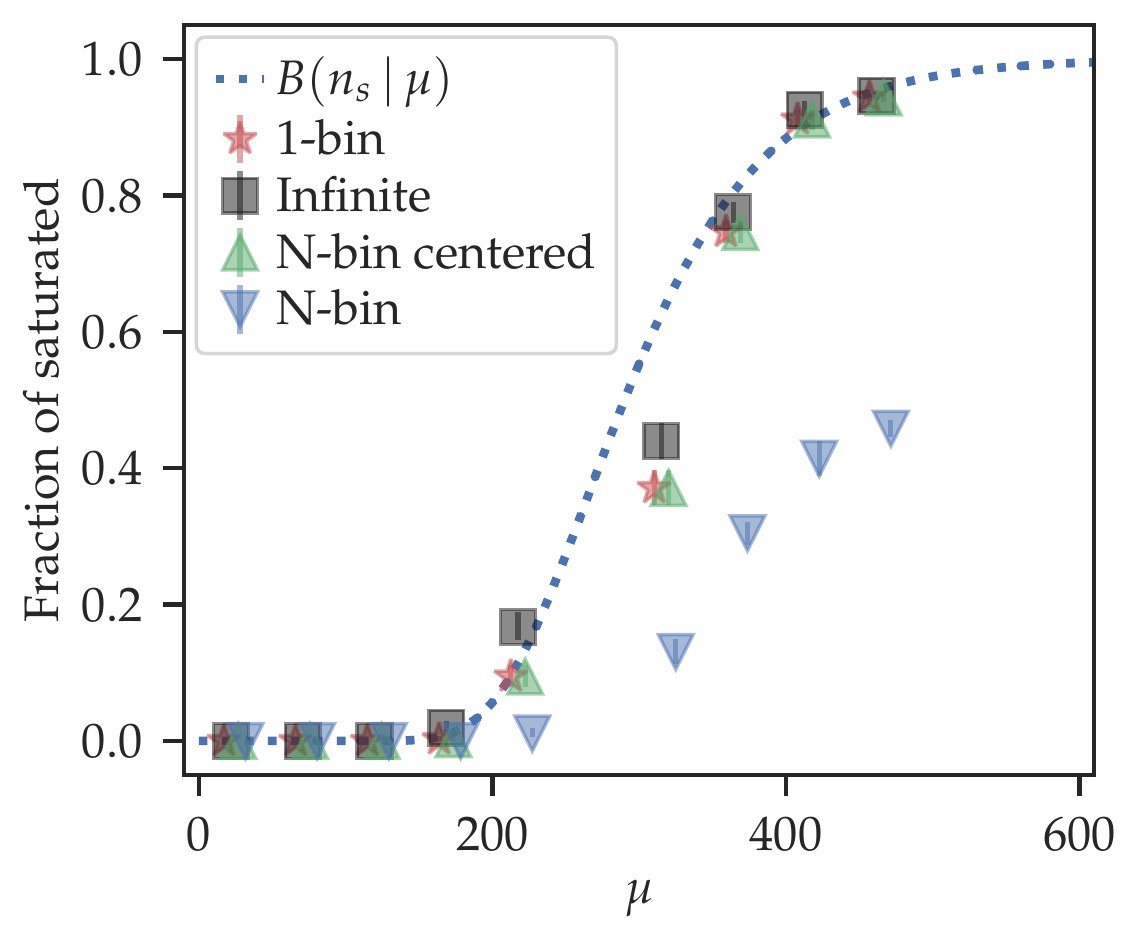}
\caption{Saturation fraction as a function of $N_\mu$ (upper panel) and as a function of $\mu$ (lower panel), as modeled (dashed and dotted lines) and from simulations (markers), for the four considered strategies. The simulations consist of proton air-showers with $17.8 \leq \log_{10}(E/\text{eV}) \leq 18.0$ and $ 18^\circ \lesssim \theta \lesssim 27^\circ$ ($0.10 \leq \sin^2\theta \leq 0.20$), with a uniform distribution in the logarithmic distance to the shower axis.}
\label{fig:sat}
\end{figure}

The expected saturation fraction and the one in simulations for each of the four counting strategies can be seen in Fig.~\ref{fig:sat}. In the example shown, we used proton air-showers with $17.8 \leq \log_{10}(E/\text{eV}) \leq 18.0$ and $ 18^\circ \lesssim \theta \lesssim 27^\circ$ ($0.10 \leq \sin^2\theta \leq 0.20$). For other energies and zenith angles, the general behavior is the same: the models are exactly the same, but the values that $\mu$ or $N_\mu$ can reach are different. We can see that the saturation fraction of the infinite window strategy follows very well the models, and that even the N-bin centered and 1-bin strategies follow them approximately too. The latter is expected for the N-bin centered strategy, because for the Auger UMD most of the signal is contained in one window, and therefore the strategy performs effectively as the infinite window. The N-bin strategy saturates at larger values. However, this is just a consequence of the design problem of the strategy, which does not account for inhibited channels.

\section{Average biases for iron air-showers}\label{sec:appiron}

\begin{figure}[!ht]
\begin{subfigure}{0.44\textwidth}
  \centering
  \includegraphics[width=\textwidth]{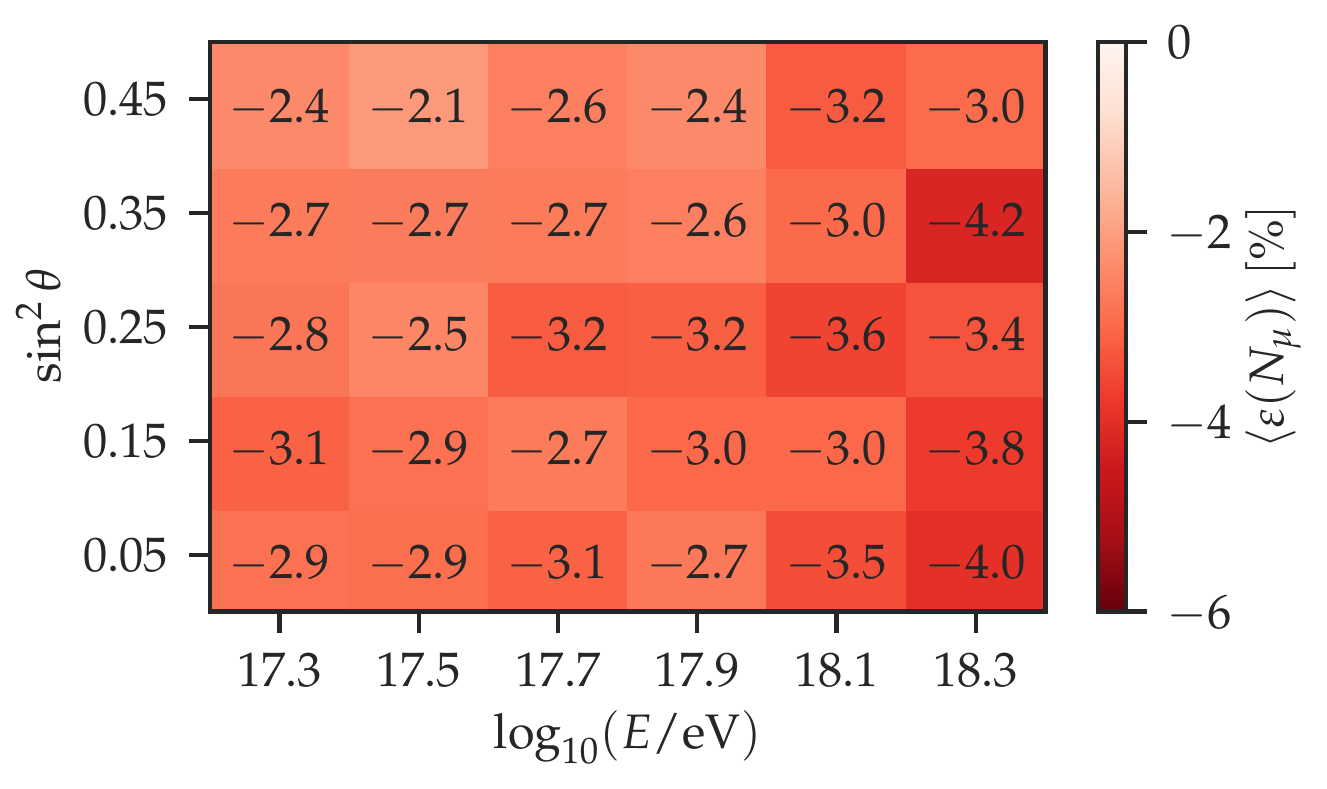}
\end{subfigure}
\begin{subfigure}{0.44\textwidth}
  \centering
  \includegraphics[width=\textwidth]{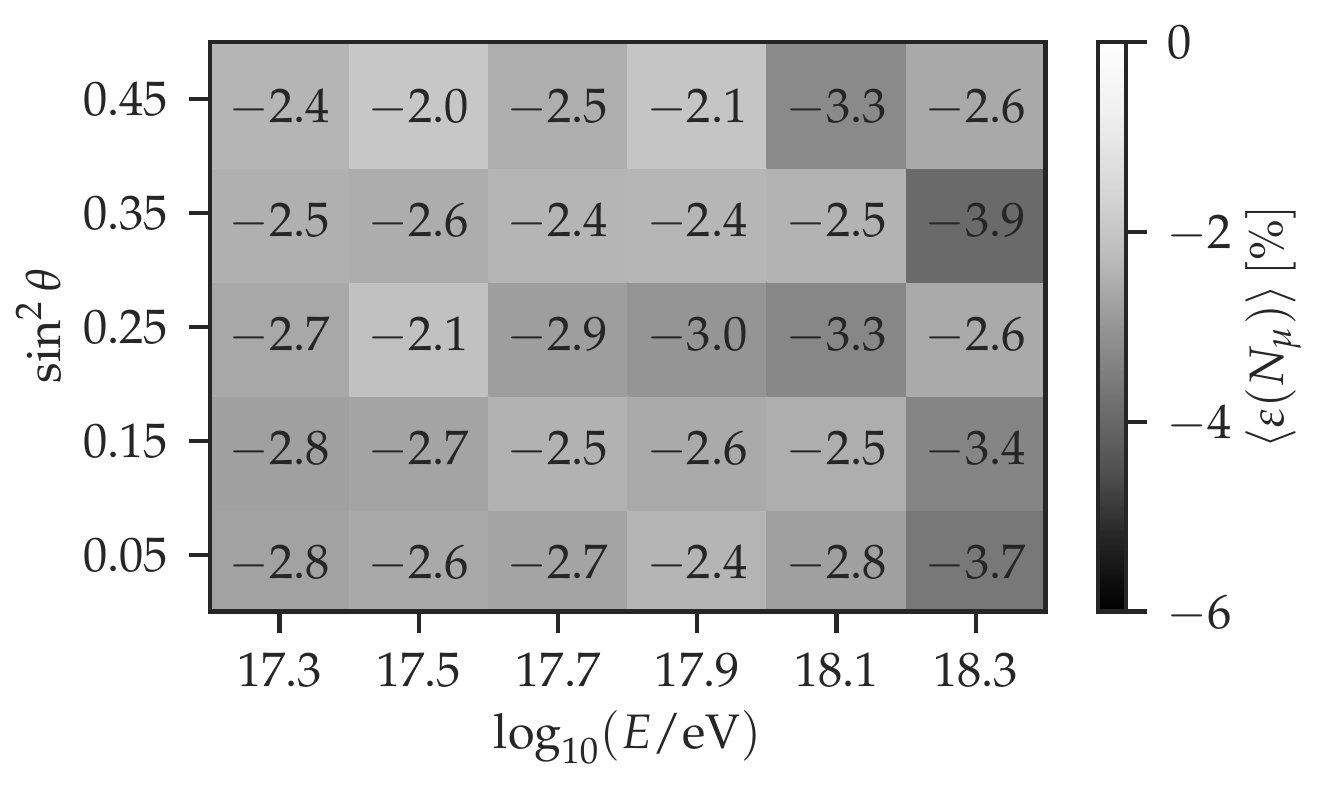}
\end{subfigure}\hspace{0.05\textwidth}
\begin{subfigure}{0.44\textwidth}
  \centering
  \includegraphics[width=\textwidth]{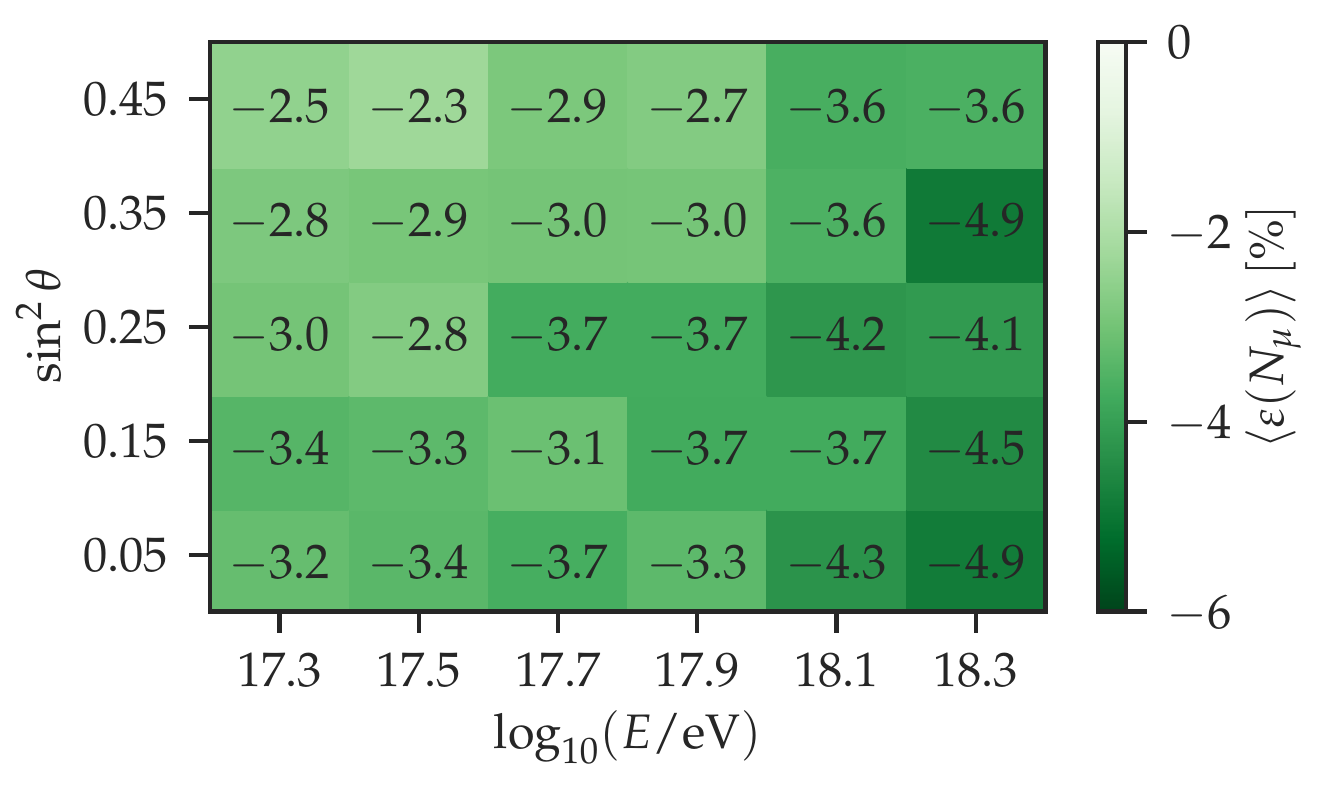}
\end{subfigure}
\begin{subfigure}{0.44\textwidth}
  \centering
  \includegraphics[width=\textwidth]{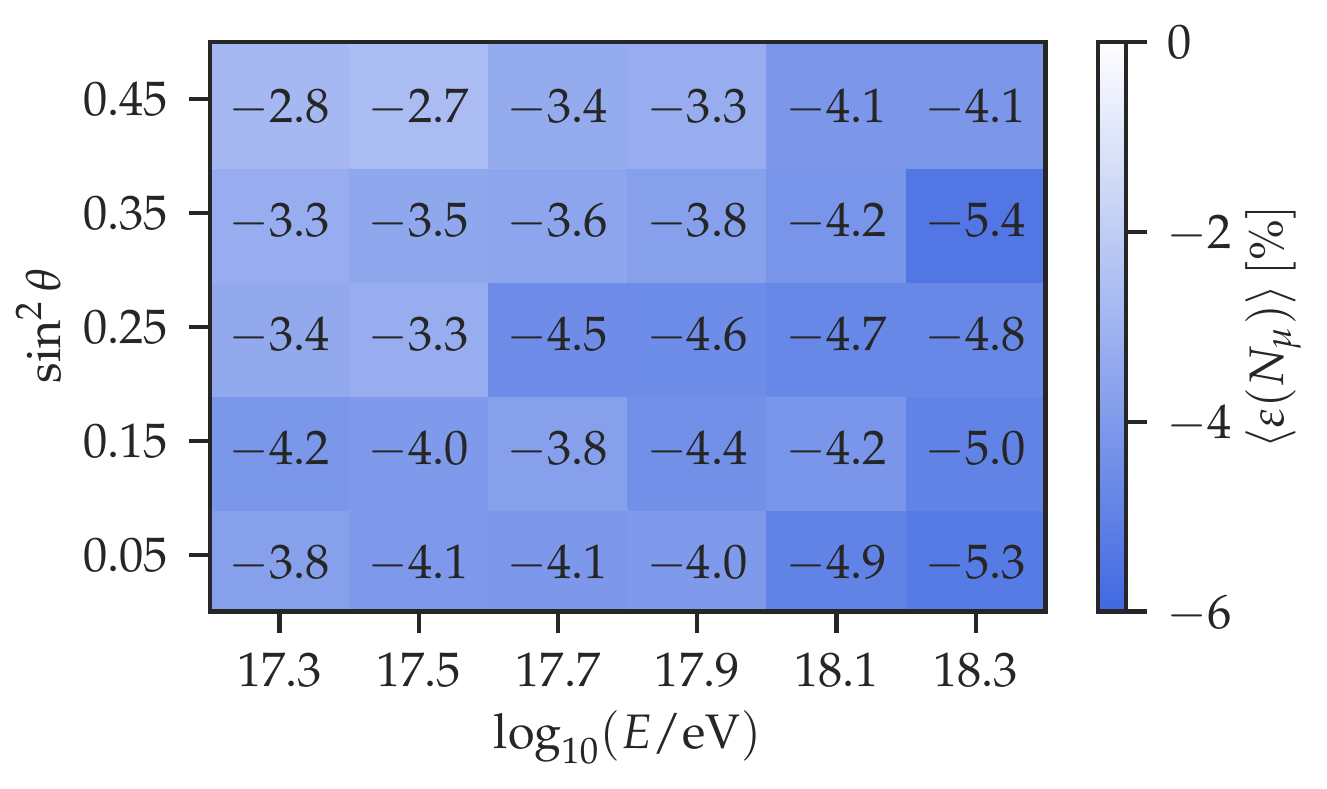}
\end{subfigure}\hspace{0.05\textwidth}
\caption{Same as Fig.~\ref{fig:biasall}, but considering iron showers.}
\label{fig:biasallFe}
\end{figure}

Figure \ref{fig:biasallFe} shows the mean $\varepsilon$ as a function of $\log_{10}(E/\text{eV})$ and $\sin^2\theta$ for all strategies, using iron showers as input, and considering distance cuts. When compared to Fig.~\ref{fig:biasall}, we observe that the mean $\varepsilon$ generally deviate slightly more from zero.

\end{appendices}

\bibliography{sn-bibliography}


\begin{thebibliography}{39}
\ifx \bisbn   \undefined \def \bisbn  #1{ISBN #1}\fi
\ifx \binits  \undefined \def \binits#1{#1}\fi
\ifx \bauthor  \undefined \def \bauthor#1{#1}\fi
\ifx \batitle  \undefined \def \batitle#1{#1}\fi
\ifx \bjtitle  \undefined \def \bjtitle#1{#1}\fi
\ifx \bvolume  \undefined \def \bvolume#1{\textbf{#1}}\fi
\ifx \byear  \undefined \def \byear#1{#1}\fi
\ifx \bissue  \undefined \def \bissue#1{#1}\fi
\ifx \bfpage  \undefined \def \bfpage#1{#1}\fi
\ifx \blpage  \undefined \def \blpage #1{#1}\fi
\ifx \burl  \undefined \def \burl#1{\textsf{#1}}\fi
\ifx \doiurl  \undefined \def \doiurl#1{\url{https://doi.org/#1}}\fi
\ifx \betal  \undefined \def \betal{\textit{et al.}}\fi
\ifx \binstitute  \undefined \def \binstitute#1{#1}\fi
\ifx \binstitutionaled  \undefined \def \binstitutionaled#1{#1}\fi
\ifx \bctitle  \undefined \def \bctitle#1{#1}\fi
\ifx \beditor  \undefined \def \beditor#1{#1}\fi
\ifx \bpublisher  \undefined \def \bpublisher#1{#1}\fi
\ifx \bbtitle  \undefined \def \bbtitle#1{#1}\fi
\ifx \bedition  \undefined \def \bedition#1{#1}\fi
\ifx \bseriesno  \undefined \def \bseriesno#1{#1}\fi
\ifx \blocation  \undefined \def \blocation#1{#1}\fi
\ifx \bsertitle  \undefined \def \bsertitle#1{#1}\fi
\ifx \bsnm \undefined \def \bsnm#1{#1}\fi
\ifx \bsuffix \undefined \def \bsuffix#1{#1}\fi
\ifx \bparticle \undefined \def \bparticle#1{#1}\fi
\ifx \barticle \undefined \def \barticle#1{#1}\fi
\bibcommenthead
\ifx \bconfdate \undefined \def \bconfdate #1{#1}\fi
\ifx \botherref \undefined \def \botherref #1{#1}\fi
\ifx \url \undefined \def \url#1{\textsf{#1}}\fi
\ifx \bchapter \undefined \def \bchapter#1{#1}\fi
\ifx \bbook \undefined \def \bbook#1{#1}\fi
\ifx \bcomment \undefined \def \bcomment#1{#1}\fi
\ifx \oauthor \undefined \def \oauthor#1{#1}\fi
\ifx \citeauthoryear \undefined \def \citeauthoryear#1{#1}\fi
\ifx \endbibitem  \undefined \def \endbibitem {}\fi
\ifx \bconflocation  \undefined \def \bconflocation#1{#1}\fi
\ifx \arxivurl  \undefined \def \arxivurl#1{\textsf{#1}}\fi
\csname PreBibitemsHook\endcsname

\bibitem{AugerPrime2016}
\begin{botherref}
\oauthor{\bsnm{Aab}, \binits{A.}}, et al.:
{The Pierre Auger Observatory Upgrade - Preliminary Design Report}
(2016)
{\href{https://arxiv.org/abs/1604.03637}{{arXiv:1604.03637}}}
{[astro-ph.IM]}
\end{botherref}
\endbibitem

\bibitem{AugerUMD2021}
\begin{barticle}
\bauthor{\bsnm{Aab}, \binits{A.}}, \betal:
\batitle{{Calibration of the underground muon detector of the Pierre Auger
  Observatory}}.
\bjtitle{Journal of Instrumentation}
\bvolume{16}(\bissue{04}),
\bfpage{04003}
(\byear{2021}).
\doiurl{10.1088/1748-0221/16/04/p04003}
\end{barticle}
\endbibitem

\bibitem{Batista2019}
\begin{botherref}
\oauthor{\bsnm{Alves~Batista}, \binits{R.}}, et al.:
{Open Questions in Cosmic-Ray Research at Ultrahigh Energies}.
Frontiers in Astronomy and Space Sciences
\textbf{6}
(2019).
\doiurl{10.3389/fspas.2019.00023}
\end{botherref}
\endbibitem

\bibitem{Mollerach2021}
\begin{barticle}
\bauthor{\bsnm{Mollerach}, \binits{S.}}:
\batitle{{Ultra-High energy cosmic rays}}.
\bjtitle{J. Phys. Conf. Ser.}
\bvolume{2156}(\bissue{1}),
\bfpage{012007}
(\byear{2021}).
\doiurl{10.1088/1742-6596/2156/1/012007}
\end{barticle}
\endbibitem

\bibitem{Engel2011}
\begin{barticle}
\bauthor{\bsnm{Engel}, \binits{R.}},
\bauthor{\bsnm{Heck}, \binits{D.}},
\bauthor{\bsnm{Pierog}, \binits{T.}}:
\batitle{{Extensive Air Showers and Hadronic Interactions at High Energy}}.
\bjtitle{Annual Review of Nuclear and Particle Science}
\bvolume{61}(\bissue{1}),
\bfpage{467}--\blpage{489}
(\byear{2011}).
\doiurl{10.1146/annurev.nucl.012809.104544}
\end{barticle}
\endbibitem

\bibitem{Blumer2009}
\begin{barticle}
\bauthor{\bsnm{Bl\"{u}mer}, \binits{J.}},
\bauthor{\bsnm{Engel}, \binits{R.}},
\bauthor{\bsnm{H\"{o}randel}, \binits{J.R.}}:
\batitle{{Cosmic rays from the knee to the highest energies}}.
\bjtitle{Progress in Particle and Nuclear Physics}
\bvolume{63}(\bissue{2}),
\bfpage{293}--\blpage{338}
(\byear{2009}).
\doiurl{10.1016/j.ppnp.2009.05.002}
\end{barticle}
\endbibitem

\bibitem{Aramburo2021}
\begin{barticle}
\bauthor{\bsnm{Ar\'amburo-Garc\'{\i}a}, \binits{A.}},
\bauthor{\bsnm{Bondarenko}, \binits{K.}},
\bauthor{\bsnm{Boyarsky}, \binits{A.}},
\bauthor{\bsnm{Nelson}, \binits{D.}},
\bauthor{\bsnm{Pillepich}, \binits{A.}},
\bauthor{\bsnm{Sokolenko}, \binits{A.}}:
\batitle{{Ultrahigh energy cosmic ray deflection by the intergalactic magnetic
  field}}.
\bjtitle{Phys. Rev. D}
\bvolume{104},
\bfpage{083017}
(\byear{2021}).
\doiurl{10.1103/PhysRevD.104.083017}
\end{barticle}
\endbibitem

\bibitem{PetersCycle}
\begin{barticle}
\bauthor{\bsnm{{Peters}}, \binits{B.}}:
\batitle{{Primary cosmic radiation and extensive air showers}}.
\bjtitle{Il Nuovo Cimento}
\bvolume{22}(\bissue{4}),
\bfpage{800}--\blpage{819}
(\byear{1961}).
\doiurl{10.1007/BF02783106}
\end{barticle}
\endbibitem

\bibitem{CombinedFitAuger}
\begin{barticle}
\bauthor{\bsnm{Aab}, \binits{A.}}, \betal:
\batitle{{Combined fit of spectrum and composition data as measured by the
  Pierre Auger Observatory}}.
\bjtitle{Journal of Cosmology and Astroparticle Physics}
\bvolume{2017}(\bissue{04}),
\bfpage{038}--\blpage{038}
(\byear{2017}).
\doiurl{10.1088/1475-7516/2017/04/038}
\end{barticle}
\endbibitem

\bibitem{Prado2018}
\begin{bchapter}
\bauthor{\bsnm{{Prado}}, \binits{R.R.}}:
\bctitle{{Tests of hadronic interactions with measurements by Pierre Auger
  Observatory}}.
In: \bbtitle{European Physical Journal Web of Conferences}.
\bsertitle{European Physical Journal Web of Conferences},
vol. \bseriesno{208},
p. \bfpage{08003}
(\byear{2019}).
\doiurl{10.1051/epjconf/201920808003}
\end{bchapter}
\endbibitem

\bibitem{AugerInclinedMuons2015}
\begin{barticle}
\bauthor{\bsnm{Aab}, \binits{A.}}, \betal:
\batitle{{Muons in air showers at the Pierre Auger Observatory: Mean number in
  highly inclined events}}.
\bjtitle{Phys. Rev. D}
\bvolume{91},
\bfpage{032003}
(\byear{2015}).
\doiurl{10.1103/PhysRevD.91.032003}
\end{barticle}
\endbibitem

\bibitem{AugerMuonFluctuations2021}
\begin{barticle}
\bauthor{\bsnm{Aab}, \binits{A.}}, \betal:
\batitle{{Measurement of the Fluctuations in the Number of Muons in Extensive
  Air Showers with the Pierre Auger Observatory}}.
\bjtitle{Phys. Rev. Lett.}
\bvolume{126},
\bfpage{152002}
(\byear{2021}).
\doiurl{10.1103/PhysRevLett.126.152002}
\end{barticle}
\endbibitem

\bibitem{Supanitsky2008}
\begin{barticle}
\bauthor{\bsnm{Supanitsky}, \binits{A.D.}},
\bauthor{\bsnm{Etchegoyen}, \binits{A.}},
\bauthor{\bsnm{Medina-Tanco}, \binits{G.}},
\bauthor{\bsnm{Allekotte}, \binits{I.}},
\bauthor{\bsnm{{G\'{o}mez Berisso}}, \binits{M.}},
\bauthor{\bsnm{Medina}, \binits{M.C.}}:
\batitle{{Underground muon counters as a tool for composition analyses}}.
\bjtitle{Astroparticle Physics}
\bvolume{29}(\bissue{6}),
\bfpage{461}--\blpage{470}
(\byear{2008}).
\doiurl{10.1016/j.astropartphys.2008.05.003}
\end{barticle}
\endbibitem

\bibitem{Arqueros2008}
\begin{barticle}
\bauthor{\bsnm{Arqueros}, \binits{F.}},
\bauthor{\bsnm{H\"{o}randel}, \binits{J.R.}},
\bauthor{\bsnm{Keilhauer}, \binits{B.}}:
\batitle{{Air fluorescence relevant for cosmic-ray detection—Summary of the
  5th fluorescence workshop, El Escorial 2007}}.
\bjtitle{Nuclear Instruments and Methods in Physics Research Section A:
  Accelerators, Spectrometers, Detectors and Associated Equipment}
\bvolume{597}(\bissue{1}),
\bfpage{1}--\blpage{22}
(\byear{2008}).
\doiurl{10.1016/j.nima.2008.08.056}.
\bcomment{Proceedings of the 5th Fluorescence Workshop}
\end{barticle}
\endbibitem

\bibitem{Book}
\begin{bbook}
\bauthor{\bsnm{{T.~K. Gaisser, R. Engel, and E. Resconi}}}:
\bbtitle{{Cosmic Rays and Particle Physics}},
\bedition{2}nd edn.
\bpublisher{Cambridge University Press},
\blocation{Cambridge}
(\byear{2016}).
\doiurl{10.1017/CBO9781139192194}
\end{bbook}
\endbibitem

\bibitem{Taboada2020}
\begin{barticle}
\bauthor{\bsnm{Taboada}, \binits{A.}}:
\batitle{{Analysis of Data from Surface Detector Stations of the AugerPrime
  Upgrade}}.
\bjtitle{PoS}
\bvolume{ICRC2019},
\bfpage{434}
(\byear{2020}).
\doiurl{10.22323/1.358.0434}
\end{barticle}
\endbibitem

\bibitem{Cataldi2021}
\begin{bchapter}
\bauthor{\bsnm{Cataldi}, \binits{G.}}:
\bctitle{{The upgrade of the Pierre Auger Observatory with the Scintillator
  Surface Detector}},
p. \bfpage{251}
(\byear{2021}).
\doiurl{10.22323/1.395.0251}
\end{bchapter}
\endbibitem

\bibitem{TelescopeArray2012}
\begin{barticle}
\bauthor{\bsnm{Abu-Zayyad}, \binits{T.}}, \betal:
\batitle{{The surface detector array of the Telescope Array experiment}}.
\bjtitle{Nuclear Instruments and Methods in Physics Research Section A:
  Accelerators, Spectrometers, Detectors and Associated Equipment}
\bvolume{689},
\bfpage{87}--\blpage{97}
(\byear{2012}).
\doiurl{10.1016/j.nima.2012.05.079}
\end{barticle}
\endbibitem

\bibitem{Hayashida1995}
\begin{barticle}
\bauthor{\bsnm{Hayashida}, \binits{N.}}, \betal:
\batitle{{Muons ({$\geqslant 1\,$}GeV) in large extensive air showers of
  energies between {$10^{16.5}\,$}eV and {$10^{19.5}\,$}eV observed at Akeno}}.
\bjtitle{J. Phys. G: Nucl. Part. Phys.}
\bvolume{21},
\bfpage{1101}--\blpage{1119}
(\byear{1995})
\end{barticle}
\endbibitem

\bibitem{KASCADEmuonsize2006}
\begin{bbook}
\bauthor{\bsnm{{KASCADE-Grande Collaboration}}},
\bauthor{\bparticle{van} \bsnm{Buren}, \binits{J.}}:
\bbtitle{{Muon size spectrum measured by KASCADE-Grande}},
pp. \bfpage{13}--\blpage{16}
(\byear{2006}).
\bcomment{51.04.01; LK 01}
\end{bbook}
\endbibitem

\bibitem{Ravignani2014}
\begin{botherref}
\oauthor{\bsnm{Ravignani}, \binits{D.}},
\oauthor{\bsnm{Supanitsky}, \binits{A.D.}}:
{A new method for reconstructing the muon lateral distribution with an array of
  segmented counters}.
Astroparticle Physics
\textbf{65}
(2014).
\doiurl{10.1016/j.astropartphys.2014.11.007}
\end{botherref}
\endbibitem

\bibitem{AugerMueller2020}
\begin{barticle}
\bauthor{\bsnm{Aab}, \binits{A.}}, \betal:
\batitle{{Direct measurement of the muonic content of extensive air showers
  between $2\times 10^{17}$ and $2\times 10^{18}~$eV at the Pierre Auger
  Observatory}}.
\bjtitle{Eur. Phys. J. C}
\bvolume{80}(\bissue{8}),
\bfpage{751}
(\byear{2020}).
\doiurl{10.1140/epjc/s10052-020-8055-y}
\end{barticle}
\endbibitem

\bibitem{Ravignani2016}
\begin{barticle}
\bauthor{\bsnm{Ravignani}, \binits{D.}},
\bauthor{\bsnm{Supanitsky}, \binits{A.D.}},
\bauthor{\bsnm{Melo}, \binits{D.}}:
\batitle{{Reconstruction of air shower muon densities using segmented counters
  with time resolution}}.
\bjtitle{Astroparticle Physics}
\bvolume{82},
\bfpage{108}--\blpage{116}
(\byear{2016}).
\doiurl{10.1016/j.astropartphys.2016.06.001}
\end{barticle}
\endbibitem

\bibitem{Mueller2018}
\begin{barticle}
\bauthor{\bsnm{{S. M\"uller for the Pierre Auger Collaboration}}}:
\batitle{{Direct Measurement of the Muon Density in Air Showers with the Pierre
  Auger Observatory}}.
\bjtitle{EPJ Web Conf.}
\bvolume{210},
\bfpage{02013}
(\byear{2019}).
\doiurl{10.1051/epjconf/201921002013}
\end{barticle}
\endbibitem

\bibitem{BottiThesis}
\begin{botherref}
\oauthor{\bsnm{Botti}, \binits{A.M.}}:
{Determination of the chemical composition of cosmic rays in the energy region
  of 5 EeV with the AMIGA upgrade of the Pierre Auger Observatory}.
PhD thesis,
KIT, Karlsruhe
(2019).
\doiurl{10.5445/IR/1000100543}
\end{botherref}
\endbibitem

\bibitem{ONeill2021}
\begin{barticle}
\bauthor{\bsnm{O{\textquotesingle}Neill}, \binits{B.}}:
\batitle{{The Classical Occupancy Distribution: Computation and
  Approximation}}.
\bjtitle{The American Statistician}
\bvolume{75}(\bissue{4}),
\bfpage{364}--\blpage{375}
(\byear{2020}).
\doiurl{10.1080/00031305.2019.1699445}
\end{barticle}
\endbibitem

\bibitem{Gaisser1985}
\begin{barticle}
\bauthor{\bsnm{Gaisser}, \binits{T.K.}},
\bauthor{\bsnm{Stanev}, \binits{T.}}:
\batitle{{Muon bundles in underground detectors}}.
\bjtitle{Nuclear Instruments and Methods in Physics Research Section A:
  Accelerators, Spectrometers, Detectors and Associated Equipment}
\bvolume{235}(\bissue{1}),
\bfpage{183}--\blpage{192}
(\byear{1985}).
\doiurl{10.1016/0168-9002(85)90260-8}
\end{barticle}
\endbibitem

\bibitem{Supanitsky2021}
\begin{barticle}
\bauthor{\bsnm{Supanitsky}, \binits{A.D.}}:
\batitle{{Estimation of the number of muons with muon counters}}.
\bjtitle{Astroparticle Physics}
\bvolume{127},
\bfpage{102535}
(\byear{2021}).
\doiurl{10.1016/j.astropartphys.2020.102535}
\end{barticle}
\endbibitem

\bibitem{Botti_2021}
\begin{barticle}
\bauthor{\bsnm{Botti}, \binits{A.M.}},
\bauthor{\bsnm{S{\'{a}}nchez}, \binits{F.}},
\bauthor{\bsnm{Roth}, \binits{M.}},
\bauthor{\bsnm{Etchegoyen}, \binits{A.}}:
\batitle{{Development and validation of the signal simulation for the
  underground muon detector of the Pierre Auger Observatory}}.
\bjtitle{Journal of Instrumentation}
\bvolume{16}(\bissue{07}),
\bfpage{07059}
(\byear{2021}).
\doiurl{10.1088/1748-0221/16/07/p07059}
\end{barticle}
\endbibitem

\bibitem{Platino2011}
\begin{barticle}
\bauthor{\bsnm{Platino}, \binits{M.}},
\bauthor{\bsnm{Hampel}, \binits{M.R.}},
\bauthor{\bsnm{Almela}, \binits{A.}},
\bauthor{\bsnm{Krieger}, \binits{A.}},
\bauthor{\bsnm{Gorbe{\~{n}}a}, \binits{D.}},
\bauthor{\bsnm{Ferrero}, \binits{A.}},
\bauthor{\bsnm{Vega}, \binits{G.D.L.}},
\bauthor{\bsnm{Lucero}, \binits{A.}},
\bauthor{\bsnm{Suarez}, \binits{F.}},
\bauthor{\bsnm{Videla}, \binits{M.}},
\bauthor{\bsnm{Wainberg}, \binits{O.}},
\bauthor{\bsnm{Etchegoyen}, \binits{A.}}:
\batitle{{{AMIGA} at the Auger Observatory: the scintillator module testing
  system}}.
\bjtitle{Journal of Instrumentation}
\bvolume{6}(\bissue{06}),
\bfpage{06006}--\blpage{06006}
(\byear{2011}).
\doiurl{10.1088/1748-0221/6/06/p06006}
\end{barticle}
\endbibitem

\bibitem{Epos}
\begin{barticle}
\bauthor{\bsnm{{T. Pierog}}},
\bauthor{\bsnm{{Iu. Karpenko}}},
\bauthor{\bsnm{{J. M. Katzy}}},
\bauthor{\bsnm{{E. Yatsenko}}},
\bauthor{\bsnm{{K. Werner}}}:
\batitle{{EPOS LHC: Test of collective hadronization with data measured at the
  CERN Large Hadron Collider}}.
\bjtitle{Phys. Rev.}
\bvolume{C92}(\bissue{3}),
\bfpage{034906}
(\byear{2015})
{\href{https://arxiv.org/abs/1306.0121}{{arXiv:1306.0121}}}
{[hep-ph]}.
\doiurl{10.1103/PhysRevC.92.034906}
\end{barticle}
\endbibitem

\bibitem{UrQMD1}
\begin{barticle}
\bauthor{\bsnm{Bass}, \binits{S.A.}},
\bauthor{\bsnm{Belkacem}, \binits{M.}},
\bauthor{\bsnm{Bleicher}, \binits{M.}},
\bauthor{\bsnm{Brandstetter}, \binits{M.}},
\bauthor{\bsnm{Bravina}, \binits{L.}},
\bauthor{\bsnm{Ernst}, \binits{C.}},
\bauthor{\bsnm{Gerland}, \binits{L.}},
\bauthor{\bsnm{Hofmann}, \binits{M.}},
\bauthor{\bsnm{Hofmann}, \binits{S.}},
\bauthor{\bsnm{Konopka}, \binits{J.}},
\bauthor{\bsnm{Mao}, \binits{G.}},
\bauthor{\bsnm{Neise}, \binits{L.}},
\bauthor{\bsnm{Soff}, \binits{S.}},
\bauthor{\bsnm{Spieles}, \binits{C.}},
\bauthor{\bsnm{Weber}, \binits{H.}},
\bauthor{\bsnm{Winckelmann}, \binits{L.A.}},
\bauthor{\bsnm{Stöcker}, \binits{H.}},
\bauthor{\bsnm{Greiner}, \binits{W.}},
\bauthor{\bsnm{Hartnack}, \binits{C.}},
\bauthor{\bsnm{Aichelin}, \binits{J.}},
\bauthor{\bsnm{Amelin}, \binits{N.}}:
\batitle{{Microscopic models for ultrarelativistic heavy ion collisions}}.
\bjtitle{Progress in Particle and Nuclear Physics}
\bvolume{41},
\bfpage{255}--\blpage{369}
(\byear{1998}).
\doiurl{10.1016/S0146-6410(98)00058-1}
\end{barticle}
\endbibitem

\bibitem{UrQMD2}
\begin{barticle}
\bauthor{\bsnm{Bleicher}, \binits{M.}},
\bauthor{\bsnm{Zabrodin}, \binits{E.}},
\bauthor{\bsnm{Spieles}, \binits{C.}},
\bauthor{\bsnm{Bass}, \binits{S.A.}},
\bauthor{\bsnm{Ernst}, \binits{C.}},
\bauthor{\bsnm{Soff}, \binits{S.}},
\bauthor{\bsnm{Bravina}, \binits{L.}},
\bauthor{\bsnm{Belkacem}, \binits{M.}},
\bauthor{\bsnm{Weber}, \binits{H.}},
\bauthor{\bsnm{Stöcker}, \binits{H.}},
\bauthor{\bsnm{Greiner}, \binits{W.}}:
\batitle{{Relativistic hadron-hadron collisions in the ultra-relativistic
  quantum molecular dynamics model}}.
\bjtitle{Journal of Physics G: Nuclear and Particle Physics}
\bvolume{25}(\bissue{9}),
\bfpage{1859}--\blpage{1896}
(\byear{1999}).
\doiurl{10.1088/0954-3899/25/9/308}
\end{barticle}
\endbibitem

\bibitem{Corsika}
\begin{botherref}
\oauthor{\bsnm{Heck}, \binits{D.}},
\oauthor{\bsnm{Knapp}, \binits{J.}},
\oauthor{\bsnm{Capdevielle}, \binits{J.N.}},
\oauthor{\bsnm{Schatz}, \binits{G.}},
\oauthor{\bsnm{Thouw}, \binits{T.}}:
{CORSIKA: A Monte Carlo code to simulate extensive air showers}.
Technical report
(1998).
\doiurl{10.5445/IR/270043064}.
51.02.03; LK 01; Wissenschaftliche Berichte, FZKA-6019 (Februar 98)
\end{botherref}
\endbibitem

\bibitem{AugerSpectrum2020}
\begin{barticle}
\bauthor{\bsnm{Aab}, \binits{A.}}, \betal:
\batitle{{Features of the Energy Spectrum of Cosmic Rays above
  $2.5\ifmmode\times\else\texttimes\fi{}{10}^{18}\text{ }\text{ }\mathrm{eV}$
  Using the Pierre Auger Observatory}}.
\bjtitle{Phys. Rev. Lett.}
\bvolume{125},
\bfpage{121106}
(\byear{2020}).
\doiurl{10.1103/PhysRevLett.125.121106}
\end{barticle}
\endbibitem

\bibitem{Cazon2005}
\begin{barticle}
\bauthor{\bsnm{Caz\'{o}n}, \binits{L.}},
\bauthor{\bsnm{V\'{a}zquez}, \binits{R.A.}},
\bauthor{\bsnm{Zas}, \binits{E.}}:
\batitle{{Depth development of extensive air showers from muon time
  distributions}}.
\bjtitle{Astroparticle Physics}
\bvolume{23}(\bissue{4}),
\bfpage{393}--\blpage{409}
(\byear{2005}).
\doiurl{10.1016/j.astropartphys.2005.01.009}
\end{barticle}
\endbibitem

\bibitem{Wu2019}
\begin{barticle}
\bauthor{\bsnm{Wu}, \binits{W.}},
\bauthor{\bsnm{He}, \binits{M.}},
\bauthor{\bsnm{Zhou}, \binits{X.}},
\bauthor{\bsnm{Qiao}, \binits{H.}}:
\batitle{{A new method of energy reconstruction for large spherical liquid
  scintillator detectors}}.
\bjtitle{Journal of Instrumentation}
\bvolume{14}(\bissue{03}),
\bfpage{03009}--\blpage{03009}
(\byear{2019}).
\doiurl{10.1088/1748-0221/14/03/p03009}
\end{barticle}
\endbibitem

\bibitem{SNOplus}
\begin{botherref}
\oauthor{\bsnm{Andringa}, \binits{S.}}, et al.:
{Current Status and Future Prospects of the SNO+ Experiment}.
Advances in High Energy Physics
\textbf{2016}
(2016)
{\href{https://arxiv.org/abs/1508.05759}{{arXiv:1508.05759}}}.
\doiurl{10.1155/2016/6194250}
\end{botherref}
\endbibitem

\bibitem{JUNO}
\begin{barticle}
\bauthor{\bsnm{Abusleme}, \binits{A.}}, \betal:
\batitle{{JUNO physics and detector}}.
\bjtitle{Progress in Particle and Nuclear Physics}
\bvolume{123},
\bfpage{103927}
(\byear{2022}).
\doiurl{10.1016/j.ppnp.2021.103927}
\end{barticle}
\endbibitem

\end{thebibliography}

\end{document}